\documentclass[11pt]{article}
\usepackage{graphicx,ifthen,color,algorithm,palatino}
\usepackage{amsmath,amsthm,amssymb,amsfonts,verbatim}
\usepackage{mathrsfs,accents,setspace,subfig}
\newtheorem{result}{\textbf{Result}}
\def\bbetatrue{\bbeta_{\mbox{\tiny true}}}
\def\sigsqtrue{\sigma^2_{\mbox{\tiny true}}}
\def\SigmaTrue{\bSigma_{\mbox{\tiny true}}}
\def\SigmadTrue{\bSigma'_{\mbox{\tiny true}}}
\def\OmegaAtwoTwo{\bOmega_4}
\def\bZdash{\bZ^{\prime}}
\def\sSigmaj{s_{\small{\bSigma,j}}}

\def\ssigsq{s_{\sigma^2}}
\def\asigsq{a_{\sigma^2}}
\def\aEps{\asigsq}
\def\MqSigmad{\bM_{q((\bSigma')^{-1})}}
\def\buall{\bu_{\mbox{\tiny all}}}
\def\mysigeps{\sigma}
\def\SolveLeastSquares{\textsc{\footnotesize SolveLeastSquares}}
\def\SolveTwoLevelSparseLeastSquares{\textsc{\footnotesize SolveTwoLevelSparseLeastSquares}}
\def\LargerSolveLeastSquares{\textsc{\normalsize SolveLeastSquares}}
\def\LargerSolveTwoLevelSparseLeastSquares{\textsc{\normalsize SolveTwoLevelSparseLeastSquares}}
\def\budash{\bu^{\prime}}
\def\tinybullet{\tiny\mbox{$\bullet$}}
\def\nidot{n_{i\tinybullet}}
\def\ndotid{n_{\tinybullet i'}}
\def\ndotdot{n_{\tinybullet\tinybullet}}
\def\veryTinyBLUP{\mbox{\fontsize{2.0mm}{1em}\selectfont {\bf BLUP}}}
\def\veryTinyMFVB{\mbox{\fontsize{2.0mm}{1em}\selectfont {\bf MFVB}}}
\def\DBLUP{\bD_{\veryTinyBLUP}}

\def\DtildeMFVB{\bDtilde_{\veryTinyMFVB}}
\def\RBLUP{\bR_{\veryTinyBLUP}}
\def\RMFVB{\bR_{\veryTinyMFVB}}
\def\oMFVB{\bo_{\veryTinyMFVB}}
\def\ruptriMFVB{\ruptri_{\veryTinyMFVB}}

\def\DuptriMFVB{\Duptri_{\veryTinyMFVB}}
\def\RuptriMFVB{\Ruptri_{\veryTinyMFVB}}
\def\ouptriMFVB{\ouptri_{\veryTinyMFVB}}

\def\smalluptri{\mbox{\fontsize{2.3mm}{1em}\selectfont{$\blacktriangle$}}}
\def\smalldntri{\mbox{\fontsize{2.3mm}{1em}\selectfont{$\blacktriangledown$}}}
\def\smallblacksquare{\mbox{\fontsize{1.5mm}{1em}\selectfont{$\blacksquare$}}}

\newcommand*{\uptri}[1]{\accentset{\stackrel{\mbox{$\smalluptri$}}{\null}}{#1}}
\newcommand*{\dntri}[1]{\accentset{\stackrel{\mbox{$\smalldntri$}}{\null}}{#1}}
\newcommand*{\myblacksquare}[1]{\accentset{\stackrel{\mbox{$\smallblacksquare$}}{\null}}{#1}}

\newcommand{\Duptri}{\uptri{\boldsymbol{D}}}
\newcommand{\Ruptri}{\uptri{\boldsymbol{R}}}
\newcommand{\ouptri}{\uptri{\boldsymbol{o}}}
\newcommand{\Cuptri}{\uptri{\boldsymbol{C}}}
\newcommand{\CTuptri}{\uptri{\boldsymbol{C}}^T}

\newcommand{\yuptri}{\uptri{\boldsymbol{y}}}
\newcommand{\ydntri}{\dntri{\boldsymbol{y}}}
\newcommand{\Xuptri}{\uptri{\boldsymbol{X}}}

\newcommand{\Xdntri}{\dntri{\boldsymbol{X}}}

\newcommand{\Zdblacksquare}{\myblacksquare{\boldsymbol{Z}}^{\prime}}
\newcommand{\Zuptri}{\uptri{\boldsymbol{Z}}}

\newcommand{\Zdntri}{\dntri{\boldsymbol{Z}}^{\prime}}

\newcommand{\ruptri}{\uptri{\boldsymbol{r}}^{\prime}}

\def\bSigmad{\bSigma^{\prime}}
\def\lambdasigsq{\lambda_{\sigma^2}}
\def\nusigsq{\nu_{\sigma^2}}
\def\sSigmadqd{s_{\small{\bSigmad,q'}}}
\def\qDens{\mathfrak{q}}
\def\pDens{\mathfrak{p}}

\def\SolveTwoLevelSparseLeastSquares{\textsc{\footnotesize SolveTwoLevelSparseLeastSquares}}

\def\veryTinyBLUP{\mbox{\fontsize{1.5mm}{1em}\selectfont {\bf BLUP}}}
\def\veryTinyMFVB{\mbox{\fontsize{1.5mm}{1em}\selectfont {\bf MFVB}}}
\def\DBLUP{\bD_{\veryTinyBLUP}}

\def\RBLUP{\bR_{\veryTinyBLUP}}
\def\RMFVB{\bR_{\veryTinyMFVB}}
\def\oMFVB{\bo_{\veryTinyMFVB}}

\def\ASigma{A_{\mbox{\tiny$\bSigma$}}}
\def\muq#1{\mu_{\qDens(#1)}}
\def\bmuq#1{\bmu_{\qDens(#1)}}
\def\MqSigma{\bM_{\qDens(\bSigma^{-1})}}

\def\idash{i^{\prime}}
\def\udashidash{\bu^{\prime}_{\idash}}
\def\mdash{m^{\prime}}
\def\Zdash{\bZ^{\prime}}

\def\thickarrow{\longleftarrow}

\def\iStt{i_{\mbox{\tiny stt}}}
\def\iEnd{i_{\mbox{\tiny end}}}

\def\Sigmad{\bSigma^{\prime}}
\def\ASigma{\bA_{\bSigma}}
\def\ASigmad{\bA_{\Sigmad}}
\def\nuSigma{\nu_{\small{\bSigma}}}
\def\nuSigmad{\nu_{\small{\Sigmad}}}
\def\qd{q^{\prime}}
\def\LambdaASigma{\bLambda_{\small{\ASigma}}}
\def\LambdaASigmad{\bLambda_{\small{\ASigmad}}}
\def\sSigmaOne{s_{\small{\bSigma,1}}}
\def\sSigmaq{s_{\small{\bSigma,q}}}
\def\sSigmadOne{s_{\small{\Sigmad,1}}}

\def\Gfull{G_{\mbox{\tiny full}}}
\def\Gdiag{G_{\mbox{\tiny diag}}}
\def\smalldot{\mbox{\fontsize{0.1mm}{0.5em}\selectfont{$\bullet$}}}
\def\nadj{{\tilde n}}
\def\bBdot{\overset{\ \smalldot}{\bB}}
\def\bveci{\boldsymbol{b}_i}
\def\Bmati{\bB_i}
\def\Bmatdoti{\bBdot_i}
\def\xveco{\bx_1}
\def\AUoo{\bA^{11}}
\def\xvectCi{\bx_{2,i}}
\def\AUttCi{\bA^{22,i}}
\def\AUotCi{\bA^{12,i}} 
\def\Bmato{\bB_1}
\def\Bmatdoto{\bBdot_1}
\def\Bmatt{\bB_2}
\def\Bmatdott{\bBdot_2}
\def\Bmatm{\bB_m}
\def\Bmatdotm{\bBdot_m}
\def\bveco{\boldsymbol{b}_1}
\def\bvect{\boldsymbol{b}_2}
\def\bvecm{\boldsymbol{b}_m}
\def\AtLev{\bA}
\def\AUotCo{\bA^{12,1}} 
\def\AUotCt{\bA^{12,2}}
\def\AUotCm{\bA^{12,m}}
\def\AUotCoT{\bA^{12,1\,T}}
\def\AUttCo{\bA^{22,1}}
\def\bigX{{\LARGE\mbox{$\times$}}}
\def\AUotCtT{\bA^{12,2\,T}}
\def\AUttCt{\bA^{22,2}}
\def\AUotCmT{\bA^{12,m\,T}}
\def\AUttCm{\bA^{22,m}}
\def\cveczi{\bc_{0i}}
\def\Cmatzi{\bC_{0i}}
\def\cvecoi{\bc_{1i}}
\def\cvecti{\bc_{2i}}
\def\Cmatoi{\bC_{1i}}
\def\Cmatti{\bC_{2i}}
\newboolean{UnBlinded}
\newboolean{DoubleSpaced}
\newboolean{ColourVersion}
\setboolean{UnBlinded}{true}
\setboolean{DoubleSpaced}{false}
\setboolean{ColourVersion}{false}
\newcommand*{\vcenteredhbox}[1]{\begin{tabular}{@{}c@{}}#1\end{tabular}}
\def\tcb#1{\textcolor{black}{#1}}
\def\tcr#1{\textcolor{black}{#1}}

\def\tcm#1{\textcolor{black}{#1}}
\definecolor{ruppertgreen}{rgb}{0,.7,.25}
\def\tcrg#1{\textcolor{black}{#1}}
\def\scoreiidj{\mbox{\texttt{score}}_{ii'j}}
\def\ageiidj{\mbox{\texttt{age}}_{ii'j}}
\def\trainiidj{\mbox{\texttt{train}}_{ii'j}}
\def\myand{\&\ }
\def\bu{\boldsymbol{u}}
\def\bx{\boldsymbol{x}}
\def\by{\boldsymbol{y}}

\def\bX{\boldsymbol{X}}

\def\bZ{\boldsymbol{Z}}
\def\bbeta{\boldsymbol{\beta}}
\def\simind{\stackrel{{\tiny \mbox{ind.}}}{\sim}}
\def\bI{\boldsymbol{I}}
\def\bSigma{\boldsymbol{\Sigma}}
\def\bzero{\boldsymbol{0}}
\def\bone{\boldsymbol{1}}
\def\bmu{\boldsymbol{\mu}}
\def\bA{\boldsymbol{A}}
\def\bLambda{\boldsymbol{\Lambda}}
\def\stackdum{\mathop{\mbox{\rm stack}}}
\def\stack#1{\stackdum_{#1}}
\def\blockdiagdum{\mathop{\mbox{\rm blockdiag}}}
\def\blockdiag#1{\blockdiagdum_{#1}}
\def\bM{\boldsymbol{M}}
\def\bO{\boldsymbol{O}}
\def\bC{\boldsymbol{C}}
\def\bD{\boldsymbol{D}}
\def\bR{\boldsymbol{R}}
\def\bo{\boldsymbol{o}}
\def\bb{\boldsymbol{b}}
\def\bB{\boldsymbol{B}}
\def\Ssc{{\mathcal S}}
\def\jump{\vskip3mm\noindent}
\def\bOmega{\boldsymbol{\Omega}}
\def\tr{\mbox{tr}}
\def\diag{\mbox{diag}}
\def\smhalf{{\textstyle{\frac{1}{2}}}}
\def\infint{\int_{-\infty}^{\infty}}
\def\bib{\vskip11pt\par\noindent\hangindent=1 true cm\hangafter=1}
\def\rest{\mbox{\rm rest}}
\def\br{\boldsymbol{r}}
\def\bQ{\boldsymbol{Q}}
\def\bc{\boldsymbol{c}}
\def\bomega{\boldsymbol{\omega}}
\def\bCtilde{{\widetilde \bC}}
\def\bDtilde{{\widetilde \bD}}
\def\bbetahat{{\widehat\bbeta}}
\def\buhat{{\widehat \bu}}
\def\Cov{\mbox{\rm Cov}}
\def\rStt{r_{\mbox{\tiny stt}}}
\def\cStt{c_{\mbox{\tiny stt}}}
\def\rEnd{r_{\mbox{\tiny end}}}
\def\cEnd{c_{\mbox{\tiny end}}}
\begin{document}

\ifthenelse{\boolean{DoubleSpaced}}{\setstretch{1.5}}{}

\thispagestyle{empty}

\centerline{\Large\bf Streamlined Variational Inference for Linear Mixed}
\vskip2mm
\centerline{\Large\bf Models with Crossed Random Effects}
\vskip7mm
\ifthenelse{\boolean{UnBlinded}}{
\centerline{\normalsize\sc By Marianne Menictas$\null^1$, Gioia Di Credico$\null^2$ and Matt P. Wand$\null^3$}
\vskip5mm
\centerline{\textit{Harvard University$\null^1$, University of Trieste$\null^2$ 
and University of Technology Sydney$\null^3$}}
}
{\null}
\vskip6mm
\centerline{14th April, 2022}

\vskip6mm

\centerline{\large\bf Abstract}
\vskip2mm

We derive streamlined mean field variational Bayes algorithms for fitting linear mixed models
with crossed random effects. In the most general situation, where the dimensions of the crossed
groups are arbitrarily large, streamlining is hindered by lack of sparseness in the underlying 
least squares system. Because of this fact we also consider a hierarchy of relaxations
of the mean field product restriction. The least stringent product restriction delivers a high 
degree of inferential accuracy. However, this accuracy must be mitigated against its higher
storage and computing demands. Faster sparse storage and computing alternatives are also provided,
but come with the price of diminished inferential accuracy. This article provides full algorithmic
details of three variational inference strategies, presents detailed empirical results on their 
pros and cons and, thus, guides the users on their choice of variational inference approach 
depending on the problem size and computing resources.


\section{Introduction}\label{sec:intro}

Linear mixed models with crossed random effects are a useful vehicle for analysis and inference
for data that are cross-classified according to two or more grouping mechanisms. One major 
application area is psychometrics in which a cohort of \emph{subjects} is assessed according
to a set of tasks or \emph{items} (e.g.\ Baayen \textit{et al.}, 2008; Jeon \textit{et al.}, 2017). 
The assessment scores are cross-classified according to subject and item. 
In such studies it is common for both the subjects and items to be treated as random 
samples from relevant populations.
For example, in a psycholinguistic study, the subjects may be a random sample from the population 
of native Greek speakers and the items may be a random sample from the population
of Greek language syllables. Other variables such as gender and stimuli type may be
treated as non-random. Mixed models with crossed random effects for subject and item and fixed effects
for variables of interest facilitate inference for Greek speakers and the 
Greek language in general rather than for the participants and syllables chosen for the study.
Other areas of psychometrics such as item response theory and Rasch analysis 
(e.g.\ Doran \textit{et al.}, 2007) benefit from crossed random effects models.
The essence of this contribution is streamlined variational inference for crossed random effects 
mixed models that scales well to the handling of very large data sets.

The term ``streamlined'' refers to the process of taking advantage of sparse structures
within the design matrices that arise in linear mixed models. The design matrices are 
often very sparse and potentially extremely large. Clever algorithms that recognize and
make use of the sparseness patterns can lead to dramatic savings in terms of storage and computing time. 
Nolan \textit{et al.} (2020) provides a systematic treatment of
streamlined variational inference for linear mixed models with two and three
levels of nesting. The group specific curves extension is dealt with 
in Menictas \textit{et al.} (2021). \ifthenelse{\boolean{UnBlinded}}
{In these articles, each involving the first and third authors of the current article,}
{In these articles} it was recognized that key variational inference updates can be embedded 
with the class of two-level sparse least squares problems (Nolan \myand Wand, 2020) 
and that this algorithmic component can be isolated into a procedure
that we call \SolveTwoLevelSparseLeastSquares. This procedure also
arises in our variational inference algorithms for crossed random 
effects in Section \ref{sec:streamVarInf}. Also, 
Nolan \textit{et al.} (2020) and Menictas \textit{et al.} (2021) 
are concerned with nested random effects models whilst this article treats
the crossed random effects situation. \tcb{The former situation is
less challenging since higher level nesting invokes hierarchical sparsity structures that 
are amenable to streamlined fitting strategies. These strategies are fully efficient in
terms of only using the non-zero entries of the design matrices. For crossed random effects
the sparsity structure, if present, is more delicate. Depending on the restrictiveness
of the variational approach and the cross-tabulation variable sizes, the cross random 
effects sparseness structure may not be amenable to fully efficient fitting and inference.}

   Throughout this article we consider two grouping mechanisms with group dimensions $m$ and $m'$.
Furthermore, we label the groups in such a way that $m\ge m'$. For example, a psycholinguistic study
involving $900$ subjects and $40$ items has group sizes $m=900$ and $m'=40$.
If a different study involved $75$ subjects and $80$ items then the $(m,m')$
labeling is reversed with respect to subjects and items and our notation is $m=80$ items
and $m'=75$ subjects. Sticking with the $m\ge m'$ notation is important, since it affects 
variational inference algorithm construction and choice. For example, if $m'$ is moderate
in size and $m$ is very large then the least squares system that underlies the least stringent
(most accurate) variational inference scheme is sparse, and streamlined computing advantages
are available. On the other hand, if $m'$ is also very large then the least stringent
algorithm is non-sparse and, depending on computing resources and run-time demands, more
stringent (less accurate) variational inference schemes may be preferred.

    The variational Bayesian inference paradigm is becoming quite a powerful one in contemporary
statistical and machine learning contexts (e.g. Blei \textit{et al.}, 2017).
Modularization variants such as variational message passing (Winn \myand Bishop, 2005;
Wand, 2017) have allowed for the development of versatile and fast inference engines
such as \textsf{Edward} (Tran \textit{et al.}, 2016) and \textsf{Infer.NET} 
(Minka \textit{et al.}, 2018). Various options concerning the stringency of mean 
field-type product restrictions allow for scalability to very large problems with 
speed being traded off against accuracy. All algorithms presented here are
purely matrix algebraic and require no root-finding or numerical integration.
Our variational inference algorithm with medium product restrictions is
able to handle hundreds of crossed random effects in tens of seconds
on contemporary laptop computers.

   The use of variational approximations for crossed random effects mixed models is an
emerging activity and, to date, there are only a few contributions of this type. The 
most prominent such contribution is Jeon \textit{et al.} (2017) which 
applied the notions of Gaussian variational approximation to frequentist 
generalized linear mixed models with crossed random effects. Jeon \textit{et al.} (2017)
concentrated on the scalar effects case and also imposed a product restriction
between the ``item'' and ``subject'' random effects. Our algorithms, which are
for approximate Bayesian inference, allow for this restriction to be removed 
albeit at the cost of increased storage and computation. We also focus on
the Gaussian response here and give a thorough treatment of this more 
straightforward case. Semiparametric mean field variational Bayes ideas 
(e.g. Nolan \myand Wand, 2017) facilitate extension to other likelihoods.

In Section \ref{sec:BayeModels} we define a general class of Gaussian 
response Bayesian crossed random effects linear mixed models.
Sections \ref{sec:varInf} and \ref{sec:streamVarInf} form the centerpiece of
the paper and explain various mean field variational Bayes strategies,
followed by listings of algorithms that facilitate streamlined implementation.
In Section \ref{sec:perfAssess} we report on the results of simulation-based
numerical studies that assess and compare the performances of these new
algorithms with respect to inferential accuracy and computing time.
Section \ref{sec:dataApplic} contains an illustration for data from
a large longitudinal education study. We summarize our findings in Section \ref{sec:conclusions}. 
An online supplement contains derivational and related details. Some
results for frequentist inference for crossed random effects are also 
given in the online supplement.

\section{Bayesian Crossed Random Effects Linear Mixed Models}\label{sec:BayeModels}

The Bayesian crossed random effects linear mixed models being considered here are such that:
\begin{equation}
  \begin{array}{c}
    \by_{i\idash} | \bbeta, \bu_{i}, \bu'_{\idash}, \sigma^2 \simind N (
    \bX_{i\idash} \bbeta + \bZ_{i\idash} \bu_{i} + \Zdash_{i\idash} \bu'_{\idash},
    \sigma^2 \bI ), \quad \bu_{i} | \bSigma \simind N(\bzero, \bSigma), \\[2ex]
    1 \le i \le m, \quad \bu'_{i'}| \bSigmad \simind N(\bzero, \bSigmad), \quad
    1 \le \idash \le \mdash, \quad \bbeta \sim N(\bmu_{\bbeta}, \bSigma_{\bbeta}). \\[2ex]
  \end{array}
\label{eq:BayesianCrossedModel}
\end{equation}
The matrices in (\ref{eq:BayesianCrossedModel}) have dimensions as follows:
\begin{equation}
\begin{array}{c}
\by_{i\idash}\ \mbox{is $n_{i\idash}\times 1$},\ \ \bX_{i\idash}\ \mbox{is $n_{i\idash}\times p$},
\ \ \bbeta\ \mbox{is $p\times1$},
\ \ \bZ_{i\idash}\ \mbox{is $n_{i\idash}\times q$},\ \ \bu_i\ \mbox{is $q \times 1$},\\[2ex]
\Zdash_{i\idash}\ \mbox{is $n_{i\idash}\times q^{\prime}$},\ \ \udashidash\ 
\mbox{is $q^{\prime} \times 1$},\ \ \bSigma \ \ \mbox{is $q \times q$}
\ \ \mbox{and}\ \ \Sigmad\ \ \mbox{is $q^{\prime} \times q^{\prime}$}.
\end{array}
\label{eq:dimCrossed}
\end{equation}
Here $n_{ii'}$ is the number of response measurements in the $(i,i')$th cell. If $n_{ii'}=0$
then each of $\by_{i\idash}$, $\bX_{i\idash}$, $\bZ_{i\idash}$ and $\Zdash_{i\idash}$
are null. \tcr{However, for upcoming matrix assembly operations it is useful to think of,
$\bZ_{i\idash}$, for example, as an $n_{i\idash}\times q$ ``matrix'' with $n_{i\idash}=0$.}

\tcb{
To aid digestibility of (\ref{eq:BayesianCrossedModel}) and (\ref{eq:dimCrossed}), consider a
generic education research study where a sample of $m$ students is followed longitudinally 
and have academic performances measured according to $m'$ items, such as those which 
quantify cognitive, literary and numeracy abilities. The items take the form of exercises 
and, for each item, a quantitative score is determined from a student's performance 
in that item's exercises. Over the duration 
of the multi-year study each of the $m$ students are scored on the $m'$ items $n$ times, 
which implies that $n_{ii'}=n$ for all $1\le i\le m$ and $1\le i'\le m'$.  Define $\scoreiidj$ to be 
the $j$th score of student $i$ for item $i'$. Let $\ageiidj$ be defined analogously, corresponding to 
age in years. Lastly, define $\trainiidj$ to be the indicator of whether the $i$th student received 
training prior to their $j$th attempt at the $i'$th item. Then a $p=3$ and $q=q'=2$ version of the 
response vector and design matrices is 
$$\by_{ii'}=\big[\scoreiidj\big]_{1\le j\le n},\quad
\bX_{ii'}=\big[1\ \ageiidj\ \trainiidj\big]_{1\le j\le n},
\quad\bZ_{ii'}=\big[1\ \ageiidj\big]_{1\le j\le n}
$$
with $\bZ'_{ii'}=\bZ_{ii'}$. According to (\ref{eq:BayesianCrossedModel}) and this set-up, 
the scores of the $i$th student on the $i'$th item are modeled to be 
{\setlength\arraycolsep{0pt}
\begin{eqnarray*}
&&y_{ii'j}|\beta_0,\beta_1,\beta_2,u_{0i},u_{1i},u'_{0i'},u'_{1i'},\sigma^2\\[1ex] 
&&\qquad \simind N\big((\beta_0+u_{0i}+u'_{0i'})+(\beta_1+u_{1i}+u'_{1i'})\ageiidj
+\beta_2\trainiidj,\sigma^2\big),\quad 1\le j\le n.
\end{eqnarray*}
}
Conditional on $\bSigma$, the $[u_{0i}\ u_{1i}]^T$ are $N(\bzero,\bSigma)$ random vectors.
The $[u'_{0\idash}\ u'_{1\idash}]^T$ are similar with $\bSigma'$ instead of $\bSigma$. It is apparent from this
that model (\ref{eq:BayesianCrossedModel}) allows for a different intercept and slope for every
subject/item combination. The heterogeneities in the intercepts and slopes correspond to 
appropriate entries of $\bSigma$ and $\bSigma'$. If the fixed effect $\beta_2$ is of primary interest
then (\ref{eq:BayesianCrossedModel}) is a parsimonious model that allows for 
subject/item heterogeneities in the age effects.
}

For the error variance $\sigma^2$ and the random effects covariance matrices
$\bSigma$ and $\bSigma'$ we consider two prior distribution families:
\begin{eqnarray*}
\mbox{(A)}&&\mbox{ordinary Inverse-Wishart priors}\\
\mbox{(B)}&&\mbox{the marginally non-informative priors proposed 
in Huang \myand Wand (2013).}
\end{eqnarray*}
In terms of the Inverse Chi-Squared and Inverse-G-Wishart distributional notation 
given in Section \ref{sec:IGWandICS}, prior specification (A) involves:
\begin{equation}
\begin{array}{l}
\sigma^2 \sim \mbox{Inverse-$\chi^2$}(\xi_{\sigma^2},\lambdasigsq),\quad
\bSigma\sim\mbox{Inverse-G-Wishart}(\Gfull,\xi_{\bSigma},\Lambda_{\bSigma}),\\[2ex] 
\bSigma'\sim\mbox{Inverse-G-Wishart}(\Gfull,\xi_{\bSigma'},\Lambda_{\bSigma'})
\end{array}
\label{eq:priorsA}
\end{equation}
for hyperparameters $\xi_{\sigma^2},\lambdasigsq>0$, $\xi_{\bSigma}>2(q-1)$, $\xi_{\bSigma'}>2(q'-1)$ and
symmetric positive definite matrices $\Lambda_{\bSigma}$ and $\Lambda_{\bSigma'}$.
Prior specification (B) involves:
\begin{equation}
  \begin{array}{c}
    \sigma^2 | a_{\sigma^2} \sim \mbox{Inverse-$\chi^2$}(\nusigsq, 1/a_{\sigma^2}), \quad
    \asigsq \sim \mbox{Inverse-$\chi^2$}(1, 1/(\nu_{\sigma^2}s_{\sigma^2}^2)), \\[2ex]
    \bSigma | \ASigma \sim \mbox{Inverse-G-Wishart}(\Gfull, \nuSigma + 2q - 2, \ASigma^{-1}), \\[2ex]
    \bSigmad | \ASigmad \sim \mbox{Inverse-G-Wishart}(\Gfull, \nuSigmad + 2 \qd - 2, \ASigmad^{-1}), \\[2ex]
    \ASigma \sim \mbox{Inverse-G-Wishart}(\Gdiag, 1, \LambdaASigma), \quad
    \LambdaASigma \equiv \left\{ \nuSigma \mbox{diag} (\sSigmaOne^2, \hdots, \sSigmaq^2) \right\}^{-1} \\[2ex]
    \ASigmad \sim \mbox{Inverse-G-Wishart}(\Gdiag, 1, \LambdaASigmad), \quad
    \LambdaASigmad \equiv \left\{ \nuSigmad \mbox{diag} (\sSigmadOne^2, \hdots, \sSigmadqd^2) \right\}^{-1}
  \end{array}
  \label{eq:priorsB}
\end{equation}
for hyperparameters $\nusigsq,\nuSigma,\nuSigmad,\sSigmaOne^2,\ldots,\sSigmaq^2,\sSigmadOne^2,\ldots,\sSigmadqd^2>0$.
As explained in Huang \myand Wand (2013), such priors allow standard deviation and
correlation parameters to have arbitrary non-informativeness.

\subsection{Additional Data Matrices}\label{sec:additDataMats}

The various streamlined mean field variational Bayes algorithms given in 
Section \ref{sec:streamVarInf} benefit from the setting up of 
additional data matrices in which the raw data in $\by_{ii'}$, 
$\bX_{ii'}$, $\bZ_{ii'}$ and $\bZ'_{ii'}$ are combined in various
ways using ``stack'' and ``blockdiag'' operators. 
These operators are defined as follows:
$$\stack{1 \le i \le d}(\bM_i)\equiv
\left[
\begin{array}{c}
\bM_1\\
\vdots\\
\bM_d
\end{array}
\right]
\quad\mbox{and}\quad
\blockdiag{1 \le i \le d}(\bM_i)\equiv
\left[
\begin{array}{cccc}
\bM_1 & \bO   & \cdots & \bO\\
\bO   & \bM_2 & \cdots & \bO\\
\vdots& \vdots& \ddots & \vdots\\
\bO   & \bO& \cdots &\bM_d 
\end{array}
\right]
$$
for matrices $\bM_1,\ldots,\bM_d$. The first of these definitions require 
that $\bM_i$, $1\le i\le d$, each have the same number of columns.
\tcr{For the null design matrices that may arise in crossed random effects
models it is convenient to adopt generalizations of regular matrix manipulations. If one of the $\bM_i$ 
is $n\times p$ where $n=0$ and $p>0$ then it is ignored by the stack operator. However, for the blockdiag operator
the column index should have an increment of $p$ before adding the next matrix. This subtledy is fully explained
in Section \ref{sec:genBlockdiag} of the online supplement.}
\tcb{To appreciate the motivation for the ``stack'' and ``blockdiag'' 
notation, consider the intercepts-only special case where $p=q=q'=1$, 
$m=m'=2$ and $n_{ii'}=1$ for $i,i'=1,2$. Then the full set of conditional 
means is contained in the vector
\begin{equation}
\left[
\begin{array}{c}
\beta_0+u_{01}+u'_{01}\\[0.7ex]
\beta_0+u_{01}+u'_{02}\\[0.7ex]
\beta_0+u_{02}+u'_{01}\\[0.7ex]
\beta_0+u_{02}+u'_{02}
\end{array}
\right]
=\left[
\begin{array}{ccccc}
1 & 1 & 0 & 1 & 0 \\[0.7ex]
1 & 1 & 0 & 0 & 1 \\[0.7ex]
1 & 0 & 1 & 1 & 0 \\[0.7ex]
1 & 0 & 1 & 0 & 1 
\end{array}
\right]
\left[
\begin{array}{c}
\beta_0\\[0.4ex]
u_{01}\\[0.4ex]
u_{02}\\[0.4ex]
u'_{01}\\[0.4ex]
u'_{02}
\end{array}
\right].
\label{eq:LedTasso}
\end{equation}
Note that the design matrix in (\ref{eq:LedTasso}) can be written as 
$$\Big[\bone_4\ \blockdiag{1\le i\le 2}\bone_2\ \stack{1\le i\le 2}\bI_2\Big].$$
where $\bone_d$ denotes the $d\times1$ vector of ones and $\bI_d$ is the $d\times d$ identity matrix. 
It is apparent from this example that such notation is very useful for handling cross random effects 
design structures. The remainder of this subsection allows for similar organization of the response and
predictor data and greatly aids succinct algorithmic description, which involve
various full conditional distributions.
} 

Our first set of additional data matrices is
$$\yuptri_i\equiv\stack{1\le i'\le m'}(\by_{ii'}),
\quad
\Xuptri_i\equiv\stack{1\le i'\le m'}(\bX_{ii'}),\quad 1\le i\le m,$$
and
$$\ydntri_{i'}\equiv\stack{1\le i\le m}(\by_{ii'}),
\quad
\Xdntri_{i'}\equiv\stack{1\le i\le m}(\bX_{ii'}),\quad 1\le i'\le m'.$$
Next define
$$\Zuptri_i\equiv\stack{1\le i'\le m'}(\bZ_{ii'}),
\  \Zdblacksquare_i\equiv\blockdiag{1\le i'\le m'}(\bZ'_{ii'}),\ 1\le i\le m,
\quad\mbox{and}\quad
\Zdntri_{i'}\equiv\stack{1\le i\le m}(\bZ'_{ii'}),\ 1\le i'\le m'.$$
Also, we define
$$\by\equiv\stack{1\le i\le m}\Big\{\stack{1\le i'\le m'}\,(\by_{ii'})\Big\}
=\stack{1\le i\le m}(\yuptri_i),\quad
\bX\equiv\stack{1\le i\le m}\Big\{\stack{1\le i'\le m'}\,(\bX_{ii'})\Big\}
=\stack{1\le i\le m}(\Xuptri_i)
$$
and
$$\bZ\equiv\left[\blockdiag{1\le i\le m}(\Zuptri_i)\ \ \stack{1\le i\le m}(\Zdblacksquare_i)\right].$$

\subsection{Additional Dimensional Notation}

The dimensions of the data matrices defined in Section \ref{sec:additDataMats} 
are such that the following notation is useful:
$$\nidot\equiv\sum_{i'=1}^{m'}n_{ii'},\quad 1\le i\le m,\quad 
\ndotid\equiv\sum_{i=1}^{m}n_{ii'},\quad 1\le i'\le m',\quad 
\quad\mbox{and}\quad
\ndotdot\equiv\sum_{i=1}^{m}\sum_{i'=1}^{m'}n_{ii'}.
$$

\section{Variational Inference}\label{sec:varInf}

The joint conditional density function of all parameters in (\ref{eq:BayesianCrossedModel}) 
with covariance priors (\ref{eq:priorsA}) is
\begin{equation}
\pDens(\bbeta,\bu,\bu',\sigma^2,\bSigma,\bSigma'|\by).
\label{eq:jointPost}
\end{equation}
where $\bu\equiv(\bu_1,\ldots,\bu_m)$ and $\bu'\equiv(\bu'_1,\ldots,\bu'_{m'})$.
Let 
\begin{equation}
\qDens(\bbeta,\bu,\bu',\sigma^2,\bSigma,\bSigma')
\label{eq:qDensFirst}
\end{equation}
be a mean field approximation of (\ref{eq:jointPost}). Several product 
restrictions can be placed on the $\qDens$-density function in 
(\ref{eq:qDensFirst}). Here we consider three such restrictions:
\begin{equation}
\qDens(\bbeta,\bu,\bu',\sigma^2,\bSigma,\bSigma')
=
\left\{
\begin{array}{ll}
\qDens(\bbeta)\qDens(\bu)\qDens(\bu')\,
\qDens(\sigma^2,\bSigma,\bSigma'),&\mbox{labeled \emph{product restriction I,}}\\[1ex]
\qDens(\bbeta,\bu)\qDens(\bu')\,\qDens(\sigma^2,\bSigma,\bSigma'),
&\mbox{labeled \emph{product restriction II,}}\\[1ex]
\qDens(\bbeta,\bu,\bu')\,\qDens(\sigma^2,\bSigma,\bSigma'),
&\mbox{labeled \emph{product restriction III.}}
\end{array}
\right.
\label{eq:PRItoIII}
\end{equation}
Product restriction I has the simplest streamlined implementation
but it sets all posterior correlations between $\bbeta$, $\bu$ and $\bu'$
to zero and, thus produces posterior distributions with overly
large variances. On the other hand, product restriction III allows
for joint posterior covariance matrix of $(\bbeta,\bu,\bu')$ 
in its $\qDens$-density to be full -- which leads to higher
inferential accuracy but more challenging computing that can
only be streamlined if $m'$ is moderate. Product restriction II
is a halfway house that recognizes the $m\ge m'$ asymmetry and
carries posterior correlations between $\bbeta$ and $\bu$,
which is the larger of $\bu$ and $\bu'$ assuming that 
$q$ and $q'$ have similar sizes. It delivers more accurate
inference than product restriction I but with similar 
computational overhead.

It should be noted that (\ref{eq:PRItoIII}) conveys the product restrictions
in their minimal forms. However, conditional independencies inherent in
(\ref{eq:BayesianCrossedModel}) mean that additional factorizations
ensue as follows:
$$
\qDens(\bbeta,\bu,\bu',\sigma^2,\bSigma,\bSigma')
=
\left\{
\begin{array}{ll}
\qDens(\bbeta)\left\{{\displaystyle\prod_{i=1}^m}\qDens(\bu_i)\right\}
\left\{{\displaystyle\prod_{i'=1}^{m'}\qDens(\bu'_{i'})}\right\}&\mbox{for product restriction I,}\\[3ex]
\qquad\times\qDens(\sigma^2)\qDens(\bSigma)\qDens(\bSigma'),&\null\\[2ex]
\qDens(\bbeta,\bu)\left\{{\displaystyle\prod_{i'=1}^{m'}\qDens(\bu'_{i'})}\right\}
\qDens(\sigma^2)\qDens(\bSigma)\qDens(\bSigma'),
&\mbox{for  product restriction II,}\\[4ex]
\qDens(\bbeta,\bu,\bu')\qDens(\sigma^2)\qDens(\bSigma)\qDens(\bSigma'),
&\mbox{for  product restriction III.}
\end{array}
\right.
$$
If, instead, the Huang \myand Wand (2013) priors are used then
conditional independencies inherent in (\ref{eq:priorsB}) lead to the
covariance matrix and auxiliary variables component of the 
joint $\qDens$-density factorizing fully as follows:
$$\qDens(\sigma^2,a_{\sigma^2},\bSigma,\ASigma,\bSigmad,\ASigmad)=
\qDens(\sigma^2)\qDens(a_{\sigma^2})\qDens(\bSigma)\qDens(\ASigma)
\qDens(\bSigmad)\qDens(\ASigmad).
$$
Under either product restrictions I, II or III, and letting 
$\buall\equiv(\bu,\bu')$, standard mean field variational Bayes steps
(e.g. Bishop, 2006; Sections 10.1--10.3) lead to the $\qDens$-density 
functions of the model parameters having the following forms:
$$
  \begin{array}{lcl}
  &&\qDens^*(\bbeta,\buall)\ \mbox{has a $N\big(\bmu_{\qDens(\bbeta,\buall)},
         \bSigma_{\qDens(\bbeta,\buall)}\big)$ distribution,}\\[1ex]
  &&\qDens^*(\sigma^2)\ \mbox{has an $\mbox{Inverse-$\chi^2$}
    \big(\xi_{\qDens(\sigma^2)},\lambda_{\qDens(\sigma^2)}\big)$ distribution,}\\[1ex]
  &&\qDens^*(\bSigma)\ \mbox{has an
    $\mbox{Inverse-G-Wishart}(\Gfull, \xi_{\qDens_{(\bSigma)}},\bLambda_{\qDens(\bSigma)})$ distribution}\\[1ex]
  \mbox{and}&&\qDens^*(\bSigmad)\ \mbox{has an
    $\mbox{Inverse-G-Wishart}(\Gfull, \xi_{\qDens_{(\bSigmad)}},\bLambda_{\qDens(\bSigmad)})$ distribution.}
  \end{array}
$$
The $\qDens$-density parameters can be obtained using a coordinate ascent iterative algorithm 
(e.g. Algorithm 1 of Ormerod \myand Wand, 2010). However, if applied na\"{\i}vely, the matrix
$\bSigma_{\qDens(\bbeta,\buall)}$ requires storage and inversion. As explained in the upcoming
Section \ref{sec:DeniseDrysdale}, this matrix is potentially prohibitively large. 
Product restrictions I, II and III lead to streamlined mean field variational Bayes algorithms
with varying degrees of storage and computational overhead.

\subsection{The $\bSigma_{\qDens(\bbeta,\buall)}$ Matrix and Product Restriction Implications}\label{sec:DeniseDrysdale}

\tcb{
The square matrix $\bSigma_{\qDens(\bbeta,\buall)}$ has $(p+mq+m'q')^2$ entries. Therefore,
a version of the Section \ref{sec:BayeModels} education study example involving $10,000$ students
is such that $\bSigma_{\qDens(\bbeta,\buall)}$  has more than $400$ million entries. However,
product restrictions I, II and III impose sparseness structures on $\bSigma_{\qDens(\bbeta,\buall)}$,
which are summarized in Table \ref{tab:SigmaSubBlocks}. Section \ref{sec:streamVarInf} is concerned 
with deriving streamlined mean field variational Bayes fitting and inference  algorithms according 
to each of the three product restrictions. Table \ref{tab:SigmaSubBlocks} provides a roadmap for 
the nature of the required results.}

\begin{table}[!ht]
\centering
\begin{tabular}{lccc}
\hline\\[-1.5ex]
sub-blocks of $\bSigma_{\qDens(\bbeta,\buall)}$ & prod. res. I  & prod. res. II  & prod. res. III \\[1.2ex]
\hline\\[-1.3ex]
$\bSigma_{\qDens(\bbeta)},\ \bSigma_{\qDens(\bu_i)}, \bSigma_{\qDens(\bu'_{i'})}$ & $\bigX$  & $\bigX$ & $\bigX$ \\[0.7ex]
$E_{\qDens}\{(\bbeta-\bmu_{\qDens(\bbeta)})(\bu_i-\bmu_{\qDens(\bu_i)})^T\}$      & $\bO$     & $\bigX$ & $\bigX$ \\[0.7ex]
$E_{\qDens}\{(\bbeta-\bmu_{\qDens(\bbeta)})(\bu'_{i'}-\bmu_{\qDens(\bu'_{i'})})^T\}$ & $\bO$ & $\bO$ & $\bigX$ \\[0.7ex]
$E_{\qDens}\{(\bu_i-\bmu_{\qDens(\bu_i)})(\udashidash-\bmu_{\qDens(\udashidash)})^T\}$& $\bO$ & $\bO$ & $\bigX$\\
\mbox{all other sub-blocks}                & $\bO$ & $\bO$ & $\bO$\\
\hline
\end{tabular}
\caption{\it The zero ($\bO$) versus non-zero ($\bigX$) status of various sub-blocks of the $\qDens(\bbeta,\buall)$ 
covariance matrix $\bSigma_{\qDens(\bbeta,\buall)}$ under product restrictions I, II and III. 
The $i$ subscript ranges over $1,\ldots,m$ and the $i'$ subscript ranges over $1,\ldots,m'$.}
\label{tab:SigmaSubBlocks} 
\end{table}

\tcb{
Under product restriction I, only the diagonal sub-blocks of $\bSigma_{\qDens(\bbeta,\buall)}$ 
given by Table \ref{tab:SigmaSubBlocks} are non-zero. These sub-blocks only have a total of 
$p^2+q^2m+(q')^2m'$ entries. A mean field variational Bayes algorithm that takes advantage
of this sparseness will scale well to very large problems.}

\tcb{
For product restriction II there are an additional $pqm$ non-zero entries in $\bSigma_{\qDens(\bbeta,\buall)}$ 
due to the $E_{\qDens}\{(\bbeta-\bmu_{\qDens(\bbeta)})(\bu_i-\bmu_{\qDens(\bu_i)})^T\}$ contributions.
The number of  non-zero entries is still linear in $m$ and $m'$, but the sparsity structure is more
delicate. The upcoming Result \ref{res:prodRestrictII} is concerned with efficient approximate
inference when such structure is present.}

\tcb{
Product restriction III is particularly mild, but involves an additional $pq'm'+q'm'qm$ potentially non-zero entries
in the $\bSigma_{\qDens(\bbeta,\buall)}$ matrix. If $m'$ is moderately sized then a type of sparseness
arises. Result \ref{res:prodRestrictIII} in the next section is motivated by this situation.
}

\section{Streamlined Variational Inference}\label{sec:streamVarInf}

Variational inference for $\sigma^2$, $\bSigma$ and $\bSigma'$ is relatively
straightforward and only moderately affected by the type of product
restriction on the effects parameters. However, there are distinct
differences among the product restrictions for updating the parameters
in $\qDens(\bbeta,\buall)$ so these are treated separately in 
each of the next three subsections. After that we treat the variance 
and covariance matrices component of the model.

\subsection{Streamlined Variational Inference for $(\bbeta,\buall)$ Under Product Restriction I}

Under product restriction I the variational inference updates are relatively simple
and can be done using standard mean field arguments. The derivational details are given in 
Section \ref{sec:PRIderiv} of the online supplement. 

Given current values of the $\qDens$-density parameters of $\sigma^2$, $\bu$ and $\bu'$
the updates for the $\qDens(\bbeta)$ parameters are:
\begin{equation}
\begin{array}{l}
\bb\longleftarrow
\left[\begin{array}{c}   
\mu_{\qDens(1/\mysigeps^2)}^{1/2}\,
\left[\by- 
{\displaystyle\stack{1\le i\le m}}\Big\{{\displaystyle\stack{1\le i'\le m'}}
\Big(\bZ_{ii'}\bmu_{\qDens(\bu_i)}+\bZ'_{ii'}\bmu_{\qDens(\bu'_{i'})}\Big)\Big\}\right]\\[2ex]
\bSigma_{\bbeta}^{-1/2}\bmu_{\bbeta}
\end{array}
\right]\\[1ex]
\bB\longleftarrow
\left[\begin{array}{c}   
\mu_{\qDens(1/\mysigeps^2)}^{1/2}\,\bX\\[2ex]
\bSigma_{\bbeta}^{-1/2}
\end{array}
\right]
\ \ \ ;\ \ \ \Ssc\longleftarrow\SolveLeastSquares\Big(\{\bb,\bB\}\Big)\\[1ex]
\bmu_{\qDens(\bbeta)}\longleftarrow \bx\ \mbox{component of}\ \Ssc
\ \ \ ;\ \ \ 
\bSigma_{\qDens(\bbeta)}\longleftarrow (\bB^T\bB)^{-1}\ \mbox{component of}\ \Ssc
\end{array}
\label{eq:bandBsimplest}
\end{equation}
where the \SolveLeastSquares\ algorithm is given by Algorithm \ref{alg:SolveLeastSquares}
in Section \ref{sec:SLSA} of the online supplement.
Then, given the current values of the $\qDens$-density parameters of 
$\bbeta$, $\bu'$, $\sigma^2$ and $\bSigma$ the updates for the parameters 
of the $\qDens(\bu_i)$, $1\le i\le m$, have similar expressions involving
the \SolveLeastSquares\ algorithm. The updates for $\qDens(\bu'_{i'})$, $1\le i'\le m'$, are analogous.
 
The full set of updates is provided by Algorithm \ref{alg:MFVBforPRI}.

\begin{algorithm}[!th]
  \begin{center}
    \begin{minipage}[t]{154mm}
      \begin{small}
        \begin{itemize}
          \setlength\itemsep{4pt}
          \item[] Data Inputs: $(\by,\bX)$,\ $\{(\yuptri_i,\Xuptri_i,\Zuptri_i):\ 1\le i\le m\},$\  
                               $\{(\ydntri_{i'},\Xdntri_{i'},\Zdntri_{i'}):\ 1\le i'\le m'\},$
          \item[]  $\qquad\qquad\quad\quad\{(\bZ_{ii'},\Zdash_{ii'}):1\le i \le m,\ 1\le i'\le m'\}$
          \item[] Hyperparameter Inputs: $\bmu_{\bbeta}(p\times1)$,
                  $\bSigma_{\bbeta}(p\times p)\ \mbox{symmetric and positive definite}$, 
                     \item[] $\qDens$-Density Inputs: $\bmu_{\qDens(\bu_{i})}$,\ $1\le i\le m$,\ 
                      $\bmu_{\qDens(\bu'_{i'})}$,\ $1\le i'\le m'$,
                   \ $\mu_{\qDens(1/\mysigeps^2)}$,\ $\bM_{\qDens(\bSigma^{-1})}(q\times q)$,
           \item[] $\qquad\qquad\qquad\qquad\bM_{\qDens((\bSigma')^{-1})}(q'\times q')\ \mbox{both symmetric and positive definite}$.
           \item[] $\bb\longleftarrow
                    \left[\begin{array}{c}   
                   \mu_{\qDens(1/\mysigeps^2)}^{1/2}\,
                  \left[\by- 
             {\displaystyle\stack{1\le i\le m}}\Big\{{\displaystyle\stack{1\le i'\le m'}}
             \Big(\bZ_{ii'}\bmu_{\qDens(\bu_i)}+\bZ'_{ii'}\bmu_{\qDens(\bu'_{i'})}\Big)\Big\}\right]\\[2ex]
             \bSigma_{\bbeta}^{-1/2}\bmu_{\bbeta}
             \end{array}
            \right]$
           \item[]$\bB\longleftarrow
           \left[\begin{array}{c}   
           \mu_{\qDens(1/\mysigeps^2)}^{1/2}\,\bX\\[2ex]
           \bSigma_{\bbeta}^{-1/2}
           \end{array}
           \right]
           $\ \ \ ;\ \ \ $\Ssc\longleftarrow\SolveLeastSquares\Big(\{\bb,\bB\}\Big)$
          \item[] $\bmu_{\qDens(\bbeta)}\longleftarrow \bx\ \mbox{component of}\ \Ssc$
                  \ \ \ ;\ \ \ 
                  $\bSigma_{\qDens(\bbeta)}\longleftarrow (\bB^T\bB)^{-1}\ \mbox{component of}\ \Ssc$
          \item[] For $i=1,\ldots,m$:
          \item[]\begin{itemize}
           \item[] $\bb\longleftarrow
                         \left[\begin{array}{c}   
                              \mu_{\qDens(1/\mysigeps^2)}^{1/2}\,
                              \Big\{\yuptri_{i}-\Xuptri_{i}\bmu_{\qDens(\bbeta)}-
                              {\displaystyle\stack{1\le i'\le m'}}
                               \Big(\bZ'_{ii'}\bmu_{\qDens(\bu'_{i'})}\Big)\Big\}\\[2ex]
                           \bzero
                      \end{array}
                    \right]
                    $
               \item[]$\bB\longleftarrow
                        \left[\begin{array}{c}   
                        \mu_{\qDens(1/\mysigeps^2)}^{1/2}\,\Zuptri_{i}\\[2ex]
                        \bM_{\qDens(\bSigma^{-1})}^{1/2}
                         \end{array}
                        \right]
                       $\ \ \ ; \ \ $\Ssc\longleftarrow\SolveLeastSquares\Big(\{\bb,\bB\}\Big)$
              \item[] $\bmu_{\qDens(\bu_{i})}\longleftarrow \bx\ \mbox{component of}\ \Ssc$
                  \ \ \ ;\ \ \ 
                 $\bSigma_{\qDens(\bu_{i})}\longleftarrow (\bB^T\bB)^{-1}\ \mbox{component of}\ \Ssc$
            \end{itemize}
           \item[] For $i'=1,\ldots,m'$:
          \item[]\begin{itemize}
           \item[] $\bb\longleftarrow
                         \left[\begin{array}{c}   
                              \mu_{\qDens(1/\mysigeps^2)}^{1/2}\,
                              \Big\{\ydntri_{i'}-\Xdntri_{i'}\bmu_{\qDens(\bbeta)}-
                              {\displaystyle\stack{1\le i\le m}}
                               \Big(\bZ_{ii'}\bmu_{\qDens(\bu_{i})}\Big)\Big\}\\[2ex]
                           \bzero
                      \end{array}
                    \right]
                    $
               \item[]$\bB\longleftarrow
                        \left[\begin{array}{c}   
                        \mu_{\qDens(1/\mysigeps^2)}^{1/2}\,\Zdntri_{i'}\\[2ex]
                        \bM_{\qDens((\bSigma')^{-1})}^{1/2}
                         \end{array}
                        \right]
                       $\ \ \ ; \ \ $\Ssc\longleftarrow\SolveLeastSquares\Big(\{\bb,\bB\}\Big)$
              \item[] $\bmu_{\qDens(\bu'_{i'})}\longleftarrow \bx\ \mbox{component of}\ \Ssc$
                  \ \ \ ;\ \ \ 
                 $\bSigma_{\qDens(\bu'_{i'})}\longleftarrow (\bB^T\bB)^{-1}\ \mbox{component of}\ \Ssc$
            \end{itemize}
          \item[] Outputs: $\bmu_{\qDens(\bbeta)},\,\bSigma_{\qDens(\bbeta)},
                   \left\{\left(\bmu_{\qDens(\bu_i)},\,\bSigma_{\qDens(\bu_i)}\right):\ 1 \le i \le m\right\},\  
                   \left\{\left(\bmu_{\qDens(\bu'_{i'})},\bSigma_{\qDens(\bu'_{i'})}\right):1 \le i'\le m'\right\}$
        \end{itemize}
      \end{small}
    \end{minipage}
  \end{center}
  \caption{\it Mean field variational Bayes algorithm for updating the 
parameters of $\qDens(\bbeta,\buall)$ under product restriction I.}
\label{alg:MFVBforPRI}
\end{algorithm}

\subsection{Streamlined Variational Inference for $(\bbeta,\buall)$ Under Product Restriction II}

Under product restriction II the updates for the $\qDens(\bu'_{i'})$ parameters are the
same as those for product restriction I. However streamlined updating of the $\qDens(\bbeta,\buall)$ 
parameters is more delicate. The problem can be embedded within the class of two-level sparse matrix 
problems as defined in Nolan \myand Wand (2020) and is encapsulated in Result \ref{res:prodRestrictII}.
Note that Result \ref{res:prodRestrictII} uses matrix sub-block notation given by (\ref{eq:AtLevInv}) 
in Section \ref{sec:STLSLS} of the online supplement. The derivation of this result is 
given in Section \ref{sec:derivProdRestrictII} of the online supplement of this article.

\begin{result} 
According to product restriction II, the mean field variational Bayes updates 
of $\bmu_{\qDens(\bbeta,\buall)}$ and each of the  sub-blocks of $\bSigma_{\qDens(\bbeta,\buall)}$ 
listed in the first row of Table \ref{tab:SigmaSubBlocks}, given the current values 
of $\bmu_{\qDens(\bu'_{i'})}$,\ $1\le i'\le m'$, are expressible as a two-level sparse 
matrix least squares problem of the form:
$$\left\Vert\bb-\bB\bmu_{\qDens(\bbeta,\buall)}
\right\Vert^2
$$
where $\bb$ and the non-zero sub-blocks of $\bB$, according to the notation
in (\ref{eq:BandbFormsReprise}) of the online supplement, are, for $1\le i\le m$,
$$\bveci\equiv\left[\begin{array}{c}   
\mu_{\qDens(1/\mysigeps^2)}^{1/2}\,
\Big\{\yuptri_i-{\displaystyle\stack{1\le i'\le m'}}
\big(\bZ'_{ii'}\bmu_{\qDens(\bu'_{i'})}\big)\Big\}\\[2ex]
m^{-1/2}\bSigma_{\bbeta}^{-1/2}\bmu_{\bbeta}\\[2ex]
\bzero
\end{array}
\right],
\qquad\Bmati\equiv\left[\begin{array}{c}   
\mu_{\qDens(1/\mysigeps^2)}^{1/2}\,\Xuptri_i\\[2ex]
m^{-1/2}\bSigma_{\bbeta}^{-1/2}\\[2ex]
\bO
\end{array}
\right]
$$
and
$$\Bmatdoti\equiv
\left[\begin{array}{c}   
\mu_{\qDens(1/\mysigeps^2)}^{1/2}\,\Zuptri_i\\[2ex]
\bO\\[1ex]
\bM_{\qDens(\bSigma^{-1})}^{1/2}
\end{array}
\right],
$$
with each of these matrices having $\nadj_i=\nidot+p+q$ rows.
The solutions are
$$\bmu_{\qDens(\bbeta)}=\xveco,\quad\bSigma_{\qDens(\bbeta)}=\AUoo$$ 
\textit{and}
$$\bmu_{\qDens(\bu_i)}=\xvectCi,\ \ \bSigma_{\qDens(\bu_i)}=\AUttCi,
\ \ E_{\qDens}\{(\bbeta-\bmu_{\qDens(\bbeta)})(\bu_i-\bmu_{\qDens(\bu_i)})^T\}=\AUotCi,
\quad 1\le i\le m,
$$
where the $\xveco$, $\xvectCi$, $\AUoo$, $\AUttCi$ and $\AUotCi$ notation
is given by (\ref{eq:AtLevInv}) in the online supplement.
\label{res:prodRestrictII}
\end{result}

Result \ref{res:prodRestrictII} gives rise to Algorithm \ref{alg:MFVBforPRII},
which provides the full set of updates of the $\qDens(\bbeta,\buall)$ parameters
under product restriction II. Note that Algorithm \ref{alg:MFVBforPRII}
makes use of the \SolveTwoLevelSparseLeastSquares\ algorithm from 
Nolan \textit{et al.} (2020) and reproduced for convenience
in Section \ref{sec:STLSLS} of the online supplement.

\begin{algorithm}[!th]
  \begin{center}
    \begin{minipage}[t]{160mm}
      \begin{small}
        \begin{itemize}
          \setlength\itemsep{4pt}
          \item[] Data Inputs: $\{(\yuptri_i,\Xuptri_i,\Zuptri_i):\ 1\le i\le m\},$\ \   
                               $\{(\ydntri_{i'},\Xdntri_{i'},\Zdntri_{i'}):\ 1\le i'\le m'\},$
          \item[]  $\qquad\qquad\quad\quad\{(\bZ_{ii'},\Zdash_{ii'}):1\le i \le m,\ 1\le i'\le m'\}$
          \item[] Hyperparameter Inputs: $\bmu_{\bbeta}(p\times1)$,
                  $\bSigma_{\bbeta}(p\times p)\ \mbox{symmetric and positive definite}$.
           \item[] $\qDens$-Density Inputs: $\bmu_{\qDens(\bu'_{i'})}$,\ $1\le i'\le m'$,
                   \ $\mu_{\qDens(1/\mysigeps^2)}$,\ $\bM_{\qDens(\bSigma^{-1})}(q\times q)$,
           \item[] $\qquad\qquad\qquad\qquad\bM_{\qDens((\bSigma')^{-1})}(q'\times q')\ \mbox{both symmetric and positive definite}$.
           \item[] For $i = 1,\ldots, m$: 
            \begin{itemize}
              \item[] $\bveci\longleftarrow \left[\begin{array}{c}   
                         \mu_{\qDens(1/\mysigeps^2)}^{1/2}\,\Big\{\yuptri_i-{\displaystyle\stack{1\le i'\le m'}}
                                \big(\bZ'_{ii'}\bmu_{\qDens(\bu'_{i'})}\big)\Big\}\\[2ex]
                         m^{-1/2}\bSigma_{\bbeta}^{-1/2}\bmu_{\bbeta}\\[2ex]
                         \bzero
                        \end{array}\right];
              \ \ \Bmati\longleftarrow \left[\begin{array}{c}   
                                          \mu_{\qDens(1/\mysigeps^2)}^{1/2}\,\Xuptri_i\\[2ex]
                                           m^{-1/2}\bSigma_{\bbeta}^{-1/2}\\[2ex]
                                           \bO
                                             \end{array}
                                           \right]$ \\[1ex]
              \item[] $\ \Bmatdoti\longleftarrow \left[\begin{array}{c}   
                                    \mu_{\qDens(1/\mysigeps^2)}^{1/2}\,\Zuptri_i\\[2ex]
                                      \bO\\[1ex]
                                      \bM_{\qDens(\bSigma^{-1})}^{1/2}
                                      \end{array}
                         \right]$
            \end{itemize}
            \item[] $\Ssc\thickarrow\SolveTwoLevelSparseLeastSquares
                    \Big(\big\{(\bveci,\Bmati,\Bmatdoti):1\le i \le m\big\}\Big)$
            \item[] $\bmu_{\qDens(\bbeta)}\thickarrow\mbox{$\xveco$ component of $\Ssc$}$
                    \ \ \ ;\ \ \ $\bSigma_{\qDens(\bbeta)}\thickarrow\mbox{$\AUoo$ component of $\Ssc$}$
             \item[] For $i = 1,\ldots, m$: 
            \begin{itemize}
             \item[] $\bmu_{\qDens(\bu_i)}\thickarrow\mbox{$\xvectCi$ component of $\Ssc$}$\ \ \ ;\ \ \ 
                       $\bSigma_{\qDens(\bu_i)}\thickarrow\mbox{$\AUttCi$ component of $\Ssc$}$
               \item[] $E_{\qDens}\{(\bbeta-\bmu_{\qDens(\bbeta)})(\bu_i-\bmuq{\bu_i})^T\}\thickarrow
                 \mbox{$\AUotCi$ component of $\Ssc$}$
            \end{itemize}
             \item[] For $i' = 1,\ldots, m'$:  
                  \begin{itemize}
                  \item[] $\bb\longleftarrow
                         \left[\begin{array}{c}   
                          \mu_{\qDens(1/\mysigeps^2)}^{1/2}\,
                          \Big\{\ydntri_{i'}-\Xdntri_{i'}\bmu_{\qDens(\bbeta)}-
                           {\displaystyle\stack{1\le i\le m}}
                           \Big(\bZ_{ii'}\bmu_{\qDens(\bu_i)}\Big)\Big\}\\[2ex]
                             \bzero
                          \end{array}
                          \right]
                          $
                \item[]$\bB\longleftarrow
                             \left[\begin{array}{c}   
                              \mu_{\qDens(1/\mysigeps^2)}^{1/2}\,\Zdntri_{i'}\\[2ex]
                              \bM_{\qDens((\bSigma')^{-1})}^{1/2}
                          \end{array}
                         \right]
                         $\ \ \ ;\ \ \ $\Ssc\longleftarrow\SolveLeastSquares\Big(\{\bb,\bB\}\Big)$
                      \item[] $\bmu_{\qDens(\bu'_{i'})}\longleftarrow \bx\ \mbox{component of}\ \Ssc$
                               \ \ \ ;\ \ \ 
                      $\bSigma_{\qDens(\bu'_{i'})}\longleftarrow (\bB^T\bB)^{-1}\ \mbox{component of}\ \Ssc$
                  \end{itemize} 
          \item[] Outputs: $\bmu_{\qDens(\bbeta)},\,\bSigma_{\qDens(\bbeta)},
                  \left\{\left(\bmu_{\qDens(\bu_{i})},\bSigma_{\qDens(\bu_{i})}\right):\ 1 \le i \le m\right\},
                  \left\{\left(\bmu_{\qDens(\bu'_{i'})},\bSigma_{\qDens(\bu'_{i'})}\right):\ 1 \le i' \le m'\right\},$
          \item[]\ \ \ \ \ \ \ \ \ \ \ \ \ \ \ $\left\{E_{\qDens}\{(\bbeta - \bmu_{\qDens(\bbeta)})
          (\bu_{i}-\bmu_{\qDens(\bu_{i})})^T:\ 1\le i\le m\right\}$
        \end{itemize}
      \end{small}
    \end{minipage}
  \end{center}
\caption{\it Mean field variational Bayes algorithm for updating the 
parameters of $\qDens(\bbeta,\buall)$ under product restriction II.}
\label{alg:MFVBforPRII}
\end{algorithm}

\subsection{Streamlined Variational Inference for $(\bbeta,\buall)$ Under Product Restriction III}

Product restriction III is such that sparse least squares systems do not arise naturally
in the same way as product restrictions I and II or the nested random effects models treated
in Lee \myand Wand (2016) and Nolan \textit{et al.} (2020).

Result \ref{res:prodRestrictIII} embeds the updates of the $\qDens(\bbeta,\buall)$ parameters
within the class of two-level sparse matrix problems as defined in Nolan \myand Wand (2020) 
and summarized in Section \ref{sec:STLSLS} of the online supplement. The updates are valid for 
any values of $m$ and $m'$. If $m'$ is moderate in size but $m$ is possibly very large then 
the system is efficient in the sense that the amount of storage and computing is linear in $m$.

\jump\noindent
\begin{result}
According to product restriction III, the mean field variational Bayes 
updates of $\bmu_{\qDens(\bbeta,\buall)}$ and each of the sub-blocks of $\bSigma_{\qDens(\bbeta,\buall)}$ 
in the first four rows of Table \ref{tab:SigmaSubBlocks} is expressible as a two-level sparse matrix least 
squares problem of the form:
$$\left\Vert\bb-\bB\bmu_{\qDens(\bbeta,\buall)}
\right\Vert^2
$$
where $\bb$ and the non-zero sub-blocks of $\bB$, according to the notation
in (\ref{eq:BandbFormsReprise}), are, for $1\le i\le m$,
$$\bveci\equiv\left[\begin{array}{c}
  \mu_{\qDens(1/\sigma^2)}^{1/2}\yuptri_i\\[2ex]
  m^{-1/2}\bSigma_{\bbeta}^{-1/2}\bmu_{\bbeta}\\[1ex]
  \bzero \\[1ex]
  \bzero
\end{array}
\right],
\quad\Bmati\equiv\left[\begin{array}{cc}
\mu_{\qDens(1/\sigma^2)}^{1/2}\Xuptri_i & \mu_{\qDens(1/\sigma^2)}^{1/2}\Zdblacksquare_i\\[2ex]
m^{-1/2}\bSigma_{\bbeta}^{-1/2} & \bO\\[1ex]
\bO & m^{-1/2}\Big(\bI_{\mdash} \otimes \bM_{\qDens((\bSigma')^{-1})}^{1/2}\Big)  \\[1ex]
\bO & \bO \end{array} \right]
$$
$$
\quad \mbox{and} \quad
\Bmatdoti\equiv
\left[\begin{array}{c}
\mu_{\qDens(1/\sigma^2)}^{1/2}\Zuptri_i\\[2ex]
\bO\\[1ex]
\bO\\[1ex]
\bM_{\qDens(\bSigma^{-1})}^{1/2}
\end{array}
\right]
$$
with each of these matrices having $\nidot+p+m'q'+q$ 
rows and with 
$\bB_{i}$ having $p + \mdash q^{\prime}$ columns and $\bBdot_{i}$ having $q$ columns. The solutions 
are, with sub-matrix labeling of $\bx$ and $\bA^{-1}$ according to (\ref{eq:AtLevInv}),
\begin{equation*}
  \begin{array}{c}
    \bmu_{\qDens(\bbeta)} = \mbox{ first $p$ rows of }\xveco, \ \ \bSigma_{\qDens(\bbeta)} = 
    \mbox{ top left $p \times p$ sub-block of }\AUoo, 
    \\[2ex] 
 \displaystyle{\stack{1\le i'\le m'}}(\bmu_{\qDens(\udashidash)}) 
     = \mbox{subsequent $(m'q')\times 1$ entries of}\ \bx_{1}\ \mbox{following $\bmu_{\qDens(\bbeta)}$}, 
    \\[2ex] 
    E_{\qDens} \{ (\bbeta - \bmu_{\qDens(\bbeta)})( \udashidash - \bmu_{\qDens(\udashidash)})^T\} =
    \mbox{ subsequent $p \times q^{\prime}$ sub-blocks of $\AUoo$ to the right of $\bSigma_{\qDens(\bbeta)}$}, 
    \\[2ex]
    \bSigma_{\qDens(\udashidash)} =\mbox{ subsequent $q' \times q'$ diagonal
    sub-blocks of $\AUoo$ following $\bSigma_{\qDens(\bbeta)}$},\ \ 1\le \idash \le \mdash,
    \\[2ex]
    \bmu_{\qDens(\bu_{i})}= \xvectCi, \ \ \bSigma_{\qDens(\bu_{i})} = \bA^{22,i}, \ \ 
    E_{\qDens} \{(\bbeta - \bmu_{\qDens(\bbeta)})(\bu_{i} - \bmu_{\qDens(\bu_{i})})^T\}= 
\mbox{first $p$ rows of $\bA^{12,i}$} 
    \\[2ex] 
   \mbox{and}{\displaystyle\stack{1\le i'\le m'}}
\Big(
[E_{\qDens}\{(\bu_{i} - \bmu_{\qDens(\bu_{i})})(\udashidash - \bmu_{\qDens(\udashidash)})^T\}]^T\Big)= 
    \mbox{ remaining $m'q'$ rows of $\bA^{12,i}$},\ \ 1\le i \le m,
  \end{array}
\end{equation*}
where the $\xveco$, $\xvectCi$, $\AUoo$, $\AUttCi$ and $\AUotCi$ notation
is given by (\ref{eq:AtLevInv}) in the online supplement.
\label{res:prodRestrictIII}
\end{result}

\ifthenelse{\boolean{ColourVersion}}{\def\myWhite{Yellow}}{\def\myWhite{White}}
\ifthenelse{\boolean{ColourVersion}}{\def\myBlack{Blue}}{\def\myBlack{Black}}
\ifthenelse{\boolean{ColourVersion}}{\def\mywhite{yellow}}{\def\mywhite{white}}
\ifthenelse{\boolean{ColourVersion}}{\def\mygrey{orange}}{\def\mygrey{grey}}
\ifthenelse{\boolean{ColourVersion}}{\def\myblack{blue}}{\def\myblack{black}}
\ifthenelse{\boolean{ColourVersion}}{\def\myApplicGrey{light blue}}{\def\myApplicGrey{grey}}
Figure \ref{fig:crossedImage} provides visualization of the strategy used by 
Result \ref{res:prodRestrictIII}. For simplicity, the values of $p$, $q$, $q'$
and $n_{ii'}$ are all set to $1$ and $m'$ is set to $2$. Each panel shows 
an image plot representation of the matrix $\bB$ according to the sparse
two-level form given by (\ref{eq:BandbFormsReprise}) but with the 
$\Bmati$ and $\Bmatdoti$ sub-blocks specific to Result \ref{res:prodRestrictIII}.
The \mywhite\ regions correspond to the two-level sparsity due to the block diagonal 
positioning of the $\Bmatdoti$, $1\le i\le $m. The \mygrey\ regions also indicate
entries, and have additional block diagonal formations, but which do not contribute
to the two-level sparsity. For moderate $m'$ and large $m$ the \myblack/\mygrey block on 
the left is small relative to the remainder of the matrix. The \SolveTwoLevelSparseLeastSquares\
algorithm, listed as Algorithm \ref{alg:SolveTwoLevelSparseLeastSquares} in Section \ref{sec:STLSLS} 
of the online supplement, affords efficient calculation of the variational inference
updates for $m$ potentially very large.

\begin{figure}[!ht]
\vcenteredhbox{\includegraphics[width=0.21\textwidth]{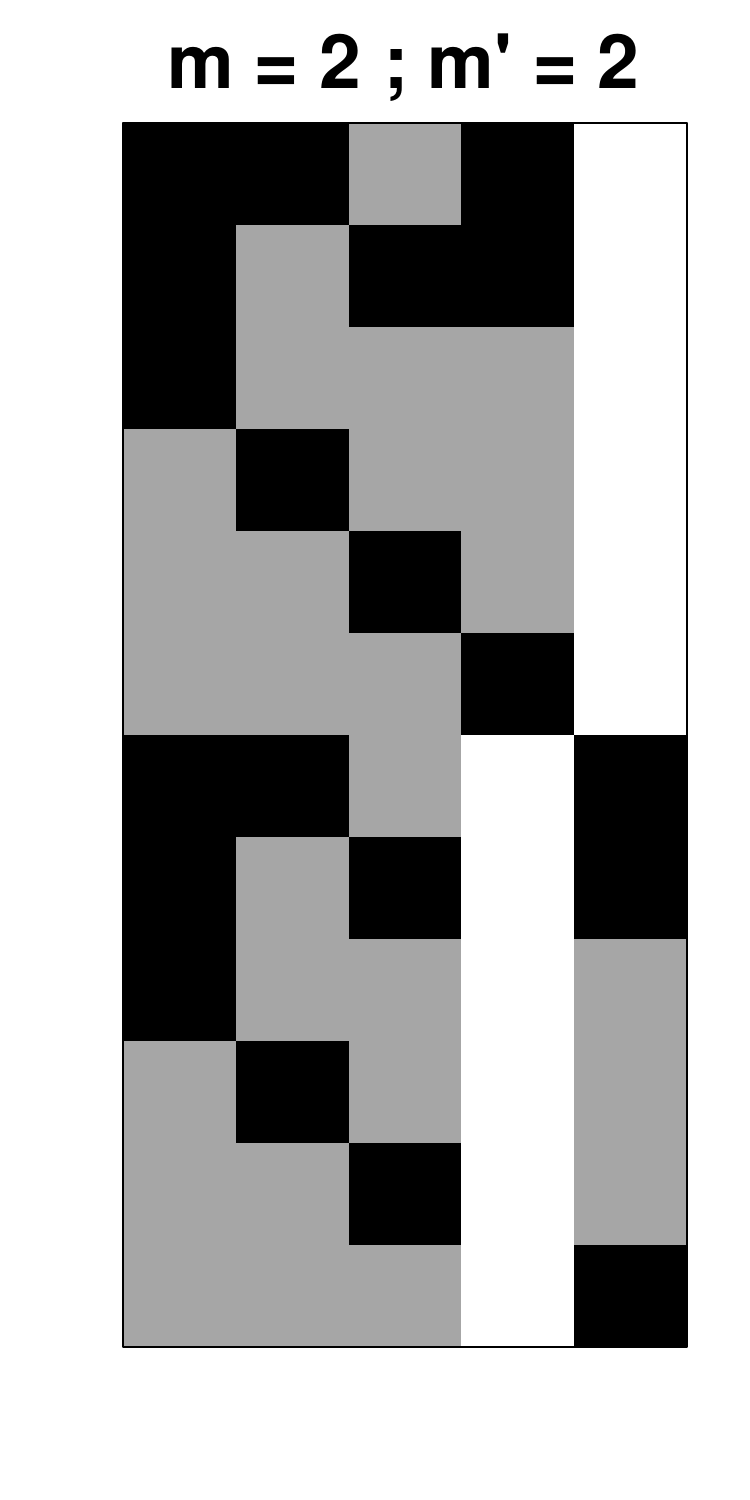}}$\ \ $
\vcenteredhbox{\includegraphics[width=0.21\textwidth]{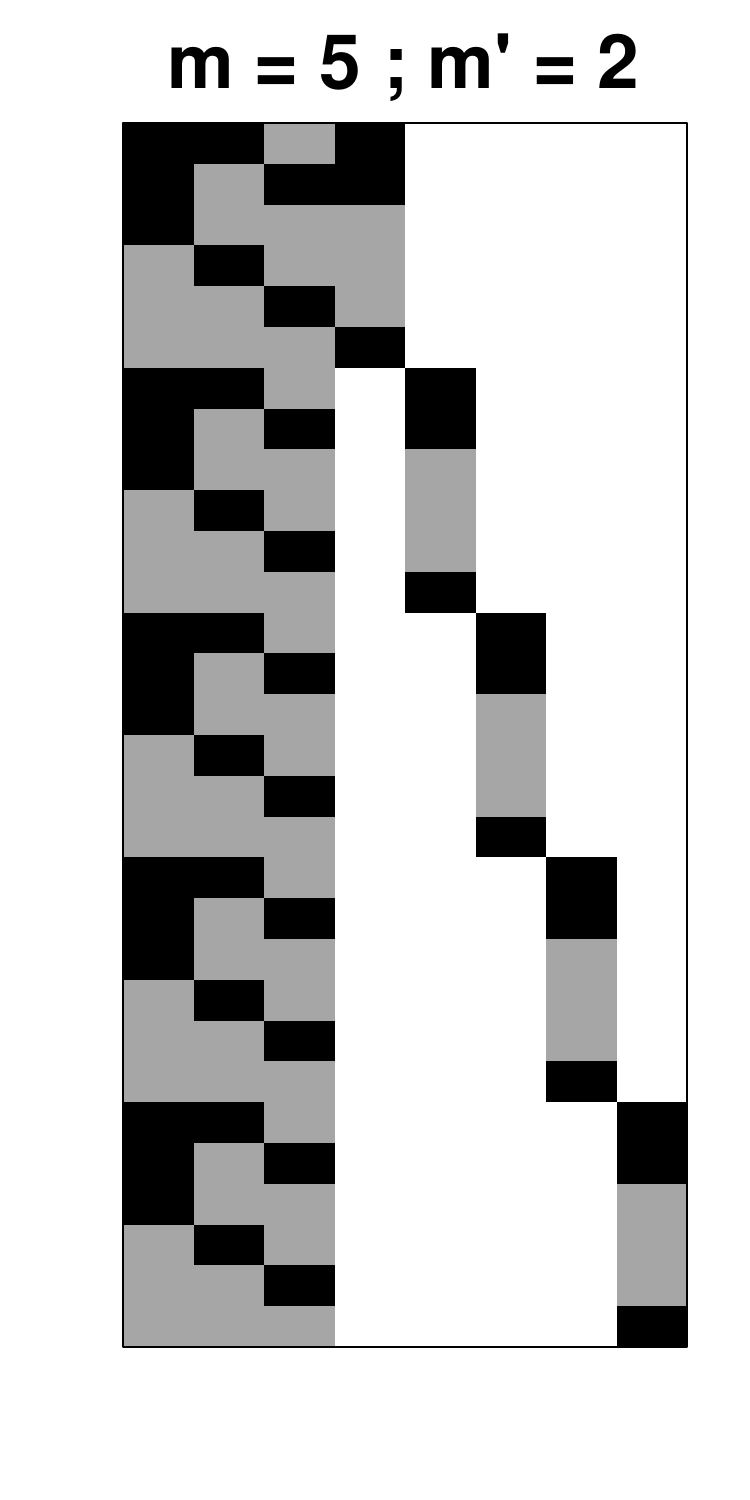}}$\ \ $
\vcenteredhbox{\includegraphics[width=0.21\textwidth]{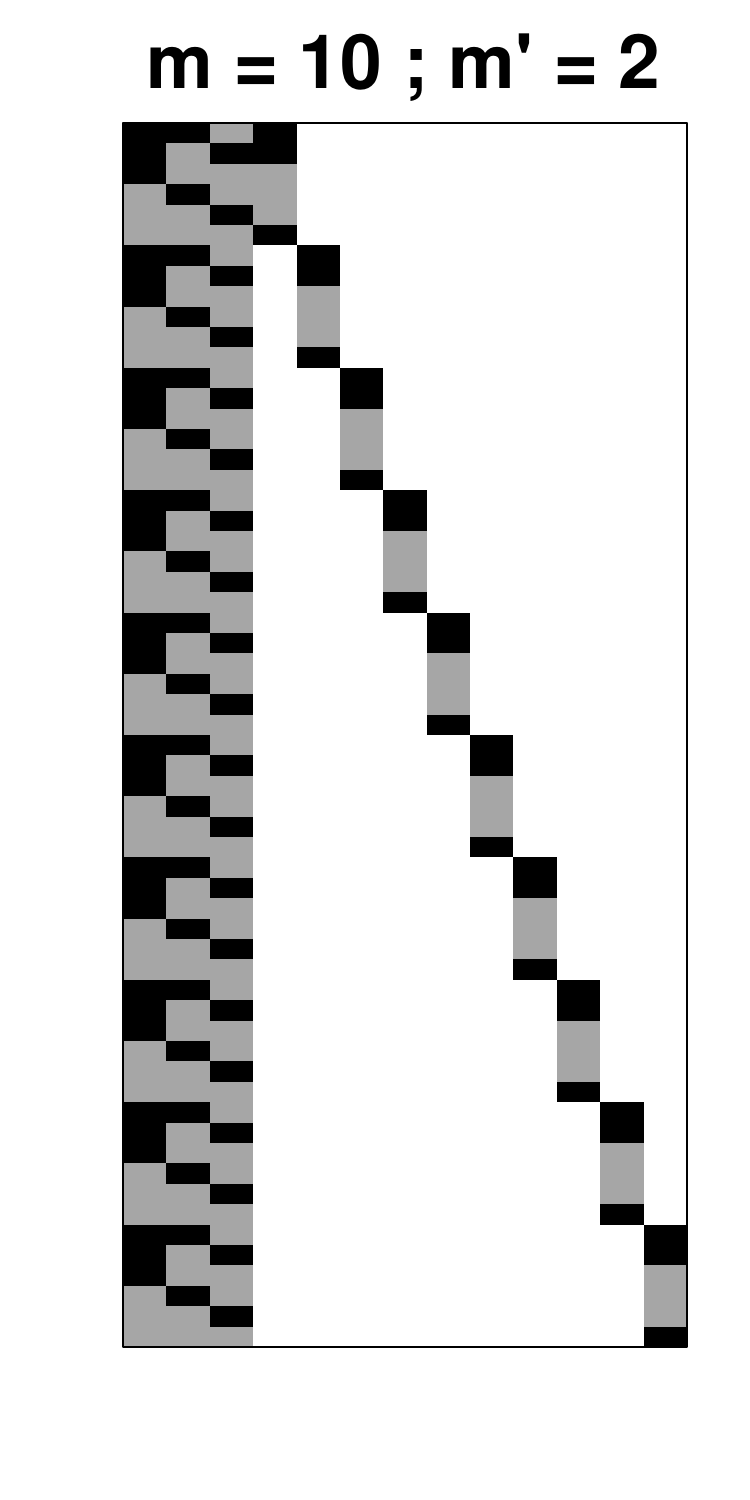}}$\ \ $
\vcenteredhbox{\includegraphics[width=0.28\textwidth]{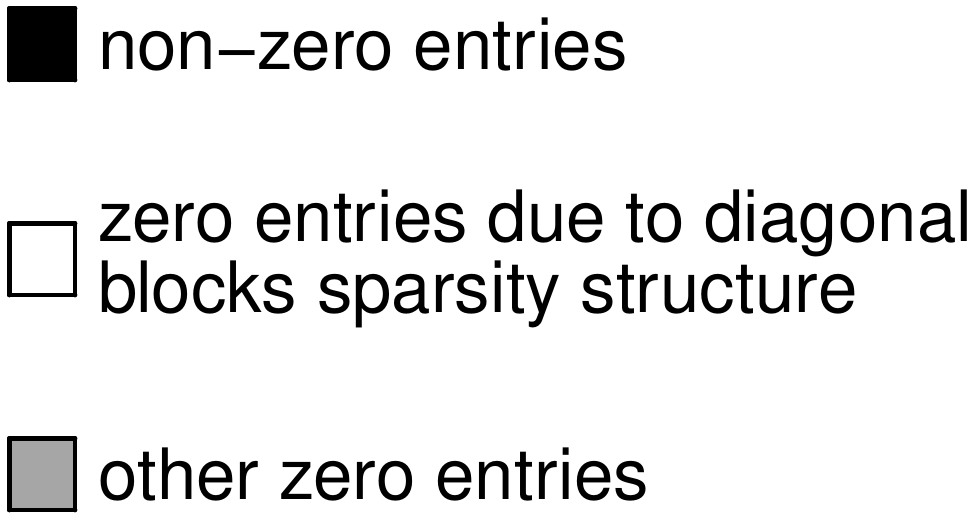}}
\caption{\it Image plot representation of the two-level sparse matrix $\bB$ 
with generic form given by (\ref{eq:BandbFormsReprise}) and with
sub-blocks as defined in Result 2. The dimension
variables are $p=q=q'=n_{ii'}=1$, $m\in\{2,5,10\}$ and $m'=2$. 
\myBlack\ indicates non-zero entries of $\bB$. 
\myWhite\ indicates zero entries corresponding
to the diagonal blocks sparsity structure of $\bB$. 
The \mygrey\ regions also correspond to zero 
entries but which do not contribute to two-level sparse 
structure.}
\label{fig:crossedImage} 
\end{figure}

An interesting future research problem concerns taking advantage of the sparseness
apparent in the \mygrey\ regions of the $\bB$ matrices displayed in 
Figure \ref{fig:crossedImage}. This is a much more subtle pattern of sparseness
compared with the two-level sparse structure corresponding to the \mywhite\ regions
in Figure \ref{fig:crossedImage} and accounting for it would require significant
additional algebraic analysis.

Algorithm \ref{alg:MFVBforPRIII} is a proceduralization of Result \ref{res:prodRestrictIII} 
and delivers the full set of updates of the $\qDens(\bbeta,\buall)$ parameters
under product restriction III. 

\begin{algorithm}[!th]
  \begin{center}
    \begin{minipage}[t]{160mm}
      \begin{small}
        \begin{itemize}
          \setlength\itemsep{4pt}
          \item[] Data Inputs: $\Big\{\left(\yuptri_i,\Xuptri_i,\Zuptri_i,\Zdblacksquare_i\right):\ 1\le i \le m\Big\}$
          \item[] Hyperparameter Inputs: $\bmu_{\bbeta}(p\times1)$,
                  $\bSigma_{\bbeta}(p\times p)\ \mbox{symmetric and positive definite}$,
                     \item[] $\qDens$-Density Inputs: $\mu_{\qDens(1/\sigma^2)}>0$,
                   $\bM_{\qDens(\bSigma^{-1})}(q\times q)$,
                   $\bM_{\qDens((\bSigma')^{-1})}(q'\times q')\ \mbox{symmetric and positive definite}$.
            \item[] For $i = 1,\ldots, m$:
            \begin{itemize}
              \item[] $\bveci\longleftarrow\left[\begin{array}{c}
                       \mu_{\qDens(1/\sigma^2)}^{1/2}\yuptri_i\\[1ex]
                       m^{-1/2}\bSigma_{\bbeta}^{-1/2}\bmu_{\bbeta}\\[1ex]
                       \bzero \\[1ex]
                       \bzero
                       \end{array}
                       \right];\ \ 
                       \Bmati\longleftarrow\left[\begin{array}{cc}
                       \mu_{\qDens(1/\sigma^2)}^{1/2}\Xuptri_i & \mu_{\qDens(1/\sigma^2)}^{1/2}\Zdblacksquare_i\\[1ex]
                       m^{-1/2}\bSigma_{\bbeta}^{-1/2} & \bO\\[1ex]
                       \bO & m^{-1/2}\Big(\bI_{\mdash} \otimes \bM_{\qDens((\bSigma')^{-1})}^{1/2}\Big)\\[1ex]
                       \bO & \bO
                       \end{array}
                       \right]$ \\[1ex]
              \item[] $\ \Bmatdoti\longleftarrow
                       \left[\begin{array}{c}
                       \mu_{\qDens(1/\sigma^2)}^{1/2}\Zuptri_i \\[1ex]
                       \bO\\[1ex]
                       \bO\\[1ex]
                       \bM_{\qDens(\bSigma^{-1})}^{1/2}
                       \end{array}
                       \right].$
            \end{itemize}
            \item[] $\Ssc\thickarrow\SolveTwoLevelSparseLeastSquares
                    \Big(\big\{(\bveci,\Bmati,\Bmatdoti):1\le i \le m\big\}\Big)$
            \item[] $\bmu_{\qDens(\bbeta)}\thickarrow\mbox{first $p$ rows of $\xveco$ component of $\Ssc$}$
            \item[] $\bSigma_{\qDens(\bbeta)}\thickarrow\mbox{top left $p \times p$ sub-block of $\AUoo$ 
                     component of $\Ssc$}$
            \item[] $\iStt \thickarrow p + 1$ 
            \item[] For $\idash = 1, \ldots, \mdash$:
            \begin{itemize}
              \setlength\itemsep{4pt}
              \item[] $\iEnd \thickarrow \iStt + q^{\prime} - 1$
              \item[] $\bmu_{\qDens(\udashidash)} \thickarrow \mbox{ sub-vector of $\bx_{1}$ component 
                      of $\Ssc$ with entries $\iStt$ to $\iEnd$}$
              \item[] $\bSigma_{\qDens(\udashidash)} \thickarrow \mbox{ diagonal sub-block of $\AUoo$ component
                       of $\Ssc$ with rows $\iStt$ to $\iEnd$}$
              \item[] $ \hspace{19mm} \mbox{and columns $\iStt$ to $\iEnd$}$
              \item[] $E_{\qDens} \{ (\bbeta - \bmu_{\qDens(\bbeta)}) (\udashidash 
                       - \bmu_{\qDens(\udashidash)})^{T} \} \thickarrow \mbox{ sub-block of $\AUoo$
                       component of $\Ssc$ with}$
              \item[] $\hspace{58mm} \mbox{rows $1$ to $p$ and columns $\iStt$ to $\iEnd$}$
              \item[] $\iStt \thickarrow \iEnd + 1$
            \end{itemize}
            \item[] For $i=1,\ldots,m$:
            \begin{itemize}
              \setlength\itemsep{4pt}
              \item[] $\bmu_{\qDens(\bu_{i})}\thickarrow\mbox{$\bx_{2, i}$ component of $\Ssc$}$\ \ \ ;\ \ \
                      $\bSigma_{\qDens(\bu_{i})}\thickarrow\mbox{$\bA^{22, i}$ component of $\Ssc$}$
              \item[] $E_{\qDens}\{(\bbeta - \bmu_{\qDens(\bbeta)})(\bu_{i}-\bmu_{\qDens(\bu_{i})})^T\}\thickarrow
                       \mbox{ sub-matrix of $\bA^{12, i}$ component of $\Ssc$ with rows $1$ to $p$}$
              \item[] $\bOmega\thickarrow \mbox{$\bA^{12, i}$ component of $\Ssc$}$\ \ \ ;\ \ \ $\iStt \thickarrow p + 1$
              \item[] For $\idash=1,\ldots,\mdash$:
              \begin{itemize}
                \setlength\itemsep{4pt}
                \item[] $\iEnd \thickarrow \iStt + q' - 1$
                \item[] $E_q\{(\bu_{i}-\bmu_{\qDens(\bu_{i})})(\udashidash - \bmu_{\qDens(\udashidash)})^T\} \thickarrow
                        \mbox{sub-matrix of $\bOmega^T$ with columns $\iStt$ to $\iEnd$}$
                \item[] $\iStt \thickarrow \iEnd + 1$
              \end{itemize}
            \end{itemize}
        \item[]\textsl{continued on a subsequent page}\ $\ldots$
        \end{itemize}
      \end{small}
    \end{minipage}
  \end{center}
  \caption{\it Mean field variational Bayes algorithm for updating the 
parameters of $\qDens(\bbeta,\buall)$ under product restriction III.}
\label{alg:MFVBforPRIII}
\end{algorithm}

\setcounter{algorithm}{2}
\begin{algorithm}[!th]
  \begin{center}
    \begin{minipage}[t]{160mm}
      \begin{small}
        \begin{itemize}
           \item[] Outputs: $\bmu_{\qDens(\bbeta)},\,\bSigma_{\qDens(\bbeta)},
                  \left\{\left(\bmu_{\qDens(\bu_{i})},\bSigma_{\qDens(\bu_{i})}\right):\ 1 \le i \le m\right\},
                  \left\{\left(\bmu_{\qDens(\bu'_{i'})},\bSigma_{\qDens(\bu'_{i'})}\right):\ 1 \le i' \le m'\right\},$
          \item[]$\left\{E_{\qDens}\{(\bbeta - \bmu_{\qDens(\bbeta)})
          (\bu_{i}-\bmu_{\qDens(\bu_{i})})^T:\ 1\le i\le m\right\},$
             $\left\{E_{\qDens}\{(\bbeta - \bmu_{\qDens(\bbeta)})
                   (\bu'_{i'}-\bmu_{\qDens(\bu'_{i'})})^T:\ 1\le i'\le m'\right\},$
          \item[] $\left\{E_{\qDens}\{(\bu_i - \bmu_{\qDens(\bu_i)})(\bu'_{i'}-\bmu_{\qDens(\bu'_{i'})})^T\}:
                                 1 \le i \le m,\ 1 \le i' \le m' \right\}$
        \end{itemize}
      \end{small}
    \end{minipage}
  \end{center}
\caption{\textbf{continued.} \textit{This is a continuation of the description of this algorithm that commences
on a preceding page.}}
\end{algorithm}

\subsection{Variational Inference for $\sigma^2$, $\bSigma$ and $\bSigma'$}

Given the current values of the $\qDens(\bbeta,\buall)$ parameters, the updates of
the parameters of $\qDens(\sigma^2)$, $\qDens(\bSigma)$ and $\qDens(\bSigma')$
are relatively simple. For example, $\sigma^2$ has the Inverse $\chi^2$ prior 
as given by (\ref{eq:priorsA}) then standard mean field variational Bayes arguments 
(e.g. Bishop, 2006; Sections 10.1--10.3) lead to $\xi_{\qDens(\sigma^2)}=\xi_{\sigma^2}+\ndotdot$ 
and 
{\setlength\arraycolsep{3pt}
\begin{eqnarray*}
\lambda_{\qDens(\sigma^2)}&=&\lambda_{\sigma^2}+E_{\qDens}\Vert\by-\bX\bbeta-\bZ\buall\Vert^2\\
&=&\lambda_{\sigma^2}+\Vert\by-\bX\bmu_{\qDens(\bbeta)}-\bZ\bmu_{\qDens(\buall)}\Vert^2
+\tr\big([\bX\ \bZ]\bSigma_{\qDens(\bbeta,\buall)}\big).
\end{eqnarray*}
}
Under product restriction I the trace term reduces to 
$$\tr\left\{[\bX\ \bZ]^T[\bX\ \bZ]\,\bSigma_{\qDens(\bbeta,\buall)}\right\}=\sum_{i=1}^m\sum_{i'=1}^{m'}
\big\{
\mbox{tr}(\bX_{ii'}^T\bX_{ii'}\bSigma_{\qDens(\bbeta)})
+\mbox{tr}(\bZ_{ii'}^T\bZ_{ii'}\bSigma_{\qDens(\bu_i)})
+\mbox{tr}({\bZ}^{\prime T}_{ii'}{\bZ}^{\prime}_{ii'}{\bSigma}_{\qDens(\bu'_{i'})})\big\}.
$$
For product restrictions II and III additional terms are present due to non-zero
cross-expectations and is reflected in the $\lambda_{\qDens(\sigma^2)}$ updates in 
Algorithm \ref{alg:crossedMFVB} given in the next sub-section.

The updates for the parameters of $\qDens(\bSigma)$ and $\qDens(\bSigma')$ uses
analogous arguments, and this is also reflected in the $\bLambda_{\qDens(\bSigma)}$
and $\bLambda_{\qDens(\bSigma')}$ updates of Algorithm \ref{alg:crossedMFVB}.

\subsection{Full Streamlined Mean Field Variational Algorithm}

We are now ready to list a full streamlined mean field variational inference algorithm,
listed as Algorithm \ref{alg:crossedMFVB}, that accounts for any of product restrictions 
I, II or III. It also allows for the covariance matrix prior specification to be 
(\ref{eq:priorsA}) or (\ref{eq:priorsB}).

\null\vfill\eject
\begin{algorithm}[H]
  \begin{center}
    \begin{minipage}[t]{154mm}
      \begin{small}
        \begin{itemize}
          \setlength\itemsep{4pt}
          \vskip1mm
          \item[] Data Inputs: $\by_{ii'}, \ \ \bX_{ii'} , \ \ \bZ_{ii'}, \ \ \Zdash_{ii'},$
                          $\quad  1\le i \le m,\ 1\le i'\le m'$.
          \item[] Hyperparameter Inputs: $\bmu_{\bbeta}(p\times1)$,
                  $\bSigma_{\bbeta}(p\times p)\ \mbox{symmetric and positive definite}$.
          \item[] \qquad\qquad\qquad\qquad\qquad\qquad If priors (\ref{eq:priorsA}): $\xi_{\sigma^2},\lambda_{\sigma^2}>0$,
                                            $\xi_{\bSigma}>2(q-1)$,\ $\xi_{\bSigma'}>2(q'-1)$, 
          \item[]$\qquad\qquad\qquad\qquad\qquad\qquad\bLambda_{\bSigma},\bLambda_{\bSigma'}$ positive definite.
          \item[] \qquad\qquad\qquad\qquad\qquad\qquad If priors (\ref{eq:priorsB}): $\nu_{\sigma^2},\ssigsq,\nuSigma,\nuSigmad,
                    \sSigmaOne,\ldots,\sSigmaq,\sSigmadOne,\ldots,\sSigmadqd> 0$.
          \item[] Product Restriction Input: Specification of product restriction I, II or III.
          \item[] $\yuptri_i\longleftarrow{\displaystyle\stack{1\le i'\le m'}}(\by_{ii'}), 
                   \ \ \Xuptri_i\longleftarrow{\displaystyle\stack{1\le i'\le m'}}(\bX_{ii'}),
                   \ \ \Zuptri_i\longleftarrow{\displaystyle\stack{1\le i'\le m'}}(\bZ_{ii'}),\quad 1\le i\le m$.
          \item[] If product restriction III then:\ $\Zdblacksquare_i\longleftarrow
                                             {\displaystyle\blockdiag{1\le i'\le m'}}(\bZ'_{ii'}),\quad 1\le i \le m$
          \item[] If product restriction I or II then:\ $\ydntri_{i'}\longleftarrow{\displaystyle\stack{1\le i\le m}}(\by_{ii'}), 
                          \ \ \Xdntri_{i'}\longleftarrow{\displaystyle\stack{1\le i\le m}}(\bX_{ii'}),$
          \item[] $\qquad\qquad\qquad\qquad\qquad\qquad\qquad\qquad
                    \Zdntri_{i'}\longleftarrow{\displaystyle\stack{1\le i\le m}}(\bZ'_{ii'}),\quad 1\le i'\le m'$.
          \item[]If product restriction I then:\ $\by\longleftarrow{\displaystyle\stack{1\le i\le m}}(\yuptri_i),\ \ 
                           \bX\longleftarrow{\displaystyle\stack{1\le i\le m}}(\Xuptri_i)$.
          \item[] If priors (\ref{eq:priorsA})
          \begin{itemize}
          \item[] $\xi_{\qDens(\sigma^2)}\thickarrow\xi_{\sigma^2}+\ndotdot$\ \ \ ;\ \ \
                  $\xi_{\qDens(\bSigma)}\thickarrow\xi_{\bSigma}+m$,\ \ \ ;\ \ \
                  $\xi_{\qDens(\bSigma')}\thickarrow\xi_{\bSigmad}+m'$
          \end{itemize}
          \item[] If priors (\ref{eq:priorsB})
          \begin{itemize}
          \item[] initialize: $\muq{1/a_{\sigma^2}} > 0$,
                              $\bM_{\qDens(\ASigma^{-1})}$, $\bM_{\qDens(\ASigmad^{-1})}$ positive definite.
          \item[] $\xi_{\qDens(\sigma^2)}\thickarrow\nu_{\sigma^2}+\ndotdot$
                  \ \ \ ;\ \ \
                  $\xi_{\qDens(\bSigma)}\thickarrow\nuSigma+2q-2+m$,\ \ \ ;\ \ \
                  $\xi_{\qDens(\bSigma')}\thickarrow\nuSigmad+2q'-2+m'$
          \item[] $\xi_{\qDens(\asigsq)}\thickarrow\nu_{\sigma^2}+1$\ \ \ ; \ \ \  
                  $\xi_{\qDens(\ASigma)}\thickarrow\nu_{\bSigma}+q$\ \ \ ; \ \ \  
                  $\xi_{\qDens(\ASigmad)}\thickarrow\nu_{\bSigma'}+q'$ 
          \end{itemize}
          \item[] Initialize: $\mu_{\qDens(1/\sigma^2)} > 0$,
                              $\bM_{\qDens(\bSigma^{-1})}$, $\bM_{\qDens((\bSigmad)^{-1})}$ positive definite.
          \item[] Cycle:
          \begin{itemize}
            \item[] If prod. restrict. I: call Algorithm \ref{alg:MFVBforPRI} 
                    to update $\bmu_{\qDens(\bbeta,\buall)}$ and relevant $\bSigma_{\qDens(\bbeta,\buall)}$ blocks
            \item[] If prod. restrict. II: call Algorithm \ref{alg:MFVBforPRII} 
                    to update $\bmu_{\qDens(\bbeta,\buall)}$ and relevant $\bSigma_{\qDens(\bbeta,\buall)}$ blocks
            \item[] If prod. restrict. III: call Algorithm \ref{alg:MFVBforPRIII} 
                    to update $\bmu_{\qDens(\bbeta,\buall)}$ and relevant $\bSigma_{\qDens(\bbeta,\buall)}$ blocks
            \item[] If priors (\ref{eq:priorsA}):\ \ \  $\lambda_{\qDens(\sigma^2)}\thickarrow\lambda_{\sigma^2}$\ \ \ ;\ \ \ 
                    $\Lambda_{\qDens(\bSigma)}\thickarrow\bLambda_{\bSigma}$\ \ \ ;\ \ \ 
                    $\Lambda_{\qDens(\bSigma')}\thickarrow\bLambda_{\bSigma'}$
            \item[] If priors (\ref{eq:priorsB}):\ \ \  $\lambda_{\qDens(\sigma^2)}\thickarrow\muq{1/a_{\sigma^2}}$\ \ \ ;\ \ \ 
                    $\Lambda_{\qDens(\bSigma)}\thickarrow\bM_{\qDens(\ASigma^{-1})}$\ \ \ ;\ \ \ 
                    $\Lambda_{\qDens(\bSigma')}\thickarrow\bM_{\qDens(\ASigmad^{-1})}$
            \item[] For $i=1,\ldots,m$:
            \begin{itemize}
              \setlength\itemsep{4pt}
              \item[] For $\idash=1,\ldots,\mdash$:
              \begin{itemize}
                \setlength\itemsep{4pt}
                \item[] $\lambda_{\qDens(\sigma^2)}\thickarrow \lambda_{\qDens(\sigma^2)}
                          +\big\Vert\by_{i\idash}-\bX_{i\idash}\bmu_{\qDens(\bbeta)}-\bZ_{i\idash}\bmu_{\qDens(\bu_i)}
                          -\Zdash_{i\idash}\bmu_{\qDens(\udashidash)}\big\Vert^2$
                \item[] $\lambda_{\qDens(\sigma^2)}\thickarrow \lambda_{\qDens(\sigma^2)}
                          +\mbox{tr}(\bX_{i\idash}^T\bX_{i\idash}\bSigma_{\qDens(\bbeta)})
                          +\mbox{tr}(\bZ_{i\idash}^T\bZ_{i\idash}\bSigma_{\qDens(\bu_i)})+\mbox{tr}({\bZ}^{\prime T}_{i\idash}
                          {\bZ}^{\prime}_{i\idash}{\bSigma}_{\qDens({\bu}^{\prime}_{\idash})})$
                \item[] If product restriction II or III:
                        \item[] $\qquad\lambda_{\qDens(\sigma^2)}\thickarrow \lambda_{\qDens(\sigma^2)}
                         +2\,\mbox{tr}\big[\bZ_{i\idash}^T\bX_{i\idash}E_{\qDens}\{(\bbeta-\bmu_{\qDens(\bbeta)})
                         (\bu_i-\bmuq{\bu_i})^T\}\big]$
                       \item[] $\qquad$If product restriction III:
                       \item[] $\qquad\qquad\lambda_{\qDens(\sigma^2)}\thickarrow \lambda_{\qDens(\sigma^2)}
                          +2\,\mbox{tr}\big[\bZ^{\prime T}_{i\idash}\bX_{i\idash}E_{\qDens}\{(\bbeta-\bmu_{\qDens(\bbeta)})
                          (\udashidash-\bmuq{\udashidash})^T\}\big]$
                       \item[] $\qquad\qquad\lambda_{\qDens(\sigma^2)}\thickarrow \lambda_{\qDens(\sigma^2)}
                          +2\,\mbox{tr}\big[\bZ^{\prime T}_{i\idash}\bZ_{i\idash}E_{\qDens}\{
                          (\bu_i-\bmuq{\bu_i})(\udashidash-\bmuq{\udashidash})^T\}\big]$
              \end{itemize}
            \end{itemize}
            \item[] \textsl{continued on a subsequent page}\ $\ldots$
            \end{itemize}
        \end{itemize}
      \end{small}
    \end{minipage}
  \end{center}
  \caption{\it Mean field variational Bayes algorithm for determining
the optimal $\qDens$-density parameters in the Bayesian crossed random
effects model under either product restriction I, II or III.}
  \label{alg:crossedMFVB}
\end{algorithm}

\setcounter{algorithm}{3}
\begin{algorithm}[H]
  \begin{center}
    \begin{minipage}[t]{154mm}
      \begin{small}
        \begin{itemize}
          \setlength\itemsep{4pt}
          \item[] 
           \begin{itemize}
             For $i=1,\ldots,m$:
            \begin{itemize}
              \setlength\itemsep{4pt}
              \item[] $\bLambda_{\qDens(\bSigma)}\thickarrow\bLambda_{\qDens(\bSigma)}+\bmu_{\qDens(\bu_i)}\bmu_{\qDens(\bu_i)}^T
                            + \bSigma_{\qDens(\bu_i)}$
            \end{itemize}
            \item[] For $i'=1,\ldots,m'$:
            \begin{itemize}
              \setlength\itemsep{4pt}
              \item[] $\bLambda_{\qDens(\bSigma')}\thickarrow\bLambda_{\qDens(\bSigma')}+\bmu_{\qDens(\bu'_{i'})}\bmu_{\qDens(\bu'_{i'})}^T
                            + \bSigma_{\qDens(\bu'_{i'})}$
            \end{itemize}
            \item[] $\muq{1/\sigma^2} \leftarrow \xi_{\qDens(\sigma^2)}/\lambda_{\qDens(\sigma^2)}$\ \ \ ;\ \ \ 
                    $\MqSigma \thickarrow(\xi_{\qDens(\bSigma)}-q+1)\,\bLambda^{-1}_{\qDens(\bSigma)}$
            \item[] $\MqSigmad \thickarrow(\xi_{\qDens(\bSigma')}-q'+1)\,\bLambda^{-1}_{\qDens(\bSigma')}$ 
            \item[] If priors (\ref{eq:priorsB}):
            \begin{itemize}
            \item[] $\lambda_{\qDens(\asigsq)}\thickarrow\muq{1/\sigma^2}+1/(\nusigsq\ssigsq^2)$\ \ \ ;\ \ \ 
                    $\muq{1/\asigsq} \thickarrow \xi_{\qDens(\asigsq)}/\lambda_{\qDens(\asigsq)}$ 
            \item[] $\bLambda_{\qDens(\ASigma)}\thickarrow 
                    \diag\big\{\mbox{diagonal}\big(\bM_{\qDens(\bSigma^{-1})}\big)\big\}+\{\nuSigma\diag(\sSigmaOne^2,\ldots,\sSigmaq^2)\}^{-1}$
            \item[] $\bLambda_{\qDens(\ASigmad)}\thickarrow 
                    \diag\big\{\mbox{diagonal}\big(\bM_{\qDens((\bSigma')^{-1})}\big)\big\}+\{\nuSigmad\diag(\sSigmadOne^2,\ldots,\sSigmadqd^2)\}^{-1}$
            \item[] $\bM_{\qDens(\ASigma^{-1})}\thickarrow \xi_{\qDens(\ASigma)}\bLambda_{\qDens(\ASigma)}^{-1}$\ \ \ ;\ \ \ 
                    $\bM_{\qDens(\ASigmad^{-1})}\thickarrow \xi_{\qDens(\ASigmad)}\bLambda_{\qDens(\ASigmad)}^{-1}$
            \end{itemize}
          \end{itemize}
          \item[] Outputs: $\bmu_{\qDens(\bbeta)},\,\bSigma_{\qDens(\bbeta)},
                  \left\{\left(\bmu_{\qDens(\bu_{i})},\bSigma_{\qDens(\bu_{i})}\right):\ 1 \le i \le m\right\},
                  \left\{\left(\bmu_{\qDens(\bu'_{i'})},\bSigma_{\qDens(\bu'_{i'})}\right):\ 1 \le i' \le m'\right\},$
          \item[] $\qquad\qquad\xi_{\qDens(\sigma^2)},\lambda_{\qDens(\sigma^2)},\xi_{\qDens(\bSigma)},\bLambda_{\qDens(\bSigma)},
                   \xi_{\qDens(\bSigmad)},\bLambda_{\qDens(\bSigmad)}.$ 
          \item[]$\qquad\qquad$ If product restriction II or III add: $\left\{E_{\qDens}\{(\bbeta - \bmu_{\qDens(\bbeta)})
          (\bu_{i}-\bmu_{\qDens(\bu_{i})})^T:\ 1\le i\le m\right\}.$
          \item[] $\qquad\qquad$ If product restriction III add: 
             $\left\{E_{\qDens}\{(\bbeta - \bmu_{\qDens(\bbeta)})(\bu'_{i'}-\bmu_{\qDens(\bu'_{i'})})^T:\ 1\le i'\le m'\right\}.$
          \item[] $\qquad\qquad\qquad\qquad\quad
                   \mbox{and}\ \left\{E_{\qDens}\{(\bu_i - \bmu_{\qDens(\bu_i)})(\bu_{i'}-\bmu_{\qDens(\bu'_{i'})})^T\}:
                                 1 \le i \le m,\ 1 \le i' \le m' \right\}.$ 
        \end{itemize}
      \end{small}
    \end{minipage}
  \end{center}
\caption{\textbf{continued.} \textit{This is a continuation of the description of this algorithm that commences
on a preceding page.}}
\end{algorithm}

\tcm{Throughout this article we confine discussion to the Gaussian response version of the 
linear mixed model with crossed random effects. Item response theory and Rasch analysis models,
which enjoy widespread use in psychometrics, have random effects structures similar to those 
given by (\ref{eq:BayesianCrossedModel}). They usually involve different conditional 
response distributions such as those corresponding to multivariate binary and multivariate 
categorical data. However, the streamlined variational inference challenges arising in random effect 
structures are independent of the likelihood. The variational message passing approach to variational 
inference (e.g. Wand, 2017, Nolan \textit{et al.}, 2020) formalizes this separation via notions 
such as factor graph fragments. The upshot is that Results \ref{res:prodRestrictII} and \ref{res:prodRestrictIII}
are still relevant to non-Gaussian crossed random effects models such as the psychometrics versions
just mentioned.}

\section{Performance Assessment and Comparison}\label{sec:perfAssess}

Any set of statistical methods for a particular problem can be 
assessed and compared on various criteria such as ease of 
implementation, time to compute and various measures of statistical 
accuracy. In this section we focus on accuracy in terms of how
close variational approximate posterior density functions are
to their exact counterparts and computational speed. The second
of these assessments and comparisons allows appreciation for the
scalability of competing approaches to very large mixed models with 
crossed random effects.

\subsection{Accuracy Assessment and Comparison}\label{sec:accAss}

We ran a simulation study to compare and assess the accuracy performance of the
three mean field variational inference schemes. The study involved simulating
100 replications of data from a version of the crossed random
effects model (\ref{eq:BayesianCrossedModel}). The dimension variables
were set to be:
$$m=100,\quad m'=20,\quad n_{ii'}=10\quad\mbox{and}\quad p=q=q'=2.$$
The true values of the parameters from which the data were generated are
\begin{equation}
\bbetatrue=\left[
\arraycolsep=1.5pt
\begin{array}{c}
0.58\\
1.89
\end{array}
\right],
\ \ 
\sigsqtrue=0.3,
\ \  
\SigmaTrue=\left[
\arraycolsep=1.5pt
\begin{array}{rr}
0.46 & -0.19 \\
-0.19 & 0.17
\end{array}
\right]
\ \mbox{and}\ \ 
\SigmadTrue=\left[
\arraycolsep=1.5pt
\begin{array}{rr}
0.3  & -0.12 \\
-0.12 & 0.25
\end{array}
\right].
\label{eq:trueParms}
\end{equation}
Each of the $\bX_{ii'}$, $\bZ_{ii'}$, $\bZ'_{ii'}$, 
$1\le i\le 100$, $1\le i'\le 20$, were $10\times2$
matrices with a column of ones and a column of 
predictor values generated to be independent
and uniformly on the unit interval. 

The priors on $\sigma^2$, $\bSigma$ and $\bSigma'$ were of
(\ref{eq:priorsB}). The hyperparameter values were 
$\bmu_{\bbeta}=\bzero$, $\bSigma_{\bbeta}=10^{10}\bI$,
$\nu_{\sigma^2}=1$, $\nu_{\bSigma}=\nu_{\bSigma'}=2$ 
and $s_{\sigma^2}=s_{\bSigma,1}=s_{\bSigma,2}
=s_{\bSigma',1}=s_{\bSigma',2}=10^5$.

For each replication we obtained approximate posterior density
functions for all model parameters and random effects using
both mean field variational Bayes and Markov chain Monte Carlo.
The mean field variational Bayes approximations were obtained
by running Algorithm \ref{alg:crossedMFVB} with each of product 
restrictions I, II and III. The number of iterations was fixed at 500.
Markov chain Monte Carlo approximate density functions
were obtained using the package \textsf{rstan} (Stan Development Team, 2021) 
within the \textsf{R} language (\textsf{R} Core Team, 2019).
One thousand warm-up samples were generated, followed by another
1000 samples retained for approximate inference.
Kernel density estimation, with direct plug-in bandwidth selection
(e.g. Wand \myand Jones, 1995; Section 3.6.1), was used to obtain 
approximate posterior density functions. 

\begin{figure}[!ht]
\centering
{\includegraphics[width=\textwidth]{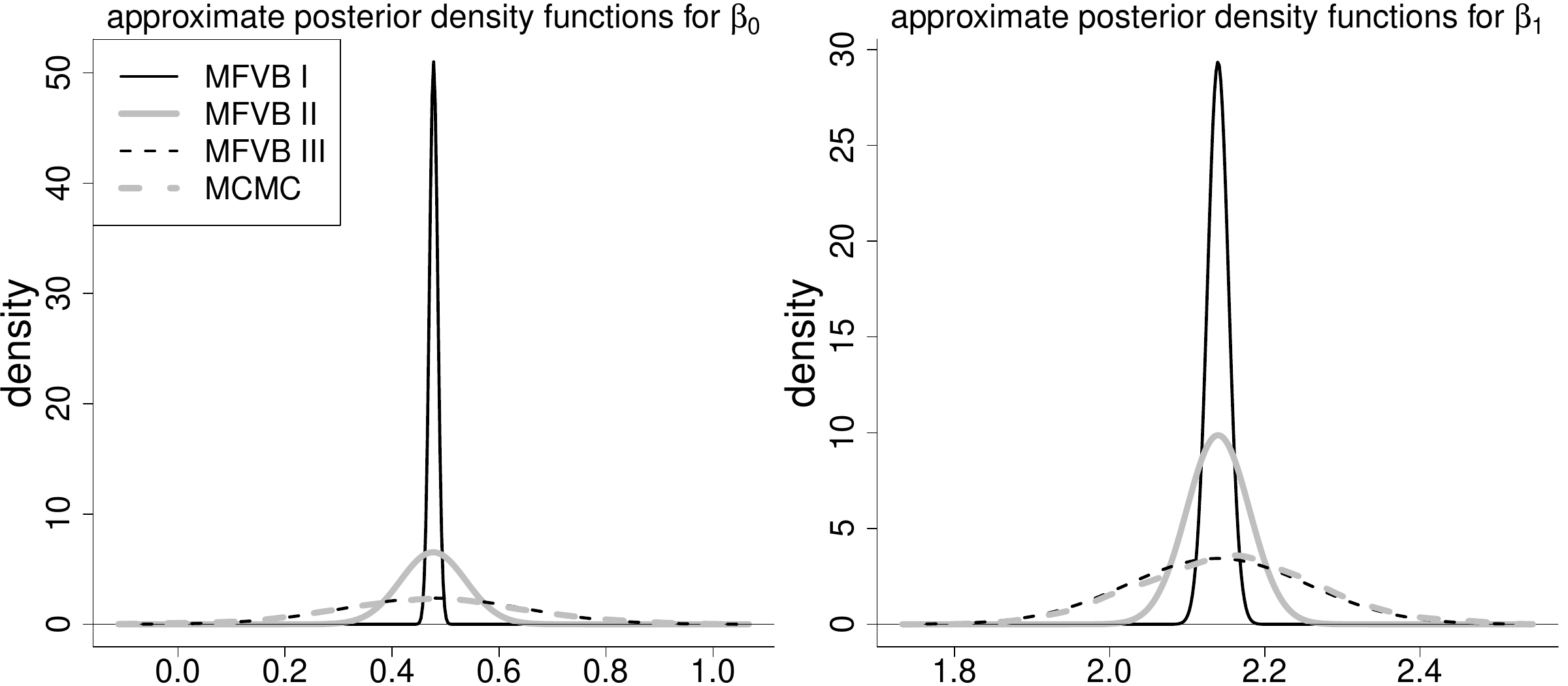}}
\caption{\it Approximate posterior density functions for $\beta_0$ and $\beta_1$,
according to three different mean field variational Bayes (MFVB) schemes and Markov chain
Monte Carlo (MCMC), for the first replication of the simulation study. The legend
uses the abbreviation ``MFVB I'' for the mean field variational Bayes according
to product restriction I. Similar abbreviations are used for the other product
restrictions.}
\label{fig:MCMCvsMFVBfixEffDensPlot} 
\end{figure}

Figure \ref{fig:MCMCvsMFVBfixEffDensPlot} compares the approximations
for the posterior distributions of the two entries of $\bbeta$. 
We denote these entries as $\beta_0$, the fixed effects intercept, 
and $\beta_1$, the fixed effects slope. The difference between
the three variational approximations is quite striking. For 
product restriction I the posterior variances are much too low,
due to posterior correlations between the entries of
$\bbeta$, $\bu$ and $\bu'$ being set to zero. However, the 
product restriction III leads to very good concordance with
the Markov chain Monte Carlo posterior densities. The 
density functions for product restriction II have intermediate
approximation quality, but appear to be closer to those
of product restriction III than those of product restriction I.

\begin{figure}[!ht]
\centering
{\includegraphics[width=\textwidth]{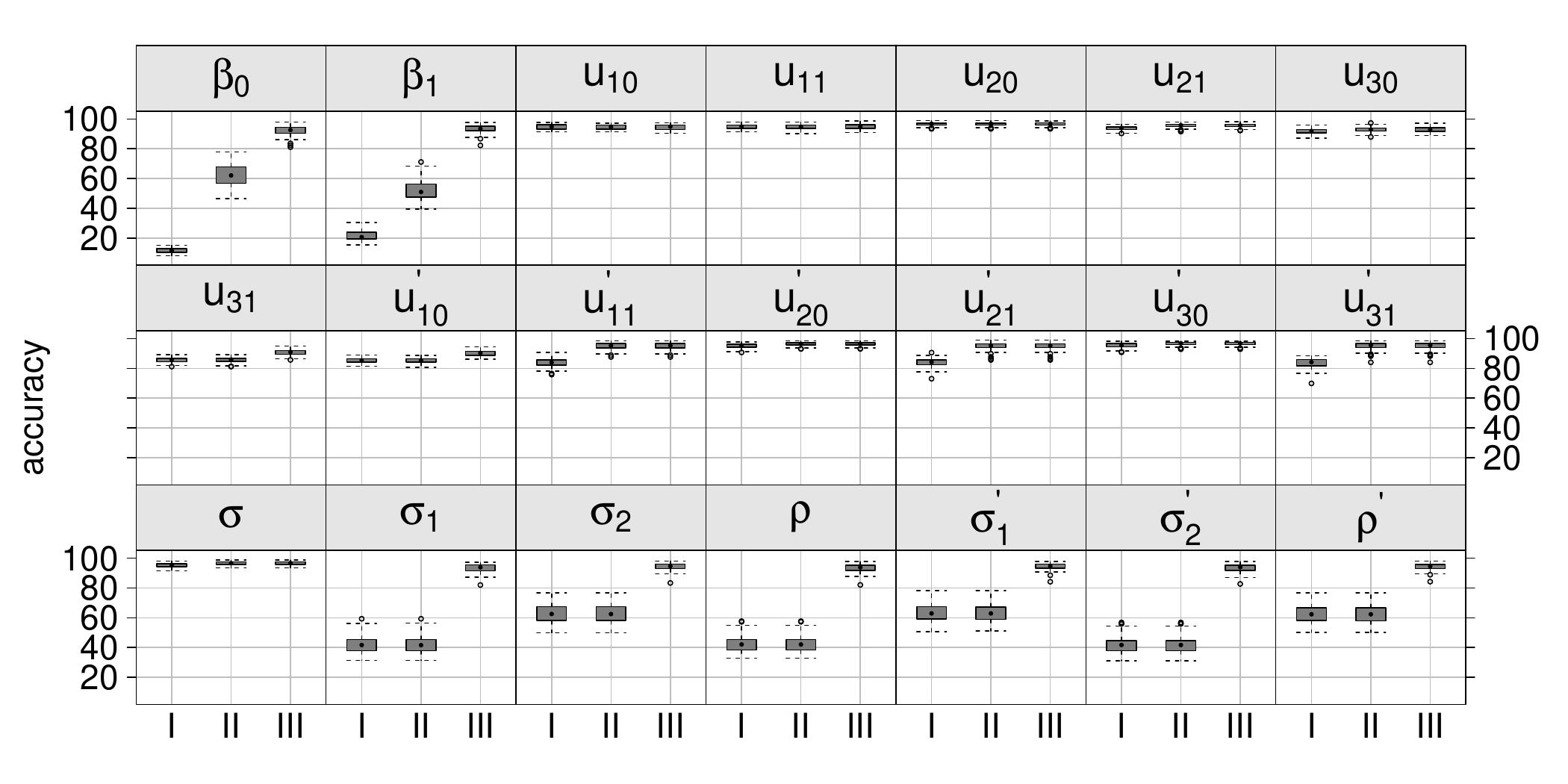}}
\caption{\it Side-by-side boxplots for the accuracy scores for 21 parameters and random effects
from the simulation study, with accuracy defined according to (\ref{eq:accurDefn}). Each panel 
corresponds to a separate parameter or random effect and contains side-by-side boxplots for product
restrictions I, II and III.}
\label{fig:simBoxpSumm} 
\end{figure}

In Figure \ref{fig:simBoxpSumm} we provide a summary of the relative performance
of product restrictions I, II and III for all model parameters and entries of
the first three $\bu_i$ and $\bu'_{i'}$ vectors using side-by-side boxplots 
of estimates of the following accuracy score for a generic target $\theta$:
\begin{equation}
\mbox{accuracy}\equiv\,
100\left\{1-\smhalf\infint\big|\qDens(\theta)-\pDens(\theta|\by)\big|\,d\theta\right\}\%.
\label{eq:accurDefn}
\end{equation}
Note that $0\%\le\mbox{accuracy}\le 100\%$ with a score of $100\%$ if 
$\qDens(\theta)$ and $\pDens(\theta|\by)$ perfectly coincide 
and a score of $0\%$ if there have no overlapping mass.
In practice $\pDens(\theta|\by)$ is replaced by a kernel density estimate
based on a large Markov chain Monte Carlo sample. Depending on tractability,
either $\qDens(\theta)$ is available in closed form or it can be estimated
from a large Monte Carlo sample from the distribution corresponding
to $\qDens(\theta)$.

Apart from the fixed effects parameters $\beta_0$ and $\beta_1$ the parameters
monitored in Figure \ref{fig:simBoxpSumm} are the error standard
deviation $\sigma$, the standard deviation and correlation parameters 
corresponding to the random effects covariance matrix $\bSigma$:
$$\sigma_1\equiv\sqrt{(\bSigma)_{11}},\quad \sigma_2\equiv\sqrt{(\bSigma)_{22}}
\quad\mbox{and}\quad \rho\equiv(\bSigma)_{12}/(\sigma_1\sigma_2)$$
and similar parameters for the random effects covariance matrix $\bSigma'$.
The random effects in Figure \ref{fig:simBoxpSumm} have notation as given by
$$\bu_i=\left[\begin{array}{c}u_{i0}\\  u_{i1}\end{array}\right],\ 1\le i\le 3,
\quad\mbox{and}\quad
\bu'_{i'}=\left[\begin{array}{c}u_{i'0}\\  u_{i'1}\end{array}\right],\ 1\le i'\le 3.
$$
%

From Figure \ref{fig:simBoxpSumm} we see that the biggest discrepancies across the three
product restrictions are for the fixed effects parameters $\beta_0$ and $\beta_1$,
which is in keeping with Figure \ref{fig:MCMCvsMFVBfixEffDensPlot}. Inferential accuracy for
the covariance matrix parameters is very good for all product restrictions and
is excellent for product restriction III. For the $\bu_i$ entries the accuracy 
of product restriction I is lower due to its ignorance of the posterior
correlations between distinct $\bu_i$ vectors. Product restrictions II and III allow
for such correlation and excellent accuracy ensues. However, for the $\bu'_{i'}$
vectors product restriction II sacrifices handling of the corresponding posterior
correlations and the drop in accuracy is quite pronounced.

\begin{figure}[H]
\centering
{\includegraphics[width=0.9\textwidth]{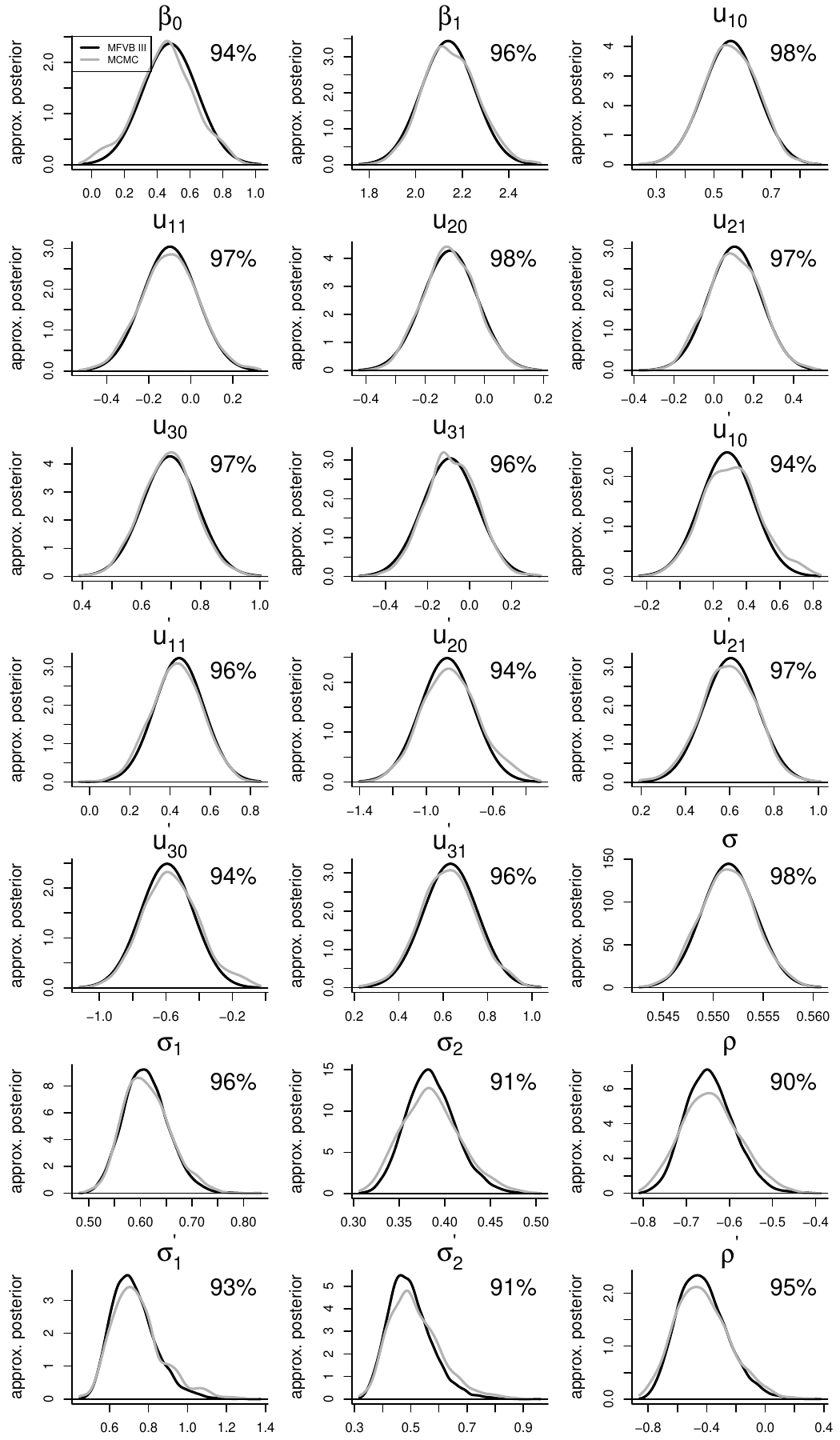}}
\caption{\it Approximate posterior density functions for the 21 parameters and random effects
for the first replication of the simulation study. The blue curves are posterior density functions
obtained using mean field variational Bayes with product restriction III and the orange
curves are based on Markov chain Monte Carlo. The accuracy percentages are 
defined according to (\ref{eq:accurDefn}).}
\label{fig:MCMCvsMFVBmultDensPlot} 
\end{figure}

Since product restriction III is the clear winner in terms of accuracy,
we show the mean field variational Bayes approximate density estimates for
the product restriction in comparison with Markov chain Monte Carlo for
the first replication in Figure \ref{fig:MCMCvsMFVBmultDensPlot}. 
The parameters and random effects subsets are the same as those used in 
Figure \ref{fig:simBoxpSumm}. Accuracy scores are also shown and,
for this data set, is always 92\% or higher. The boxplots in 
Figure \ref{fig:simBoxpSumm} indicate that excellent accuracy
is typical for this particular simulation setting.

\tcb{
The excellent accuracy under product restriction III is tied to the orthogonality 
between $(\bbeta,\bu,\bu')$  and $(\sigma^2,\bSigma,\bSigma')$ from likelihood theory,
h-likelihood theory and best prediction for the frequentist version of (\ref{eq:BayesianCrossedModel}). 
Section 3.1 of Menictas \myand Wand (2013) provides a detailed account of this phenomenon for 
a similar model. The approximately non-informative priors used in this section's
empirical studies imply that the approximate Bayesian inference is close to 
what would be obtained using frequentist paradigms. Since the product density forms of  
product restriction III separate orthogonal quantities, there is little loss in accuracy 
compared with the unrestricted case. On the other hand, there is no such orthogonality
within the components of $(\bbeta,\bu,\bu')$. Hence, product restrictions I and II
pay a price for imposing their product density constraints.
}

\subsection{Speed Assessment and Comparison}

We ran another simulation study that recorded computing times for data
generated according to the model as in the previous subsection's
simulation study -- but with increasing crossed random effects 
dimensions. Specifically, the data were generated according to 
(\ref{eq:trueParms}) with $n_{ii'}=10$ but with 
$$m\in\{100,200,400,800\}\quad\mbox{and}\quad m'=m/5.$$
We then simulated 10 replications of the data for each 
$(m,m')$ combination and recorded the computational times
for fitting via mean field variational Bayes with product restrictions II
and III. 

The mean field variational Bayes computations were performed using
Algorithm \ref{alg:crossedMFVB}, with calls to Algorithms \ref{alg:MFVBforPRII},
\ref{alg:MFVBforPRIII}, \ref{alg:SolveLeastSquares} and \ref{alg:SolveTwoLevelSparseLeastSquares}.
All five algorithms were implemented in the fast \texttt{Fortran 77} language.
The number of mean field variational Bayes iterations was fixed at 100.
All computations were carried out on the third author's \textsf{MacBook Air} laptop,
which has a 2.2 gigahertz processor and 8 gigabytes of random access memory.

Table \ref{tab:timeResults} lists the average and standard deviation times in seconds.  

\begin{table}[!ht]
\centering
\begin{tabular}{lll}
\hline\\[-1.5ex]
$(m,m')$        & MFVB II   & MFVB III \\[1.0ex]
\hline\\[-1.2ex]
(100,20)    &  0.267 (0.0267) &    4.93  (0.082)  \\
(200,40)    &  1.44  (0.0996) &    66.8  (1.40)     \\
(400,80)    &  8.92  (0.587)  &  1130    (8.48)     \\
(800,160)   & 54.7   (1.54)   &  21300   (41.0)     \\
\hline
\end{tabular}
\caption{\it Average (standard deviation) time in seconds for each method,
in the speed assessment study. ``MFVB II''  is short for the mean field variational 
Bayes according to product restriction II and ``MFVB III'' is defined similarly.}
\label{tab:timeResults} 
\end{table}

Table \ref{tab:timeResults} shows that mean field variational Bayes with 
product restriction II scales very well to large crossed random effects
problems with less than a minute required for the largest $(m,m')=(800,160)$
case and less than 10 seconds required for the second largest $(m,m')=(400,80)$
situation. The highly accurate Mean field variational Bayes 
with product restriction III computes in a few seconds for $(m,m')=(100,20)$
and about a minute for $(m,m')=(200,40)$. But eventually it gets 
affected by the quadratic dependence on  $(m,m')$ and the average computing
time up to about 6 hours for $(m,m')=(800,160)$, which is about 400 times
slower than for product restriction II. As we have seen in Figure \ref{fig:simBoxpSumm}, 
the accuracy of product restriction III is higher
than that of product restriction II. Despite their limitation
to a few settings, Figure \ref{fig:simBoxpSumm} and Table \ref{tab:timeResults} 
provides valuable guidance regarding the accuracy versus run-time trade-off
for mean field variational Bayes approaches to approximate inference
for linear mixed models with crossed random effects.

\tcrg{
A more challenging problem is that of meaningful timing comparisons with Markov chain
Monte Carlo alternatives to the Algorithms \ref{alg:MFVBforPRI}--\ref{alg:crossedMFVB} 
streamlined mean field variational Bayes strategies. Firstly, there is the issue that 
the elapsed computation time for mean field variational Bayes approach is governed by 
the number of iterations, whereas for Markov chain Monte Carlo approaches it is sample 
size. Ideally notions of convergence could be used to arrive at comparable stopping rules. 
But this has its own difficulties due to factors such as tolerance choice and chain stickiness.
The accuracy comparisons in Figures \ref{fig:MCMCvsMFVBfixEffDensPlot}  
and \ref{fig:MCMCvsMFVBmultDensPlot} involved the default Markov chain 
Monte Carlo implementation used by the \textsf{rstan} package. For the first three sample size
pairs of Table \ref{tab:timeResults} \textsf{rstan}, with warm-up and kept sample sizes of $1000$,  
required between $50$ and $400$ times the computational time compared with
mean field variational Bayes with product restriction III and failed to compute for
the fourth sample size pair. However, it is well-known that general purpose Bayesian inference 
engines such as \text{rstan} tend to be considerably slower than fit-for-purpose code. 
We implemented the plain block Gibbs sampling algorithm for model (\ref{eq:BayesianCrossedModel})
in a low-level language. As expected, this was much faster than \textsf{rstan} with respect to
number of draws per second. However, plain block Gibbs sampling exhibited extremely poor mixing 
for the fixed effects parameters with lag 1 autocorrelation values as high as $0.99$. For tests 
involving warm-up and kept sample sizes of $1000$ the \emph{effective} sample size, according 
to the definition used by the \textsf{rstan} package, was as low as $5$ for the components of 
$\bbeta$. The \textsf{rstan} effective sample sizes are much higher, typically by a factor of $10$ 
or more. This chain stickiness problem with plain block Gibbs sampling implies a degradation 
in the quality of its Bayesian inference which stymies fair timing comparisons.
Recent work by Papaspiliopoulos \textit{et al.} (2020) provided a theoretical explanation of the poor 
performance of plain Gibbs sampling for cross random effects models and proposed a remedy
for models similar to (\ref{eq:BayesianCrossedModel}). This new work may lead to competitive scalable 
alternatives to this article's streamlined mean field variational Bayes approaches for model 
(\ref{eq:BayesianCrossedModel}).
}

\subsection{Conclusions from Comparison Studies}

Our first conclusion based on the studies described in this section is that
product restriction I should not be used for streamlined variational
inference since it is much less accurate than product restriction II
without any significant speed and storage advantages. 
Even though the asymmetry of product restriction II is slightly 
disconcerting, it is better to bear with it in the interest of 
having the fixed effects posterior density functions approximated
more accurately.

The choice between product restrictions II and III depends on the 
size of the problem, availability of computing resources and the
need for speed in the application at hand. If speed is not important
then product restriction III is preferable due to its high
inferential accuracy. Product restriction II is a fallback 
for extremely large problems.

\section{Illustration for Data From a Large Longitudinal Education Study}\label{sec:dataApplic}

We now provide illustration for data from the National Education Longitudinal Study 
which was launched in the United States in early 1988. Details of the study are
given in  Thurgood \textit{et.al.} (2003). The data are publicly available from the U.S. 
National Center for Education Statistics. Our illustration focuses 
on students within their last 5 years of secondary education. The data involve 
longitudinal measurements on 8,564 students with each student 
having his or her academic ability assessed according to 24 items.
The full list of items is given in Table \ref{tab:itemDefinitions} of
the online supplement and includes, for example, test scores
in reading, mathematics and science. All data scores are expressed
in percentage form. Other variables such as gender and parental education 
levels were also recorded.

We did not conduct a full and thorough analysis of these data and avoid exploring matters
such as careful variable creation and model selection. Instead, we consider
an illustrative Bayesian mixed model with a very large number of crossed
random effects.

The model we considered is, for $1\le i\le 8,442$ and $1\le i'\le 24$,
\begin{equation}
\begin{array}{l}
\by_{ii'}|\beta_0,\ldots,\beta_{5},u_{i0},u_{i1},u'_{i'0},u'_{i'1},\sigma^2\simind
    N\Big(\beta_0+u_{i0}+{u'}_{i'0}+(\beta_1+u_{i1}+{u'}_{i'1})\bx_{1,ii'}
      \\[3ex]
\ \ + \beta_{2}\bx_{2,ii'}+ \ldots +\beta_{5} \bx_{5,ii'},\sigma^2 \bI\Big),\
    \left[
    \begin{array}{c}
    u_{i0}\\
    u_{i1}
    \end{array}
    \right]\Big|\bSigma\simind N(\bzero,\bSigma), 
    \ 
    \left[
    \begin{array}{c}
    {u'}_{i'0}\\
    {u'}_{i'1}
    \end{array}
    \right]\Big|\bSigma'\simind N(\bzero,\bSigma')
\end{array}
\label{eq:NELSmodel}
\end{equation}
where $\by_{ii'}$ is the $n_{ii'}\times 1$ vector of scores for the $i$th student
and $i'$th item. The $n_{ii'}\times 1$ predictor vectors $\bx_{1,ii'},\ldots,\bx_{5,ii'}$ are 
$n_{ii'}\times 1$ vectors containing measurements for the $(i,i')$th student/item pair 
on values of the variables $x_1,\ldots,x_5$ which are defined as follows:
{\setlength\arraycolsep{1pt}
\begin{eqnarray*}
x_1&=&\mbox{year of study (either 1, 3 or 5)},\\[0.5ex]
x_2&=&\mbox{indicator that the student is male},\\[0.5ex]
x_3&=&\mbox{indicator that the student spent at least 30 hours per week on homework},\\[0.5ex]
x_4&=&\mbox{indicator that the student's father has at least a high school education, and }\\[0.5ex]
x_5&=&\mbox{indicator that the student's mother has at least a high school education}.
\end{eqnarray*}
}
The priors were set to be
$$
\begin{array}{c}
\beta_0,\ldots,\beta_5 \ \simind N(0,10^{10}),\quad
\mysigeps^2|\asigsq\sim\mbox{Inverse-$\chi^2$}(1,1/\asigsq),\quad
\asigsq\sim\mbox{Inverse-$\chi^2$}(1,10^{-10}),\\[1ex]
\bSigma|\ASigma\sim\mbox{Inverse-G-Wishart}\big(\Gfull,4,\ASigma^{-1}\big),
\ASigma\sim\mbox{Inverse-G-Wishart}(\Gdiag,1,\frac{2}{10^{10}}\bI_2),\\[1ex]
\bSigmad|\ASigmad\sim\mbox{Inverse-G-Wishart}\big(\Gfull,4,\ASigmad^{-1}\big)\ 
\mbox{and}\ \ASigmad\sim\mbox{Inverse-G-Wishart}(\Gdiag,1,\frac{2}{10^{10}}\bI_2).
\end{array}
$$
The response data was transformed to the unit interval for Bayesian
analysis with these priors. The parameters were then back-transformed
to match the original response scale. In addition, to make the 
Gaussian assumption more plausible, we only considered fields with 
test scores between 1\% and 99\% inclusive.
We fit model (\ref{eq:NELSmodel}) using mean field variational Bayes under 
product restriction III with \texttt{Fortran 77} implementation of Algorithm \ref{alg:MFVBforPRIII} with 100 iterations.
Again, we used the third author's \textsf{MacBook Air} laptop with its 2.2 gigahertz processor 
and 8 gigabytes of random access memory and the fit took just under 5 minutes. 

\begin{figure}[!ht]
\centering
{\includegraphics[width=\textwidth]{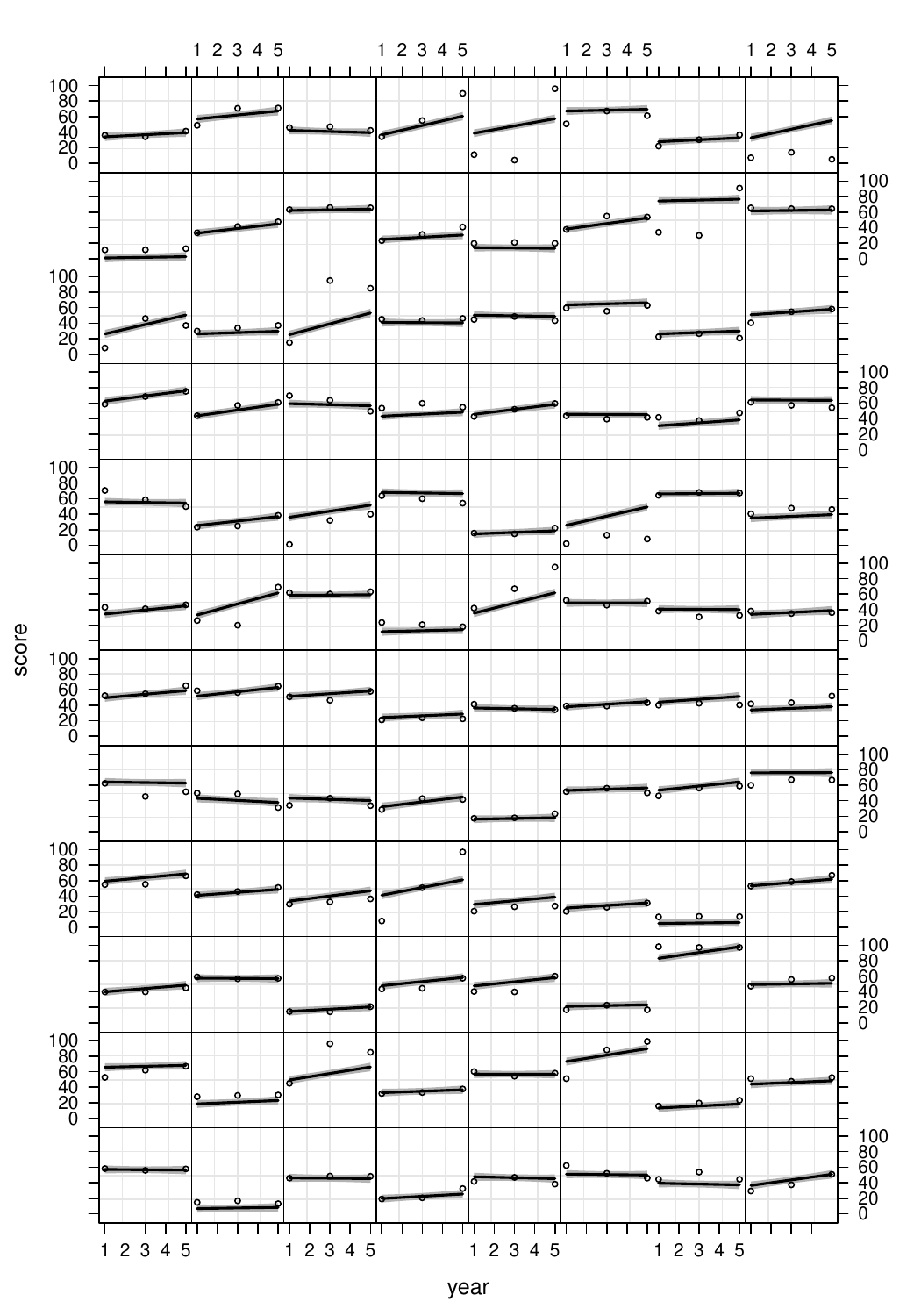}}
\caption{\it Fitted lines for 96 randomly chosen student/item pairs 
from the streamlined mean field variational Bayes analysis of data 
from the National Education Longitudinal Study of 1988. The
other predictors are set to their average values.
The \myApplicGrey\ shading corresponds to pointwise 95\% credible intervals.}
\label{fig:NELSyearFits}
\end{figure}

Figure \ref{fig:NELSyearFits} shows 96 randomly chosen of the random line year
effects, corresponding to posterior means, with each of $x_2,\ldots,x_5$ 
fixed at their average values and the horizontal and vertical ranges set to be
the same for each panel. Shading corresponds to pointwise 95\% credible intervals.
Strong heterogeneity in the year effects across subject/item pairs is apparent,
although it should be noted that Figure \ref{fig:NELSyearFits} represents
only about 0.05\% of all such effects.

%

\begin{figure}[!ht]
\centering
\includegraphics[width=0.8\textwidth]{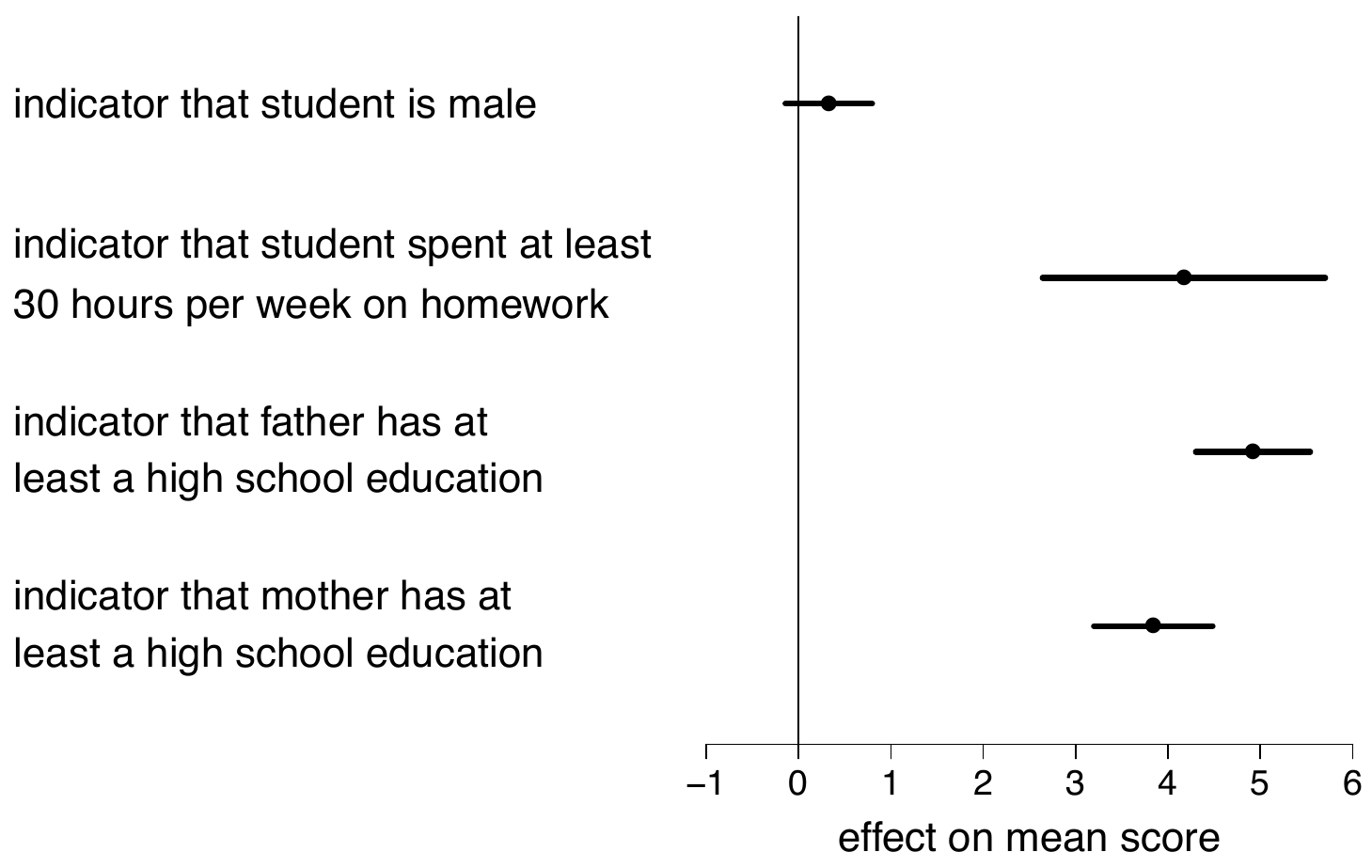}
\caption{\it Approximate posterior means (solid dots)
and 95\% credible intervals (line segments) for $\beta_2,\ldots,\beta_5$
for the mean field variational Bayes, with product restriction III,
fit of (\ref{eq:NELSmodel}) to data from
the National Education Longitudinal Study of 1988.}
\label{fig:NELSeffects}
\end{figure}

Figure \ref{fig:NELSeffects} provides a graphical summary of the
effects of $x_2,\ldots,x_5$. Each line segment corresponds to
an approximate 95\% credible interval for the corresponding
coefficient. The mean field variational Bayes posterior means
are shown as solid dots. For example, having a father
with at least a high school education leads to an elevation
of about 5\% in mean test score. The homework 
and education-related predictors are seen to be highly
significant, whereas gender is not significant.

\tcm{It is apparent from Figure \ref{fig:NELSyearFits} that inclusion of crossed random effects 
that is crucial for} \tcm{well grounded estimation and inference concerning the fixed effects, provided
by Figure \ref{fig:NELSeffects}. An ordinary least squares analysis with the crossed random
effects omitted would ignore the pronounced heterogeneities in the age effects and their 
subject/item interactions and result in imprecise inference for the Figure \ref{fig:NELSeffects}
effects. Ordinary least squares also ignores within-subject and within-item correlations of the 
scores, whereas such correlations are accounted for by model (\ref{eq:BayesianCrossedModel}).
}

\section{Conclusions}\label{sec:conclusions}

We have derived and evaluated three streamlined variational inference algorithms
for Gaussian response linear mixed models with crossed random effects, with differing 
product restriction stringencies.  It is concluded that the most stringent algorithm, 
labeled mean field variational Bayes with product restriction I, should be eliminated
from contention which leaves product restriction II and product restriction III.
Mean field variational Bayes with product restriction II is shown to be
scalable to very large numbers of crossed random effects.  
Mean field variational Bayes with product restriction III is less scalable but
highly accurate. Our numerical results provide valuable guidance for
use of our algorithms in terms of accuracy and run-time trade-offs. For moderate
problems product restriction III delivers fast and accurate inference. 
For increasingly large problems, product restriction II offers a scalable
alternative.

\ifthenelse{\boolean{UnBlinded}}{
\section*{Acknowledgements}

This research was partially supported by Australian Research Council
Discovery Project DP140100441 and U.S. National Institutes of Health 
grants R01AA23187, P50DA039838 and U01CA229437. We are grateful
for comments from Doug Bates, Emanuele Degani and Omar Ghattas.
}
{\null}

\section*{References}

\bib
Atay-Kayis, A. \myand Massam, H. (2005).
A Monte Carlo method for computing marginal likelihood
in nondecomposable Gaussian graphical models.
\textit{Biometrika}, {\bf 92}, 317--335.

\bib
Baayen, R.H., Davidson, D.J. \myand Bates, D.M. (2008).
Mixed-effects modeling with crossed random effects for 
subjects and items. {\it Journal of Memory and Language},
{\bf 59}, 390--412.

\bib
Bishop, C.M. (2006). \textit{Pattern Recognition and Machine Learning},
New York: Springer.

\bib
Blei, D.M., Kucukelbir, A. \myand McAuliffe, J.D. (2017).
Variational inference: a review of statisticians.
{\it Journal of the American Statistical Association}, 
{\bf 112}, 859--877. 

\bib
Doran, H., Bates, D.M., Bliese, P. \myand Dowling, M. (2007).
Estimating the multilevel Rasch model: with the \textsf{lme4} package.
\textit{Journal of Statistical Software}, {\bf 20}, Issue 2, 1--18.

\bib
Papaspiliopoulos, O., Roberts, G.O. and Zanella, G. (2020).
Efficient parameterisations for normal linear mixed models.
\textit{Biometrika}, \textbf{107}, 25--40.

\bib
Jeon, M., Rijmen, F. \myand Rabe-Hesketh, S. (2017).
A variational maximization-maximization algorithm for
generalized linear mixed models with crossed random effects.
\textit{Psychometrika}, {\bf 3}, 693--716.

\bib
Huang, A. \myand Wand, M.P. (2013).
Simple marginally noninformative prior distributions 
for covariance matrices. \textit{Bayesian Analysis}, 
{\bf 8}, 439--452.

\bib
Lee, C.Y.Y. \myand Wand, M.P. (2016).
Streamlined mean field variational Bayes for longitudinal
and multilevel data analysis. \textit{Biometrical Journal},
{\bf 58}, 868--895.

\bib
Menictas, M., Nolan, T.H., Simpson, D.G. \myand Wand, M.P. (2021).
Streamlined variational inference for higher level group-specific curve models.
\textit{Statistical Modelling}, \textbf{21}, 479--519.

\bib
Menictas, M. and Wand, M.P. (2013).
Variational inference for marginal longitudinal semiparametric regression. 
\textit{Stat}, {\bf 2}, 61--71.

\bib
Minka, T., Winn, J., Guiver, Y., Fabian, D. \myand Bronskill, J. (2018).
\textsf{Infer.NET 0.3}, Microsoft Research Cambridge,
\texttt{http://dotnet.github.io/infer}

\bib
Nolan, T.H., Menictas, M. \myand Wand, M.P. (2020).
Streamlined computing for variational inference with higher level 
random effects. \textit{Journal of Machine Learning Research}, 
\textbf{21(157)}, 1--62. 

\bib
Nolan, T.H. \myand Wand, M.P. (2017).
Accurate logistic variational message passing: 
algebraic and numerical details. \textit{Stat}, {\bf 6},

\bib
Nolan, T.H. \myand Wand, M.P. (2020). Solutions
to multilevel sparse matrix problems. 
\textit{ANZIAM Journal}, {\bf 62}, 18--41.

\bib
Ormerod, J.T. and Wand, M.P. (2010).
Explaining variational approximations.
{\it The American Statistician},
{\bf 64}, 140--153.

\bib
\textsf{R} Core Team (2019). \textsf{R}: A language and
environment for statistical computing. \textsf{R} Foundation
for Statistical Computing, Vienna, Austria.
\texttt{https://www.R-project.org/}.

\bib
\textsf{Stan} Development Team (2021). \textsf{RStan}: the \textsf{R}
interface to \textsf{Stan}. \textsf{R} package version 2.26.3.
\texttt{https://mc-stan.org/}.

\bib
Thurgood, L., Walter, E., Carter, G., Henn, S.,
Huang, G., Nooter, D., Smith, W., Cash, R.W. \myand
Salvucci, S. (2003). National Education Longitudinal Study
of 1988 (NELS: 88). In M. Seastrom, T. Phan \myand M. Cohen (editors).
\textit{NCES Handbook of Survey Methods} (pp. 53-66).
Washington D.C.: U.S. Department of Education, National
Center for Education Studies.

\bib
Tran, D., Kucukelbir, A., Dieng, A.B., Rudolph, M., Liang, D.
\myand Blei, D.M. (2016). \textit{Edward: A library for
probabilistic modeling, inference, and criticism.}
Unpublished manuscript available at \textit{https://arxiv.org/abs/1610.09787}.

\bib
Wand, M. P. \myand Jones, M.C. (1995). {\it Kernel Smoothing}.
London: Chapman and Hall.

\bib
Wand, M.P. (2017).
Fast approximate inference for arbitrarily large semiparametric
regression models via message passing (with discussion).
\textit{Journal of the American Statistical Association}, 
{\bf 112}, 137--168.

\bib
Winn, J. \myand Bishop, C.M. (2005). Variational message passing.
\textit{Journal of Machine Learning Research}, {\bf 6}, 661--694.

\vfill\eject
%
%
\renewcommand{\theequation}{S.\arabic{equation}}
\renewcommand{\thesection}{S.\arabic{section}}
\renewcommand{\thetable}{S.\arabic{table}}
\renewcommand{\thealgorithm}{S.\arabic{algorithm}}
\renewcommand{\theresult}{S.\arabic{result}}
\setcounter{equation}{0}
\setcounter{table}{0}
\setcounter{section}{0}
\setcounter{page}{1}
\setcounter{footnote}{0}
\setcounter{algorithm}{0}
\setcounter{result}{0}

\cleardoublepage

\null
\vskip5mm
\centerline{\Large Online Supplement for:}
\vskip3mm
\centerline{\Large\bf Streamlined Variational Inference for Linear Mixed}
\vskip2mm
\centerline{\Large\bf Models with Crossed Random Effects}
\vskip7mm
\ifthenelse{\boolean{UnBlinded}}{
\centerline{\normalsize\sc By Marianne Menictas$\null^1$, Gioia Di Credico$\null^2$ and Matt P. Wand$\null^3$}
\vskip5mm
\centerline{\textit{Harvard University$\null^1$, University of Trieste$\null^2$ 
and University of Technology Sydney$\null^3$}}
}
{\null}
\vskip6mm

\section{The Inverse G-Wishart and Inverse $\chi^2$ Distributions}\label{sec:IGWandICS}

The Inverse G-Wishart corresponds to the matrix inverses of random matrices that have
a \emph{G-Wishart} distribution (e.g. Atay-Kayis \myand Massam, 2005). 
For any positive integer $d$, let $G$ be an undirected graph with $d$ nodes 
labeled $1,\ldots,d$ and set $E$ consisting of sets of pairs of nodes that 
are connected by an edge. We say that the symmetric $d\times d$ matrix $\bM$ 
\emph{respects} $G$ if 
$$\bM_{ij}=0\quad\mbox{for all}\quad \{i,j\}\notin E.$$
A $d\times d$ random matrix $\bX$ has an Inverse G-Wishart distribution
with graph $G$ and parameters $\xi>0$ and symmetric $d\times d$ 
matrix $\bLambda$, written
$$\bX\sim\mbox{Inverse-G-Wishart}(G,\xi,\bLambda)$$
if and only if the density function of $\bX$ satisfies
$$\pDens(\bX)\propto |\bX|^{-(\xi+2)/2}\exp\{-\smhalf\tr(\bLambda\,\bX^{-1})\}$$
over arguments $\bX$ such that $\bX$  is symmetric and positive definite 
and $\bX^{-1}$ respects $G$. Two important special cases are 
$$G=\Gfull\equiv\mbox{totally connected $d$-node graph},$$
for which the Inverse G-Wishart distribution coincides with the ordinary
Inverse Wishart distribution, and
$$G=\Gdiag\equiv\mbox{totally disconnected $d$-node graph},$$
for which the Inverse G-Wishart distribution coincides with
a product of independent Inverse Chi-Squared random variables.
The subscripts of $\Gfull$ and $\Gdiag$ reflect the 
fact that $\bX^{-1}$ is a full matrix and $\bX^{-1}$ is
a diagonal matrix in each special case.

The $G=\Gfull$ case corresponds to the ordinary Inverse Wishart
distribution. However, with modularity in mind, we will
work with the more general Inverse G-Wishart
family throughout this article.

In the $d=1$ special case the graph $G=\Gfull=\Gdiag$ 
and the Inverse G-Wishart distribution reduces to the
Inverse Chi-Squared distributions. We write
$$x\sim\mbox{Inverse-$\chi^2$}(\xi,\lambda)$$
for this  $\mbox{Inverse-G-Wishart}(\Gdiag,\xi,\lambda)$
special case with $d=1$ and $\lambda>0$ scalar.

\section{The Generalized Blockdiag Operator}\label{sec:genBlockdiag}

If $\bM_1$, $\bM_2$ and $\bM_3$ are each $2\times2$ matrices 
then a well-established notation is 
$$\blockdiag{1\le i\le 3}(\bM_i)=
\left[
\begin{array}{ccc}
\bM_1   & \bO_{2\times2}  & \bO_{2\times2} \\[1ex]
\bO_{2\times2}   & \bM_2  & \bO_{2\times2} \\[1ex]
\bO_{2\times2}   & \bO_{2\times2}  & \bM_3
\end{array}
\right]
$$
where $\bO_{2\times2}$ denotes the $2\times2$ matrix of zeroes.

Suppose instead that $\bM_2$ is $n\times 2$ where $n=0$. Then $\bM_2$ is null but here
we allow its column dimension to be a positive integer. The \emph{generalized}
blockdiag operator is such that
$$\blockdiag{1\le i\le 3}(\bM_i)=
\left[
\begin{array}{ccc}
\bM_1   & \bO_{2\times2}  & \bO_{2\times2} \\[1ex]
\bO_{2\times2}   & \bO_{2\times2}  & \bM_3
\end{array}
\right].
$$
The key aspect is that, after positioning $\bM_1$, the column index is incremented by $2$
before positioning $\bM_3$. This is due to $\bM_2$ being a ``matrix'' having  
\emph{generalized dimension} $0\times2$.

The generalized blockdiag operator is useful for describing the design matrices that arise 
from model (\ref{eq:BayesianCrossedModel}). Suppose that $m=5$, $m'=3$, $q'=2$ and that 
the $n_{i\idash}$ values are as given by Table \ref{tab:niidashExamp}.

\begin{table}[!ht]
\centering
\begin{tabular}{lccc}
\hline\\[-1.5ex]
            & $\idash=1$   & $\idash=2$  & $\idash=3$ \\[1.0ex]
\hline\\[-1.2ex]
$i=1$    &  2  &  4  &  0  \\
$i=2$    &  0  &  3  &  1  \\
$i=3$    &  0  &  0  &  6  \\
$i=4$    &  7  &  2  &  9  \\
$i=5$    &  5  &  0  &  8  \\
\hline
\end{tabular}
\caption{\it The $n_{i\idash}$ values for an illustrative example concerning the use of
the generalized blockdiag operator to describe cross random effects design matrices.}
\label{tab:niidashExamp} 
\end{table}
\noindent
Then, according to the definitions given in Section \ref{sec:additDataMats}
and the rules of the generalized blockdiag operator:
{\setlength\arraycolsep{3pt}
\begin{eqnarray*}
\Zdblacksquare_1&=&\blockdiag{1\le i'\le 3}(\bZ'_{1i'})=
\left[\begin{array}{ccc}            
\bZ'_{11}     & \bO_{2\times 2}  & \bO_{2\times2} \\[1ex]
\bO_{4\times2}& \bZ'_{12} & \bO_{4\times2}
\end{array}
\right],\\[1ex]
\Zdblacksquare_2&=&\blockdiag{1\le i'\le 3}(\bZ'_{2i'})=
\left[\begin{array}{ccc}            
\bO_{3\times2}     & \bZ'_{22}      & \bO_{3\times2} \\[1ex]
\bO_{1\times2}     & \bO_{1\times2} & \bZ'_{23}
\end{array}
\right],\\[1ex]
\Zdblacksquare_3&=&\blockdiag{1\le i'\le 3}(\bZ'_{3i'})=
\left[\begin{array}{ccc}            
\bO_{6\times2}     & \bO_{6\times2} & \bZ'_{33}
\end{array}
\right],\\[1ex]
\Zdblacksquare_4&=&\blockdiag{1\le i'\le 3}(\bZ'_{4i'})=
\left[\begin{array}{ccc}            
\bZ'_{41}     & \bO_{7\times 2}  & \bO_{7\times2} \\[1ex]
\bO_{2\times2}& \bZ'_{42} & \bO_{2\times2}        \\[1ex]
\bO_{9\times2}& \bO_{9\times2} & \bZ'_{43} 
\end{array}
\right]\\[1ex]
\mbox{and}\quad
\Zdblacksquare_5&=&\blockdiag{1\le i'\le 3}(\bZ'_{5i'})=
\left[\begin{array}{ccc}            
\bZ'_{51}     & \bO_{5\times 2}  & \bO_{5\times2} \\[1ex]
\bO_{8\times2}& \bO_{8\times2} & \bZ'_{53} 
\end{array}
\right].
\end{eqnarray*}
}

Algorithm \ref{alg:genBlockdiag} provides full details of the generalized blockdiag operator 
for general input matrices, with some possibly having generalized dimension for which $0$ is
allowed.

\begin{algorithm}[!th]
\begin{center}
\begin{minipage}[t]{160mm}
\begin{small}
\begin{itemize}
\item[] Inputs: $\{\bM_i : 1\le i\le d\}$ where $\bM_i$ is an $r_i\times c_i$ generalized dimensioned matrix 
\item[] $\qquad\qquad$ for integers $r_i\ge0$ and $c_i\ge0$.
\begin{itemize}
\item[] $\bA\thickarrow \mbox{the}\ \left(\displaystyle{\sum_{i=1}^d} r_i\right)
\times\left(\displaystyle{\sum_{i=1}^d} c_i\right)$ 
\mbox{matrix of zeroes}
\item[] $\rStt\thickarrow 1\ \ \ ;\ \ \ \cStt\thickarrow 1$
\item[] for $i=1,\ldots,m$
\begin{itemize}
\item[] $\rEnd\thickarrow\rStt + r_i-1$\ \ \ ;\ \ \ $\cEnd\thickarrow\cStt + c_i-1$
\item[] if $r_i>0$ and $c_i>0$
\begin{itemize}
\item[] (rows $\rStt$ to $\rEnd$ and columns $\cStt$ to $\cEnd$ of $\bA$) $\thickarrow\bM_i$
\end{itemize}
\item[] $\rStt\thickarrow\rEnd + 1$\ \ \ ;\ \ \ $\cStt\thickarrow\cEnd + 1$
\end{itemize}
\end{itemize}
\item[] Output: $\bA$
\end{itemize}
\end{small}
\end{minipage}
\end{center}
\caption{\textit{The generalized blockdiag algorithm.}}
\label{alg:genBlockdiag}
\end{algorithm}

\section{Derivation of the $\qDens(\bbeta,\buall)$ Parameters Updates Under Product Restriction I}
\label{sec:PRIderiv}

The full conditional distribution of $\bbeta$ is 
$$
\pDens(\bbeta|\rest)\propto\pDens(\by|\bbeta,\bu,\budash,\sigma^2)\pDens(\bbeta).
$$
Note that $\pDens(\by|\bbeta,\bu,\budash,\sigma^2)$ can be expressed as the 
$N\left(\bX\bbeta,\sigma^2\bI\right)$
density function in the vector
$$\by-{\displaystyle\stack{1\le i\le m}}\Big\{{\displaystyle\stack{1\le i'\le m'}}
\Big(\bZ_{ii'}\bu_i+\bZ'_{ii'}\bu'_{i'}\Big)\Big\}
$$
Also, $\pDens(\bbeta)$ is the $N\left(\bmu_{\bbeta},\bSigma_{\bbeta} \right)$
density function in the vector $\bbeta$. Then, under product restriction I, 
standard quadratic form manipulations lead to the optimal $\qDens$-density function of $\bbeta$
being that of the $N(\bmu_{\qDens(\bbeta)},\bSigma_{\qDens(\bbeta)})$ distribution 
with updates 
$$\bmu_{\qDens(\bbeta)}\longleftarrow (\mu_{\qDens(\sigma^2)}\bX^T\bX+\bSigma_{\bbeta}^{-1})^{-1}
(\mu_{\qDens(\sigma^2)}\bX^T\br+\bSigma_{\bbeta}^{-1}\bmu_{\bbeta}^{-1})
\ \mbox{and}\ 
\bSigma_{\qDens(\bbeta)}\longleftarrow (\mu_{\qDens(\sigma^2)}\bX^T\bX+\bSigma_{\bbeta}^{-1})^{-1}.
$$
where
$$\br\equiv\by- 
             {\displaystyle\stack{1\le i\le m}}\Big\{{\displaystyle\stack{1\le i'\le m'}}
             \Big(\bZ_{ii'}\bmu_{\qDens(\bu_i)}+\bZ'_{ii'}\bmu_{\qDens(\bu'_{i'})}\Big)\Big\}.
$$
If $\bb$ and $\bB$ are defined according to the updates in (\ref{eq:bandBsimplest}) then
simple algebra shows that
$$\bB^T\bb=\mu_{\qDens(\sigma^2)}\bX^T\br+\bSigma_{\bbeta}^{-1}\bmu_{\bbeta}^{-1}\quad\mbox{and}\quad
\bB^T\bB=\mu_{\qDens(\sigma^2)}\bX^T\bX+\bSigma_{\bbeta}^{-1}.
$$
Therefore, the $\bmu_{\qDens(\bbeta)}$ update corresponds to the least
squares solution $\bx=(\bB^T\bB)^{-1}\bB^T\bb$ and the update of 
$\bSigma_{\qDens(\bbeta)}$ corresponds to $(\bB^T\bB)^{-1}$.

Analogous arguments can be used to justify the updates for the parameters of 
$\qDens(\bu_i)$, $1\le i\le m$, and $\qDens(\bu'_{i'})$, $1\le i'\le m'$.

\section{The \LargerSolveLeastSquares\ Algorithm}\label{sec:SLSA}

The \SolveLeastSquares\ is concerned with solving the least
squares problem
$$\min_{\bx}\Vert\bb-\bB\bx\Vert^2.$$
which has solution $\bx=(\bB^T\bB)^{-1}\bB^T\bb$.
The matrix $(\bB^T\bB)^{-1}$ is also of intrinsic interest.
In next subsection a version of this problem is solved for the
situation where $\bB$ has two-level sparse structure. In this
subsection there is no sparseness structure imposed on $\bB$.

\begin{algorithm}[!th]
\begin{center}
\begin{minipage}[t]{154mm}
\begin{small}
\begin{itemize}
\setlength\itemsep{4pt}
\item[] Inputs: $\big\{\bb(\nadj\times1),\ \bB(\nadj\times p)\big\}$
\setlength\itemsep{4pt}
\item[] Decompose $\bB=\bQ\left[\begin{array}{c}   
\bR\\
\bzero
\end{array}
\right]$ such that $\bQ^{-1}=\bQ^T$ and $\bR$ is upper-triangular.
\item[] $\bc\thickarrow\bQ^T\bb$\ \ \ ;\ \ \ $\bc_1\thickarrow\mbox{first $p$ rows of}\ \bc$
\item[] $\bx\thickarrow\bR^{-1}\bc_1$\ \ \ ;\ \ \ $(\bB^T\bB)^{-1}\thickarrow\bR^{-1}\bR^{-T}$
\item[] Output: $\Big(\bx,(\bB^T\bB)^{-1}\Big)$
\end{itemize}
\end{small}
\end{minipage}
\end{center}
\caption{\SolveLeastSquares\ \textit{for solving the least squares problem: 
minimise $\Vert\bb-\bB\,\bx\Vert^2$ in $\bx$ and obtaining $(\bB^T\bB)^{-1}$.}}
\label{alg:SolveLeastSquares} 
\end{algorithm}

\section{The \LargerSolveTwoLevelSparseLeastSquares\ Algorithm}\label{sec:STLSLS}

The \SolveTwoLevelSparseLeastSquares\ algorithm 
solves a sparse version of the the least squares problem:
$$\min_{\bx}\Vert\bb-\bB\bx\Vert^2$$
which has solution $\bx=\bA^{-1}\bB^T\bb$ where $\bA=\bB^T\bB$ where $\bB$ 
and $\bb$ have the following structure:
\begin{equation}
\bB\equiv
\left[
\arraycolsep=2.2pt\def\arraystretch{1.6}
\begin{array}{c|c|c|c|c}
\setstretch{4.5}
\Bmato            &\Bmatdoto           &\bO   &\cdots&\bO\\ 
\hline
\Bmatt            &\bO              &\Bmatdott&\cdots&\bO\\ 
\hline
\vdots            &\vdots           &\vdots           &\ddots&\vdots\\
\hline
\Bmatm &\bO       &\bO              &\cdots           &\Bmatdotm
\end{array}
\right]
\quad\mbox{and}\quad
\bb=\left[
\arraycolsep=2.2pt\def\arraystretch{1.6}
\begin{array}{c}
\setstretch{4.5}
\bveco  \\ 
\hline
\bvect \\ 
\hline
\vdots \\
\hline
\bvecm \\
\end{array}
\right].
\label{eq:BandbFormsReprise}
\end{equation}
The sub-vectors of $\bx$ and the sub-matrices of $\AtLev$ corresponding to its
non-zero blocks of are labeled as follows:
\begin{equation}
\bx=
\left[
\arraycolsep=2.2pt\def\arraystretch{1.6}
\begin{array}{c}
\setstretch{4.5}
\bx_1\\
\hline
\bx_{2,1}\\
\hline
\bx_{2,2}\\
\hline
\vdots\\
\hline
\bx_{2,m}
\end{array}
\right]
\quad\mbox{and}\quad
\AtLev^{-1}=
\left[
\arraycolsep=2.2pt\def\arraystretch{1.6}
\begin{array}{c|c|c|c|c}
\setstretch{4.5}
\AUoo     & \AUotCo & \AUotCt  &\ \ \cdots\ \ &\AUotCm \\
\hline
\AUotCoT & \AUttCo & \bigX      & \cdots   & \bigX    \\
\hline
\AUotCtT & \bigX     & \AUttCt  & \cdots   & \bigX    \\ 
\hline
\vdots    & \vdots  & \vdots   & \ddots   & \vdots   \\
\hline
\AUotCmT & \bigX     & \bigX      & \cdots   &\AUttCm \\ 
\end{array}
\right]
\label{eq:AtLevInv}
\end{equation}
with $\bigX$ denoting sub-blocks that are not of interest.
The \SolveTwoLevelSparseLeastSquares\ algorithm is given
in Algorithm \ref{alg:SolveTwoLevelSparseLeastSquares}.

\begin{algorithm}[!th]
\begin{center}
\begin{minipage}[t]{154mm}
\begin{small}
\begin{itemize}
\setlength\itemsep{4pt}
\item[] Inputs: $\big\{\big(\bveci(\nadj_i\times1),
\ \Bmati(\nadj_i\times p),\ \Bmatdoti(\nadj_i\times q)\big):\ 1\le i\le m\big\}$
\item[] $\bomega_3\thickarrow\mbox{NULL}$\ \ \ ;\ \ \ $\bOmega_4\thickarrow\mbox{NULL}$
\item[] For $i=1,\ldots,m$:
\begin{itemize}
\setlength\itemsep{4pt}
\item[] Decompose $\Bmatdoti=\bQ_i\left[\begin{array}{c}   
\bR_i\\
\bzero
\end{array}
\right]$ such that $\bQ_i^{-1}=\bQ_i^T$ and $\bR_i$ is upper-triangular.
\item[] $\cveczi\thickarrow\bQ_i^T\bveci\ \ \ ;\ \ \ \Cmatzi\thickarrow\bQ_i^T\Bmati$
\item[] $\cvecoi\thickarrow\mbox{first $q$ rows of}\ \cveczi$\ \ \ ;\ \ \ 
$\cvecti\thickarrow\mbox{remaining rows of}\ \cveczi$\ \ \ ;\ \ \ 
$\bomega_3\thickarrow
\left[
\begin{array}{c}
\bomega_3\\
\cvecti
\end{array}
\right]$
\item[]$\Cmatoi\thickarrow\mbox{first $q$ rows of}\ \Cmatzi$\ \ \ ;\ \ \ 
$\Cmatti\thickarrow\mbox{remaining rows of}\ \Cmatzi$\ \ \ ;\ \ \  
$\bOmega_4\thickarrow
\left[
\begin{array}{c}
\bOmega_4\\
\Cmatti
\end{array}
\right]$
\end{itemize}
\item[] Decompose $\OmegaAtwoTwo=\bQ\left[\begin{array}{c}   
\bR\\
\bzero
\end{array}
\right]$ such that $\bQ^{-1}=\bQ^T$ and $\bR$ is upper-triangular.
\item[] $\bc\thickarrow\mbox{first $p$ rows of $\bQ^T\bomega_3$}$
\ \ \ ;\ \ \ $\xveco\thickarrow\bR^{-1}\bc$\ \ \ ;\ \ \ 
$\AUoo\thickarrow\bR^{-1}\bR^{-T}$
\item[] For $i=1,\ldots,m$:
\begin{itemize}
\setlength\itemsep{4pt}
\item[] $\xvectCi\thickarrow\bR_i^{-1}(\bc_{1i}-\Cmatoi\xveco)$\ \ \ ;\ \ \ 
$\AUotCi\thickarrow\,-\AUoo(\bR_i^{-1}\Cmatoi)^T$
\item[] $\AUttCi\thickarrow\bR_i^{-1}(\bR_i^{-T} - \Cmatoi\AUotCi)$
\end{itemize}
\item[] Output: $\Big(\xveco,\AUoo,\big\{\big(\xvectCi,\AUttCi,\AUotCi):\ 1\le i\le m\big\}\Big)$
\end{itemize}
\end{small}
\end{minipage}
\end{center}
\caption{\SolveTwoLevelSparseLeastSquares\ \textit{for solving the two-level sparse matrix
least squares problem: minimise $\Vert\bb-\bB\,\bx\Vert^2$ in $\bx$ and sub-blocks of $\bA^{-1}$
corresponding to the non-zero sub-blocks of $\bA=\bB^T\bB$. The sub-block notation is
given by (\ref{eq:BandbFormsReprise}) and (\ref{eq:AtLevInv}).}}
\label{alg:SolveTwoLevelSparseLeastSquares} 
\end{algorithm}

\section{Derivation of Result \ref{res:prodRestrictII}}\label{sec:derivProdRestrictII}

The full conditional density function of $(\bbeta,\bu)$ satisfies
$$
\pDens(\bbeta,\bu|\rest)\propto\pDens(\by|\bbeta,\bu,\budash,\sigma^2)\pDens(\bbeta,\bu|\bSigma).
$$
Note that $\pDens(\by|\bbeta,\bu,\budash,\sigma^2)$ can be expressed as the 
$$N\left(\Cuptri\left[\begin{array}{c}     
\bbeta\\
\bu
\end{array}
\right],\sigma^2\bI\right)
$$
density function in the vector $\ruptri$, where
$$\Cuptri\equiv\left[\stack{1\le i\le m}(\Xuptri_i)\ \ \blockdiag{1\le i\le m}(\Zuptri_i)\right]
\quad\mbox{and}\quad
\ruptri\equiv\stack{1\le i\le m}\Big\{\yuptri_i-\stack{1\le i'\le m'}(\bZ'_{ii'}\bu'_{i'})\Big\}.$$
Also, 
$$\pDens(\bbeta,\bu|\bSigma)\quad\mbox{is the}\quad
N\left(\left[\begin{array}{c}    
\bmu_{\bbeta}\\
\bzero
\end{array}
\right],            
\left[
\begin{array}{cc}
\bSigma_{\bbeta} & \bO \\
\bO              & \bI_m\otimes \bSigma
\end{array}
\right]
\right)
$$
density function in the vector $(\bbeta,\bu)$. Then, under product restriction II, 
standard quadratic form manipulations lead to the optimal $\qDens$-density function of $(\bbeta,\bu)$
being that of the $N(\bmu_{\qDens(\bbeta,\bu)},\bSigma_{\qDens(\bbeta,\bu)})$ distribution 
with updates 
$$\bmu_{\qDens(\bbeta,\bu)}\longleftarrow(\CTuptri\RuptriMFVB^{-1}\Cuptri+\DuptriMFVB)^{-1}
(\CTuptri\RuptriMFVB^{-1}\ruptriMFVB+\ouptriMFVB)
\quad\mbox{and}\quad
\bSigma_{\qDens(\bbeta,\bu)}\longleftarrow(\CTuptri\RMFVB^{-1}\Cuptri+\DuptriMFVB)^{-1}.$$
Here $\RuptriMFVB\equiv\mu_{\qDens(1/\sigma^2)}^{-1}\bI$, 
$$\DuptriMFVB\equiv\left[
\begin{array}{cc}
\bSigma_{\bbeta}^{-1} & \bO \\
\bO                   & \bI_m\otimes\bM_{\qDens(\bSigma^{-1})}
\end{array}
\right],\quad
\ouptriMFVB\equiv \left[
\begin{array}{c}
\bSigma_{\bbeta}^{-1}\bmu_{\bbeta}\\
\bzero                  
\end{array}
\right]
$$
and
$$\ruptriMFVB\equiv\stack{1\le i\le m}\Big\{\yuptri_i
-\stack{1\le i'\le m'}\big(\bZ'_{ii'}\bmu_{\qDens(\bu'_{i'})}\big)\Big\}.$$

If $\bb$ and $\bB$ are defined according to (\ref{eq:BandbFormsReprise}) and the matrices 
$\bb_i$, $\Bmati$ and $\Bmatdoti$ are defined as in Result \ref{res:prodRestrictII}
then 
$$\bB^T\bb=\CTuptri\RuptriMFVB^{-1}\ruptriMFVB+\ouptriMFVB\quad\mbox{and}\quad
\bB^T\bB=\CTuptri\RuptriMFVB^{-1}\Cuptri+\DuptriMFVB.
$$
Therefore, with this assignment of $\bb_i$, $\Bmati$ and $\Bmatdoti$,  
the $\bmu_{\qDens(\bbeta,\buall)}$ update corresponds to the least
squares solution $\bx=(\bB^T\bB)^{-1}\bB^T\bb$ and the updates
of the sub-blocks of $\bSigma_{\qDens(\bbeta,\buall)}$ listed 
in the first two rows of Table \ref{tab:SigmaSubBlocks} correspond to the sub-blocks of 
$(\bB^T\bB)^{-1}$ in the positions where $\bB^T\bB$ has non-zero
sub-blocks.

\section{Derivation of Result \ref{res:prodRestrictIII}}\label{sec:derivProdRestrictIII}

Result \ref{res:prodRestrictIII} uses the following re-ordering of the overall design
matrix:
$$\bCtilde\equiv\left[\bX\ \stack{1\le i\le m}(\Zdblacksquare_i)\ \ \blockdiag{1\le i\le m}(\Zuptri_i)\ \ \right].$$
rather than $\bC\equiv[\bX\ \bZ]$ in the generalized ridge regression 
expressions of Section \ref{sec:varInf}. This re-ordering involves the $\qDens$-density
parameters of $\bu'$ preceding those of $\bu$ and is brought about by our $m\ge m'$
convention throughout this article and the requirement that the potentially very large
$$\blockdiag{1\le i\le m}(\Zuptri_i)$$
appears on the right for embedding within the two-level sparse least squares 
infrastructure of Nolan \myand Wand (2020) and Nolan \textit{et al.} (2019).
The re-ordering means that the updates for $\bmu_{\qDens(\bbeta,\bu',\bu)}$ 
and $\bSigma_{\qDens(\bbeta,\bu',\bu)}$ are
$$
\begin{array}{l}
\bmu_{\qDens(\bbeta,\bu',\bu)}\longleftarrow(\bCtilde^T\RMFVB^{-1}\bCtilde+\DtildeMFVB)^{-1}
(\bCtilde^T\RMFVB^{-1}\by + \oMFVB)
\quad \mbox{and}\quad\\[1ex]
\bSigma_{\qDens(\bbeta,\bu',\bu)}\longleftarrow(\bCtilde^T\RMFVB^{-1}\bCtilde+\DtildeMFVB)^{-1}
\end{array}
$$
where
$$   
\DtildeMFVB\equiv\left[
\begin{array}{ccc}
      \bSigma_{\bbeta}^{-1} & \bO & \bO \\[1ex]
      \bO & \bI_{m'}\otimes\bM_{\qDens((\bSigmad)^{-1})}& \bO \\[1ex]
      \bO & \bO & \bI_m \otimes \bM_{\qDens(\bSigma^{-1})} 
\end{array}
\right]
$$
has the $\bM_{\qDens((\bSigmad)^{-1})}$ matrices appearing before the $\bM_{\qDens(\bSigma^{-1})}$
matrices due to the switch in the ordering of the random effects vectors.

If $\bb$ and $\bB$ are defined according to (\ref{eq:BandbFormsReprise}) with the matrices 
$\bb_i$, $\Bmati$ and $\Bmatdoti$ defined as in Result \ref{res:prodRestrictIII}
then straightforward matrix algebra can be used to show that 
$$\bB^T\bb=\bCtilde^T\RMFVB^{-1}\by + \oMFVB\quad\mbox{and}\quad
\bB^T\bB=\bCtilde^T\RMFVB^{-1}\bCtilde+\DtildeMFVB.
$$
Therefore, with this assignment of $\bb_i$, $\Bmati$ and $\Bmatdoti$,  
the $\bmu_{\qDens(\bbeta,\bu',\bu)}$ update corresponds to the least
squares solution $\bx=(\bB^T\bB)^{-1}\bB^T\bb$ and the updates
of the sub-blocks of $\bSigma_{\qDens(\bbeta,\bu',\bu)}$ listed 
in the first four rows of Table \ref{tab:SigmaSubBlocks}  correspond to the sub-blocks of 
$(\bB^T\bB)^{-1}$ in the positions where $\bB^T\bB$ has non-zero
sub-blocks.

\section{Marginal Log-Likelihood Lower Bound and Derivation}\label{sec:lowerBound}

The logarithmic form of the variational lower bound on the marginal log-likelihood,
corresponding to model (\ref{eq:BayesianCrossedModel}) with prior specification $($B$)$ 
and product restriction III is
\begin{equation*}
  \begin{array}{lcl}
    \log \underline{\pDens} (\by; \qDens) & = & E_{\qDens} \left\{ \log \pDens
    \left( \by,
         \bbeta,\bu,\bu',\aEps,\ASigma,\ASigmad,\sigma^2,\bSigma,\bSigma' \right)
         \right.
    \\[1ex]
    & & \left. \quad \quad
        - \log \qDens^{*} \left( \bbeta,\bu,\bu',\aEps,\ASigma,\ASigmad,\sigma^2,
          \bSigma,\bSigma' \right) \right\}
    \\[2ex]
    & = & E_{\qDens} \left\{ \pDens \left( \by | \bbeta, \bu, \bu', \sigma^2 \right) \right\}
          + E_{\qDens} \left\{ \log \pDens \left( \bbeta, \bu, \bu' | \bSigma, \bSigma' \right) \right\}
    \\[1ex]
     & & - E_{\qDens} \left\{ \log \qDens^{*} \left( \bbeta, \bu, \bu' \right) \right\}
         + E_{\qDens} \left\{ \log \pDens \left( \sigma^2 | \aEps \right) \right\}
         - E_{\qDens} \left\{ \log \qDens^{*} \left( \sigma^2 \right) \right\}
    \\[1ex]
    & & + E_{\qDens} \left\{ \log \pDens \left( \aEps \right) \right\}
        - E_{\qDens} \left\{ \log \qDens^{*} \left( \aEps \right) \right\}
        + E_{\qDens} \left\{ \log \pDens \left( \bSigma | \ASigma \right) \right\}
    \\[1ex]
    & & - E_{\qDens} \left\{ \log \qDens^{*} \left( \bSigma \right) \right\}
        + E_{\qDens} \left\{ \log \pDens \left( \ASigma \right) \right\}
        - E_{\qDens} \left\{ \log \qDens^{*} \left( \ASigma \right) \right\}
    \\[1ex]
    & & + E_{\qDens} \left\{ \log \pDens \left( \bSigma' | \ASigmad \right) \right\}
        - E_{\qDens} \left\{ \log \qDens^{*} \left( \bSigma' \right) \right\}
        + E_{\qDens} \left\{ \log \pDens \left( \ASigmad \right) \right\}
    \\[1ex]
    & & - E_{\qDens} \left\{ \log \qDens^{*} \left( \ASigmad \right) \right\}.
  \end{array}
\end{equation*}
The first of the $\log \underline{\pDens} (\by; \qDens)$ terms is
\begin{equation*}
  \begin{array}{lcl}
    E_{\qDens} \left\{ \pDens \left( \by | \bbeta, \bu, \bu', \sigma^2 \right) \right\}
    & = &
        -\smhalf \ndotdot \log(2 \pi) - \smhalf \ndotdot E_{\qDens} \left\{ \log (\sigma^2) \right\}
    \\[1ex]
    & & -\smhalf \mu_{\qDens (1/{\sigma^{2}})} \displaystyle{\sum_{i=1}^{m} \sum_{i'=1}^{m'}}
        \left\{ \left| \left| \by_{ii'} - \bX_{ii'} \bmu_{\qDens (\bbeta)} - \bZ_{ii'} \bmu_{\qDens(\bu_{i})}
        - \bZdash_{ii'} \bmu_{\qDens(\budash_{i'})}  \right| \right|^2  \right.
    \\[3ex]
    & & \left. \quad \quad + \mbox{tr} \left( \bX_{ii'}^{T}\bX_{ii'} \bSigma_{\qDens(\bbeta)} \right)
        + \mbox{tr} \left( \bZ_{ii'}^{T}\bZ_{ii'} \bSigma_{\qDens(\bu_{i})} \right) \right.
    \\[1ex]
    & & \left. \quad \quad
        + \mbox{tr} \left( \left(\bZ'_{ii'}\right)^{T}\bZ'_{ii'} \bSigma_{\qDens(\bu'_{i'})} \right)
        \right.
    \\[2ex]
    & & \left. \quad \quad + 2 \mbox{tr} \left[ \bZ_{ii'}^{T} \bX_{ii'} E_{\qDens} \{
        ( \bbeta - \bmu_{\qDens(\bbeta)} ) ( \bu_{i} - \bmu_{\qDens(\bu_{i})} )^{T}
        \} \right] \right.
    \\[2ex]
    & & \left. \quad \quad
        + 2 \mbox{tr} \left[ \left( {\bZ'}_{ii'} \right)^{T} \bX_{ii'} E_{\qDens} \{
        (\bbeta - \bmu_{\qDens(\bbeta)})(\bu'_{i'} - \bmu_{\qDens(\bu'_{i'})} )^{T}
        \} \right] \right.
    \\[2ex]
    & & \left. \quad \quad
        + 2 \mbox{tr} \left[ \bZ_{ii'}^{T} \bZ'_{ii'} E_{\qDens} \{
        (\bu_{i} - \bmu_{\qDens(\bu_{i})}) (\bu'_{i'} -
        \bmu_{\qDens(\bu'_{i'})})^{T} \} \right] \right\}.
  \end{array}
\end{equation*}
Under product restrictions I and II, $E_{\qDens} \left\{ \pDens \left( \by | \bbeta,
\bu, \bu', \sigma^2 \right) \right\}$ simplify further as we have
$$
  E_{\qDens} \{ (\bbeta - \bmu_{\qDens(\bbeta)})(\bu'_{i'} - \bmu_{\qDens(\bu'_{i'})} )^{T} \}
  = \bO, \quad 1 \le i' \le m',
$$
and
$$
  E_{\qDens} \{ (\bu_{i} - \bmu_{\qDens(\bu_{i})}) (\bu'_{i'} -
  \bmu_{\qDens(\bu'_{i'})})^{T} \} = \bO, \quad 1 \le i \le m, 1 \le i' \le m'.
$$
Under product restriction I we also have
$$
   E_{\qDens} \{ ( \bbeta - \bmu_{\qDens(\bbeta)} ) ( \bu_{i} - \bmu_{\qDens(\bu_{i})} )^{T} \}
   = \bO, \quad 1 \le i \le m.
$$
The second of the $\log\underline{\pDens}(\by;\qDens)$ terms is
$$
  E_{\qDens}[\log\{\pDens(\bbeta, \bu, \bu'\ | \ \bSigma, \bSigma') \}]
  = E_{\qDens}[\log\{\pDens(\bbeta) \} + \log \{ \pDens( \bu | \bSigma)\} + \log\{\pDens(\bu' | \Sigmad)\}]
$$
{\setlength\arraycolsep{3pt}
\begin{eqnarray*}
  &=& -\smhalf (p + mq + m'q') \log (2 \pi) - \smhalf \log | \bSigma_{\bbeta} | - \frac{m}{2} E_{\qDens} \left\{ \log | \bSigma |
      \right\} - \frac{m'}{2} E_{\qDens} \left\{ \log | \bSigma' | \right\}
  \\
  & &
      - \smhalf \mbox{tr} \left( \bSigma_{\bbeta}^{-1} \left\{
      \left( \bmu_{\qDens (\bbeta)} - \bmu_{\bbeta} \right) \left( \bmu_{\qDens (\bbeta)} - \bmu_{\bbeta} \right)^T
      + \bSigma_{\qDens (\bbeta)} \right\} \right)
  \\
  & &
      -\smhalf \mbox{tr} \left( \bM_{\qDens (\bSigma^{-1})} \left\{ \displaystyle{\sum_{i=1}^{m}} \left(
      \bmu_{\qDens (\bu_{i})} \bmu_{\qDens (\bu_{i})}^T + \bSigma_{\qDens (\bu_{i})} \right) \right\} \right)
  \\
  & & -\smhalf \mbox{tr} \left( \bM_{\qDens \left(\left(\Sigmad\right)^{-1} \right)} \left\{ \displaystyle{\sum_{i'=1}^{m'}} \left(
      \bmu_{\qDens (\bu'_{i'})} \bmu_{\qDens (\bu'_{i'})}^T + \bSigma_{\qDens (\bu'_{i'})} \right) \right\} \right).
\end{eqnarray*}
}
The third of the $\log\underline{\pDens}(\by;\qDens)$ terms is the negative of
{\setlength\arraycolsep{3pt}
\begin{eqnarray*}
  E_{\qDens}[\log\{\qDens(\bbeta, \bu, \bu')\}]
  & = & -\smhalf (p + mq + m'q') - \smhalf (p+mq+m'q') \log (2 \pi) - \smhalf \log | \bSigma_{\qDens(\bbeta, \bu, \bu')}|.
\end{eqnarray*}
}
The fourth of the $\log\underline{\pDens}(\by;\qDens)$ terms is
{\setlength\arraycolsep{3pt}
\begin{eqnarray*}
  E_{\qDens}[\log\{\pDens(\sigma^2| \asigsq)\}]
  &=&E_{\qDens}\left(\log\left[\frac{\{1/(2\asigsq)\}^{\nusigsq/2}}{\Gamma(\nusigsq/2)}(\sigma^2)^{-(\nusigsq/2)-1}\exp\{-1/(2\asigsq \sigma^2)\}\right]\right)\\[1ex]
  &=&-\smhalf\nusigsq\,E_{\qDens}\{\log(2\asigsq)\}-\log\{\Gamma(\smhalf\nusigsq)\}-(\smhalf\nusigsq+1)E_{\qDens}\{\log(\sigma^2)\}
  \\
  & & -\smhalf\mu_{\qDens(1/\asigsq)}\mu_{\qDens(1/\sigma^2).}
\end{eqnarray*}
}
The fifth of the $\log\underline{\pDens}(\by;\qDens)$ terms is the negative of
{\setlength\arraycolsep{3pt}
\begin{eqnarray*}
E_{\qDens}[\log\{\qDens(\sigma^2)\}]
&=&E_{\qDens}\left(\log\left[\frac{\{\lambda_{\qDens(\sigma^2)}/2\}^{\xi_{\qDens(\sigma^2)}/2}}{\Gamma(\xi_{\qDens(\sigma^2)}/2)}(\sigma^2)^{-(\xi_{\qDens(\sigma^2)}/2)-1}\exp\{-\lambda_{\qDens(\sigma^2)}/(2\sigma^2)\}\right]\right)\\[1ex]
&=&\smhalf\xi_{\qDens(\sigma^2)}\,\log(\lambda_{\qDens(\sigma^2)}/2)-\log\{\Gamma(\smhalf\xi_{\qDens(\sigma^2)})\}-(\smhalf\xi_{\qDens(\sigma^2)}+1)E_{\qDens}\{\log(\sigma^2)\}
\\
& & -\smhalf\lambda_{\qDens(\sigma^2)}\mu_{\qDens(1/\sigma^2)}.
\end{eqnarray*}
}
The sixth of the $\log\underline{\pDens}(\by;\qDens)$ terms is
{\setlength\arraycolsep{3pt}
\begin{eqnarray*}
E_{\qDens}[\log\{\pDens(\asigsq)\}]
&=&E_{\qDens}\left(\log\left[\frac{\{1/(2\nusigsq \ssigsq^2)\}^{1/2}}{\Gamma(1/2)}\asigsq^{-(1/2)-1}\exp\{-1/(2\nusigsq \ssigsq^2 \asigsq)\}\right]\right)\\[1ex]
&=&-\smhalf\,\log(2\nusigsq \ssigsq^2)-\log\{\Gamma(\smhalf)\}-(\smhalf+1)E_{\qDens}\{\log(\asigsq)\}-\{1/(2\nusigsq \ssigsq^2)\}\mu_{\qDens(1/\asigsq)}.
\end{eqnarray*}
}
The seventh of the $\log\underline{\pDens}(\by;\qDens)$ terms is the negative of
{\setlength\arraycolsep{3pt}
\begin{eqnarray*}
E_{\qDens}[\log\{\qDens(\asigsq)\}]
&=&E_{\qDens}\left(\log\left[\frac{\{\lambda_{\qDens(\asigsq)}/2\}^{\xi_{\qDens(\asigsq)}/2}}{\Gamma(\xi_{\qDens(\asigsq)}/2)}(\asigsq)^{-(\xi_{\qDens(\asigsq)}/2)-1}\exp\{-\lambda_{\qDens(\asigsq)}/(2\asigsq)\}\right]\right)\\[1ex]
&=&\smhalf\xi_{\qDens(\asigsq)}\,\log(\lambda_{\qDens(\asigsq)}/2)-\log\{\Gamma(\smhalf\xi_{\qDens(\asigsq)})\}-(\smhalf\xi_{\qDens(\asigsq)}+1)E_{\qDens}\{\log(\asigsq)\}
\\
& & -\smhalf\lambda_{\qDens(\asigsq)} \mu_{\qDens(1/\asigsq)}.
\end{eqnarray*}
}
The eighth of the $\log\underline{\pDens}(\by;\qDens)$ terms is
{\setlength\arraycolsep{3pt}
\begin{eqnarray*}
E_{\qDens}[\log\{\pDens(\bSigma | \ASigma)\}]
&=& E_{\qDens}\left( \frac{|\ASigma|^{-\smhalf (\nuSigma + q - 1)} |\bSigma|^{-\smhalf(\nuSigma + 2q)}}{2^{\frac{q}{2}(\nuSigma + 2q - 1)} \pi^{\frac{q}{4}(q-1)} \prod_{j=1}^{q} \Gamma(\smhalf (\nuSigma + 2q - j))}
    \exp\{ -\smhalf \mbox{tr} (\ASigma^{-1} \bSigma^{-1}) \} \right)
    \\
    & = & - \smhalf (\nuSigma + q - 1) E_{\qDens} \{ \log | \ASigma | \} - \smhalf (\nuSigma + 2q) E_{\qDens}
    \{ \log | \bSigma | \}
    \\
    & & - \smhalf \mbox{tr} ( \bM_{\qDens(\ASigma^{-1})} \bM_{\qDens(\bSigma^{-1})} )
    - \frac{q}{2} (\nuSigma+2q - 1) \log(2) - \frac{q}{4} (q-1) \log(\pi)
    \\
    & & - \sum_{j=1}^{q} \log \Gamma (\smhalf (\nuSigma + 2q - j)).
\end{eqnarray*}
}
The ninth of the $\log\underline{\pDens}(\by;\qDens)$ terms is the negative of
{\setlength\arraycolsep{3pt}
\begin{eqnarray*}
E_{\qDens}[\log\{\qDens(\bSigma)\}]
&=& E_{\qDens} \left( \frac{|\bLambda_{\qDens(\bSigma)}|^{\smhalf (\xi_{\qDens(\bSigma)} - q + 1)} |\bSigma|^{-\smhalf(\xi_{\qDens(\bSigma)} + 2)}}{2^{\frac{q}{2}(\xi_{\qDens(\bSigma)} + 1)} \pi^{\frac{q}{4}(q-1)} \prod_{j=1}^{q} \Gamma(\smhalf (\xi_{\qDens(\bSigma)} + 2 - j))}
    \exp\{ -\smhalf \mbox{tr} (\bLambda_{\qDens(\bSigma)} \bSigma^{-1}) \} \right)
    \\
    & = & \smhalf (\xi_{\qDens(\bSigma)} - q + 1) \log |\bLambda_{\qDens(\bSigma)}| - \smhalf (\xi_{\qDens(\bSigma)} + 2)
    E_{\qDens} \{ \log |\bSigma| \} - \smhalf \mbox{tr} (\bLambda_{\qDens(\bSigma)} \bM_{\qDens(\bSigma^{-1})})
    \\
    & & -\frac{q}{2} (\xi_{\qDens(\bSigma)} + 1) \log (2) - \frac{q}{4} (q-1) \log(\pi)
    - \sum_{j=1}^{q} \log \Gamma (\smhalf (\xi_{\qDens(\bSigma)} + 2 - j)).
\end{eqnarray*}
}
The tenth of the $\log\underline{\pDens}(\by;\qDens)$ terms is
{\setlength\arraycolsep{3pt}
\begin{eqnarray*}
E_{\qDens}[\log\{\pDens(\ASigma)\}]
&=& E_{\qDens}\left( \frac{|\bLambda_{\ASigma}|^{\smhalf (2-q)} |\ASigma|^{-3/2}}{2^{q} \pi^{\frac{q}{4}(q-1)} \prod_{j=1}^{q} \Gamma(\smhalf (3 - j))}
    \exp\{ -\smhalf \mbox{tr} (\bLambda_{\ASigma} \ASigma^{-1}) \} \right)
    \\
    & = & - \smhalf q (2-q) \log (\nuSigma) - \smhalf (2-q) \sum_{j=1}^{q} \log (\sSigmaj^2)
    - \frac{3}{2} E_{\qDens} \{ \log |\ASigma| \}
    \\
    & & - \smhalf \sum_{j=1}^{q} 1/(\nuSigma \sSigmaj^2) \left( \bM_{\qDens(\ASigma^{-1})} \right)_{jj}
    - q \log (2) - \frac{q}{4} (q-1) \log(\pi)
    \\
    & & - \sum_{j=1}^{q} \log \Gamma (\smhalf (3-j)).
\end{eqnarray*}
}
The eleventh of the $\log\underline{\pDens}(\by;\qDens)$ terms is the negative of
{\setlength\arraycolsep{3pt}
\begin{eqnarray*}
E_{\qDens}[\log\{\qDens(\ASigma)\}]
&=& E_{\qDens} \left( \frac{|\bLambda_{\qDens(\ASigma)}|^{\smhalf (\xi_{\qDens(\ASigma)} - q + 1)} |\ASigma|^{-\smhalf(\xi_{\qDens(\ASigma)} + 2)}}{2^{\frac{q}{2}(\xi_{\qDens(\ASigma)} + 1)} \pi^{\frac{q}{4}(q-1)} \prod_{j=1}^{q} \Gamma(\smhalf (\xi_{\qDens(\ASigma)} + 2 - j))}
    \exp\{ -\smhalf \mbox{tr} (\bLambda_{\qDens(\ASigma)}^{-1} \ASigma^{-1}) \} \right)
    \\
    & = & \smhalf (\xi_{\qDens(\ASigma)} - q + 1) \log |\bLambda_{\qDens(\ASigma)}| - \smhalf (\xi_{\qDens(\ASigma)} + 2)
    E_{\qDens} \{ \log |\ASigma| \} - \smhalf \mbox{tr} (\bLambda_{\qDens(\ASigma)} \bM_{\qDens(\ASigma^{-1})})
    \\
    & & -\frac{q}{2} (\xi_{\qDens(\ASigma)} + 1) \log (2) - \frac{q}{4} (q-1) \log(\pi)
    - \sum_{j=1}^{q} \log \Gamma (\smhalf (\xi_{\qDens(\ASigma)} + 2 - j)).
\end{eqnarray*}
}
The remaining four terms of $\log\underline{\pDens}(\by;\qDens)$ are
{\setlength\arraycolsep{3pt}
\begin{eqnarray*}
E_{\qDens}[\log\{\pDens(\Sigmad | \ASigmad)\}]
& = & - \smhalf (\nuSigmad + q' - 1) E_{\qDens} \{ \log | \ASigmad | \} - \smhalf (\nuSigmad + 2q') E_{\qDens}
    \{ \log | \Sigmad | \}
    \\
    & & - \smhalf \mbox{tr} ( \bM_{\qDens( \left( \ASigmad \right)^{-1})} \bM_{\qDens(\left( \bSigma \right)^{-1})} )
    - \frac{q'}{2} (\nuSigmad+2q' - 1) \log(2) - \frac{q'}{4} (q'-1) \log(\pi)
    \\
    & & - \sum_{j=1}^{q'} \log \Gamma (\smhalf (\nuSigmad + 2q' - j)),
\end{eqnarray*}
}
the negative of
{\setlength\arraycolsep{3pt}
\begin{eqnarray*}
E_{\qDens}[\log\{\qDens(\Sigmad)\}]
& = & \smhalf (\xi_{\qDens(\Sigmad)} - q' + 1) \log |\bLambda_{\qDens(\Sigmad)}| - \smhalf (\xi_{\qDens(\Sigmad)} + 2)
    E_{\qDens} \{ \log |\Sigmad| \} - \smhalf \mbox{tr} (\bLambda_{\qDens(\Sigmad)} \bM_{\qDens(\left(\Sigmad\right)^{-1})})
    \\
    & & -\frac{q'}{2} (\xi_{\qDens(\Sigmad)} + 1) \log (2) - \frac{q'}{4} (q'-1) \log(\pi)
    - \sum_{j=1}^{q'} \log \Gamma (\smhalf (\xi_{\qDens(\Sigmad)} + 2 - j)),
\end{eqnarray*}
}
{\setlength\arraycolsep{1pt}
\begin{eqnarray*}
E_{\qDens}[\log\{\pDens(\ASigmad)\}]
& = & - \smhalf q' (2-q') \log (\nuSigmad) - \smhalf (2-q') \sum_{j'=1}^{q'} \log (\sSigmaj^2)
    - \frac{3}{2} E_{\qDens} \{ \log |\ASigma| \}
    \\
    & & - \smhalf \sum_{j=1}^{q'} 1/(\nuSigmad \sSigmaj^2) \left( \bM_{\qDens(\left(\ASigmad \right)^{-1})} \right)_{jj}
    - q' \log (2) - \frac{q'}{4} (q'-1) \log(\pi)
    \\
    & & - \sum_{j=1}^{q'} \log \Gamma (\smhalf (3-j)),
\end{eqnarray*}
}
and the negative of
{\setlength\arraycolsep{3pt}
\begin{eqnarray*}
E_{\qDens}[\log\{\qDens(\ASigmad)\}]
& = & \smhalf (\xi_{\qDens(\ASigmad)} - q' + 1) \log |\bLambda_{\qDens(\ASigmad)}| - \smhalf (\xi_{\qDens(\ASigmad)} + 2)
    E_{\qDens} \{ \log |\ASigmad| \}
    \\
    & & - \smhalf \mbox{tr} (\bLambda_{\qDens(\ASigmad)} \bM_{\qDens(\left(\ASigmad \right)^{-1})})
    -\frac{q'}{2} (\xi_{\qDens(\ASigmad)} + 1) \log (2) - \frac{q'}{4} (q'-1) \log(\pi)
    \\
    & & - \sum_{j=1}^{q'} \log \Gamma (\smhalf (\xi_{\qDens(\ASigmad)} + 2 - j)).
\end{eqnarray*}
}
In the summation of each of these $\log\underline{\pDens}(\bx;q)$ terms, note that the coefficient of $E_{\qDens}\{\log(\sigma^2)\}$ is
$$-\smhalf\,\ndotdot-\smhalf\nusigsq-1+\smhalf\xi_{\qDens(\sigma^2)}+1=-\smhalf\,\ndotdot-\smhalf\nusigsq-1+\smhalf(\nusigsq+\ndotdot)+1=0.$$
The coefficient of $E_{\qDens}\{\log(\asigsq)\}$ is
$$-\smhalf\nusigsq-(\smhalf+1)+\smhalf\xi_{\qDens(\asigsq)}+1=-\smhalf\nusigsq-(\smhalf+1)+\smhalf(\nusigsq+1)+1=0.$$
The coefficient of $E_{\qDens}\{\log|\bSigma|\}$ is
$$-\frac{m}{2} - \smhalf(\nuSigma + 2q) + \smhalf(\xi_{\qDens(\bSigma)} + 2) =
  -\smhalf(m + \nuSigma + 2q) + \smhalf(m + \nuSigma + 2q)=0.$$
The coefficient of $E_{\qDens}\{\log|\ASigma|\}$ is
$$- \smhalf(\nuSigma + q - 1) - \frac{3}{2} + \smhalf(\xi_{\qDens(\ASigma)} + 2) =
  -\smhalf(\nuSigma + q + 2) + \smhalf(\nuSigma + q + 2)=0.$$
The coefficient of $E_{\qDens}\{\log|\Sigmad|\}$ is
$$-\frac{m'}{2} - \smhalf(\nuSigmad + 2q') + \smhalf(\xi_{\qDens(\Sigmad)} + 2) =
  -\smhalf(m' + \nuSigmad + 2q') + \smhalf(m' + \nuSigmad + 2q')=0.$$
The coefficient of $E_{\qDens}\{\log|\ASigmad|\}$ is
$$- \smhalf(\nuSigmad + q' - 1) - \frac{3}{2} + \smhalf(\xi_{\qDens(\ASigmad)} + 2) =
  -\smhalf(\nuSigmad + q' + 2) + \smhalf(\nuSigmad + q' + 2)=0.$$
Therefore, the terms in $E_{\qDens}\{\log(\sigma^2)\}$, $E_{\qDens}\{\log(a)\}$, $E_{\qDens}\{\log|\bSigma|\}$ and $E_{\qDens}\{\log|\ASigma|\}$
can be dropped and we then have
$$\log\underline{\pDens}(\by;\qDens)= \sum_{i=1}^{15} T_{i} $$
where
{\setlength\arraycolsep{3pt}
\begin{eqnarray*}
T_1&=&-\smhalf \ndotdot \log(2 \pi)\\
    & & -\smhalf \mu_{\qDens (1/{\sigma^{2}})} \displaystyle{\sum_{i=1}^{m} \sum_{i'=1}^{m'}}
        \left\{ \left| \left| \by_{ii'} - \bX_{ii'} \bmu_{\qDens (\bbeta)} - \bZ_{ii'} \bmu_{\qDens(\bu_{i})}
        - \bZdash_{ii'} \bmu_{\qDens(\budash_{i'})}  \right| \right|^2  \right.
    \\
    & & \left. \quad \quad + \mbox{tr} \left( \bX_{ii'}^{T}\bX_{ii'} \bSigma_{\qDens(\bbeta)} \right)
        + \mbox{tr} \left( \bZ_{ii'}^{T}\bZ_{ii'} \bSigma_{\qDens(\bu_{i})} \right)
        + \mbox{tr} \left( \left(\bZ'_{ii'}\right)^{T}\bZ'_{ii'} \bSigma_{\qDens(\bu'_{i'})} \right) \right.
    \\
    & & \left. \quad \quad + 2 \mbox{tr} \left[ \bZ_{ii'}^{T} \bX_{ii'} E_{\qDens} \left\{
        \left( \bbeta - \bmu_{\qDens(\bbeta)} \right) \left( \bu_{i} - \bmu_{\qDens(\bu_{i})} \right)^{T}
        \right\} \right] \right.
    \\
    & & \left. \quad \quad
        + 2 \mbox{tr} \left[ \left(\bZ'_{ii'} \right)^{T} \bX_{ii'} E_{\qDens} \left\{
        \left( \bbeta - \bmu_{\qDens(\bbeta)} \right) \left( \bu'_{i'} - \bmu_{\qDens(\bu'_{i'})} \right)^{T}
        \right\} \right] \right.
    \\
    & & \left. \quad \quad
        + 2 \mbox{tr} \left[ \bZ_{ii'}^{T} \bZ'_{ii'} E_{\qDens} \left\{
        \left( \bu_{i} - \bmu_{\qDens(\bu_{i})} \right) \left( \bu'_{i'} - \bmu_{\qDens(\bu'_{i'})} \right)^{T}
        \right\} \right] \right\},\\[1ex]
T_2&=& -\smhalf (p + mq + m'q') \log (2 \pi) - \smhalf \log | \bSigma_{\bbeta} |
   \\
   & & - \smhalf \mbox{tr} \left( \bSigma_{\bbeta}^{-1} \left\{
       \left( \bmu_{\qDens (\bbeta)} - \bmu_{\bbeta} \right) \left( \bmu_{\qDens (\bbeta)} - \bmu_{\bbeta} \right)^T
       + \bSigma_{\qDens (\bbeta)} \right\} \right)
   \\
   & & -\smhalf \mbox{tr} \left( \bM_{\qDens (\bSigma^{-1})} \left\{ \displaystyle{\sum_{i=1}^{m}} \left(
       \bmu_{\qDens (\bu_{i})} \bmu_{\qDens (\bu_{i})}^T + \bSigma_{\qDens (\bu_{i})} \right) \right\} \right)
   \\
   & & -\smhalf \mbox{tr} \left( \bM_{\qDens \left(\left(\Sigmad\right)^{-1} \right)} \left\{ \displaystyle{\sum_{i'=1}^{m'}} \left(
       \bmu_{\qDens (\bu'_{i'})} \bmu_{\qDens (\bu'_{i'})}^T + \bSigma_{\qDens (\bu'_{i'})} \right) \right\} \right),\\[1ex]
T_3&=& \smhalf (p + mq+ m'q') + \smhalf (p+mq+m'q') \log (2 \pi) + \smhalf \log \left| \bSigma_{\qDens(\bbeta, \bu, \bu')} \right|,\\[1ex]
T_4&=& \smhalf\nusigsq \log(2)-\log\{\Gamma(\smhalf\nusigsq)\}-\smhalf\mu_{\qDens(1/\asigsq)}\mu_{\qDens(1/\sigma^2)},\\[1ex]
T_5&=& -\smhalf\xi_{\qDens(\sigma^2)}\,\log(\lambda_{\qDens(\sigma^2)}/2)+\log\{\Gamma(\smhalf\xi_{\qDens(\sigma^2)})\}
      +\smhalf\lambda_{\qDens(\sigma^2)}\mu_{\qDens(1/\sigma^2)},\\[1ex]
T_6&=& -\smhalf\,\log(2\nusigsq \ssigsq^2)-\log\{\Gamma(\smhalf)\}-\{1/(2\nusigsq \ssigsq^2)\}\mu_{\qDens(1/\asigsq)}
\end{eqnarray*}
}
{\setlength\arraycolsep{3pt}
\begin{eqnarray*}
T_7&=& -\smhalf\xi_{\qDens(\asigsq)}\,\log(\lambda_{\qDens(\asigsq)}/2)+\log\{\Gamma(\smhalf\xi_{\qDens(\asigsq)})\}+\smhalf\lambda_{\qDens(\asigsq)} \mu_{\qDens(1/\asigsq)},\\[1ex]
T_8&=& - \smhalf \mbox{tr} ( \bM_{\qDens(\ASigma^{-1})} \bM_{\qDens(\bSigma^{-1})} )
       - \frac{q}{2} (\nuSigma+2q - 1) \log(2) - \frac{q}{4} (q-1) \log(\pi)
       \\
       & & - \sum_{j=1}^{q} \log \Gamma (\smhalf (\nuSigma + 2q - j)),\\[1ex]
T_9&=& - \smhalf (\xi_{\qDens(\bSigma)} - q + 1) \log |\bLambda_{\qDens(\bSigma)}|
       + \smhalf \mbox{tr} (\bLambda_{\qDens(\bSigma)} \bM_{\qDens(\bSigma^{-1})})
       +\frac{q}{2} (\xi_{\qDens(\bSigma)} + 1) \log (2),
       \\
   & & + \frac{q}{4} (q-1) \log(\pi) + \sum_{j=1}^{q} \log \Gamma (\smhalf (\xi_{\qDens(\bSigma)} + 2 - j)), \\[1ex]
T_{10}&=& - \smhalf q (2-q) \log (\nuSigma) - \smhalf (2-q) \sum_{j=1}^{q} \log (\sSigmaj^2)
        - \smhalf \sum_{j=1}^{q} 1/(\nuSigma \sSigmaj^2) \left( \bM_{\qDens(\ASigma^{-1})} \right)_{jj}
        \\
    & & - q \log (2) - \frac{q}{4} (q-1) \log(\pi) - \sum_{j=1}^{q} \log \Gamma (\smhalf (3-j)), \\[1ex]
 T_{11}&=& - \smhalf (\xi_{\qDens(\ASigma)} - q + 1) \log |\bLambda_{\qDens(\ASigma)}|
          + \smhalf \mbox{tr} (\bLambda_{\qDens(\ASigma)} \bM_{\qDens(\ASigma^{-1})})
          \\
      & & +\frac{q}{2} (\xi_{\qDens(\ASigma)} + 1) \log (2) + \frac{q}{4} (q-1) \log(\pi)
          + \sum_{j=1}^{q} \log \Gamma (\smhalf (\xi_{\qDens(\ASigma)} + 2 - j)),
\end{eqnarray*}
}
{\setlength\arraycolsep{3pt}
\begin{eqnarray*}
  T_{12}&=& - \smhalf \mbox{tr} ( \bM_{\qDens(\ASigmad^{-1})} \bM_{\qDens(\left(\Sigmad\right)^{-1})} )
         - \frac{q'}{2} (\nuSigmad+2q' - 1) \log(2) - \frac{q'}{4} (q'-1) \log(\pi)
         \\
         & & - \sum_{j=1}^{q'} \log \Gamma (\smhalf (\nuSigmad + 2q' - j)),\\[1ex]
  T_{13}&=& - \smhalf (\xi_{\qDens(\Sigmad)} - q' + 1) \log |\bLambda_{\qDens(\Sigmad)}|
         + \smhalf \mbox{tr} (\bLambda_{\qDens(\Sigmad)} \bM_{\qDens(\left(\Sigmad\right)^{-1})})
         +\frac{q'}{2} (\xi_{\qDens(\Sigmad)} + 1) \log (2),
         \\
     & & + \frac{q'}{4} (q'-1) \log(\pi) + \sum_{j=1}^{q'} \log \Gamma (\smhalf (\xi_{\qDens(\Sigmad)} + 2 - j)), \\[1ex]
  T_{14}&=& - \smhalf q' (2-q') \log (\nuSigmad) - \smhalf (2-q') \sum_{j=1}^{q'} \log (\sSigmaj^2)
          - \smhalf \sum_{j=1}^{q'} 1/(\nuSigmad \sSigmaj^2) \left( \bM_{\qDens(\left(\ASigmad \right)^{-1})} \right)_{jj}
          \\
      & & - q' \log (2) - \frac{q'}{4} (q'-1) \log(\pi) - \sum_{j=1}^{q'} \log \Gamma (\smhalf (3-j)) \\[1ex]
  \mbox{and}\quad T_{15}&=& - \smhalf (\xi_{\qDens(\ASigmad)} - q' + 1) \log |\bLambda_{\qDens(\ASigmad)}|
            + \smhalf \mbox{tr} (\bLambda_{\qDens(\ASigmad)} \bM_{\qDens(\left(\ASigmad\right)^{-1})})
            \\
        & & +\frac{q'}{2} (\xi_{\qDens(\ASigmad)} + 1) \log (2) + \frac{q'}{4} (q'-1) \log(\pi)
            + \sum_{j=1}^{q'} \log \Gamma (\smhalf (\xi_{\qDens(\ASigmad)} + 2 - j)).
\end{eqnarray*}
}
Note that the component of $\log\underline{\pDens}(\by;\qDens)$ which does not get updated during the coordinate
ascent iterations, except for the irreducible $\log\Gamma$ terms, and which we will call `const` is:
{\setlength\arraycolsep{3pt}
\begin{eqnarray*}
  \mbox{const}&\equiv&-\smhalf \ndotdot \log (2 \pi) - \smhalf (p + mq + m'q') \log(2 \pi)
  - \smhalf \log |\bSigma_{\bbeta}| + \smhalf (p + mq + m'q')
  \\
  & & + \smhalf (p + mq + m'q') \log (2 \pi) - \smhalf
  \nusigsq \log (2) + \smhalf (\xi_{\qDens(\sigma^2)}) \log (2) - \smhalf \log (2 \nusigsq \ssigsq^2)
  \\
  & & - \smhalf q (\nuSigma + 2q - 1)\log (2) - \frac{q}{2} (q-1) \log (\pi)
  + \smhalf q (\xi_{\qDens(\bSigma)} + 1) \log (2) + \frac{q}{2} (q-1) \log (\pi)
  \\
  & &
  - \smhalf q (2 - q) \log (\nuSigma) - \smhalf (2-q) \sum_{j=1}^{q} \log (\sSigmaj^2) - q \log (2)
  + \smhalf q (\xi_{\qDens(\ASigma)} + 1) \log(2)
  \\
  & &  - \smhalf q' (\nuSigmad + 2q' - 1)\log (2) - \frac{q'}{2} (q'-1) \log (\pi)
       + \smhalf q' (\xi_{\qDens(\Sigmad)} + 1) \log (2)
  \\
  & &  + \frac{q'}{2} (q'-1) \log (\pi)
       - \smhalf q' (2 - q') \log (\nuSigmad) - \smhalf (2-q') \sum_{j=1}^{q'} \log (\sSigmaj^2) - q' \log (2)
  \\
  & & + \smhalf q' (\xi_{\qDens(\ASigmad)} + 1) \log(2) - \log \Gamma (\smhalf)
  \\[2ex]
  & = & - \smhalf \left( \ndotdot + 1 \right) \log( \pi) - \smhalf \log |\Sigma_{\bbeta}|
  + \smhalf (p + mq + m'q') - \smhalf \log (\nusigsq) - \smhalf \log (\ssigsq^2)
  \\[2ex]
  & & + \smhalf \left\{ q (\nuSigma + q + m - 1) + q' (\nuSigmad + q' + m' - 1) - 1 \right\} \log (2)
  \\
  & &  - \smhalf q (2 - q) \log (\nuSigma) - \smhalf (2-q) \sum_{j=1}^{q} \log (\ssigsq^2)
       - \smhalf q' (2 - q') \log (\nuSigmad) - \smhalf (2-q') \sum_{j=1}^{q'} \log (\ssigsq^2)
\end{eqnarray*}
}
Our final $\log\underline{\pDens}(\by;\qDens)$ expression is then

{\setlength\arraycolsep{0.5pt}
\begin{eqnarray*}
  \log\underline{\pDens}(\by;\qDens) & = & - \smhalf \left( \ndotdot + 1 \right) \log( \pi) - \smhalf \log |\Sigma_{\bbeta}|
  + \smhalf (p + mq + m'q') - \smhalf \log (\nusigsq) - \smhalf \log (\ssigsq^2)
  \\
  & & + \smhalf \left\{ q (\nuSigma + q + m - 1) + q' (\nuSigmad + q' + m' - 1) - 1 \right\} \log (2)
  - \smhalf q (2 - q) \log (\nuSigma)
  \\
  & &
  - \smhalf (2-q) \sum_{j=1}^{q} \log (\ssigsq^2)
  - \smhalf q' (2 - q') \log (\nuSigmad) - \smhalf (2-q') \sum_{j=1}^{q'} \log (\ssigsq^2)
  -\log\{\Gamma(\smhalf\nusigsq)\}
  \\
  & & - \smhalf \mbox{tr} \left( \bSigma_{\bbeta}^{-1} \left\{
  \left( \bmu_{\qDens (\bbeta)} - \bmu_{\bbeta} \right) \left( \bmu_{\qDens (\bbeta)} - \bmu_{\bbeta} \right)^T
  + \bSigma_{\qDens (\bbeta)} \right\} \right) + \smhalf \log | \bSigma_{\qDens(\bbeta, \bu, \bu')} |
  \\
  & & -\smhalf \mbox{tr} \left( \bM_{\qDens (\bSigma^{-1})} \left\{ \displaystyle{\sum_{i=1}^{m}} \left(
  \bmu_{\qDens (\bu_{i})} \bmu_{\qDens (\bu_{i})}^T + \bSigma_{\qDens (\bu_{i})} \right) \right\} \right)
   -\smhalf\mu_{\qDens(1/\asigsq)}\mu_{\qDens(1/\sigma^2)}
  \\
  & &
  -\smhalf \mbox{tr} \left( \bM_{\qDens \left(\left(\Sigmad\right)^{-1}\right)} 
  \left\{ \displaystyle{\sum_{i'=1}^{m'}} \left(
  \bmu_{\qDens (\bu'_{i'})} \bmu_{\qDens (\bu'_{i'})}^T + \bSigma_{\qDens (\bu'_{i'})} \right) \right\} \right)
  -\{1/(2\nusigsq \ssigsq^2)\}\mu_{\qDens(1/\asigsq)}
  \\
  & &
   - \smhalf\xi_{\qDens(\sigma^2)}\,\log(\lambda_{\qDens(\sigma^2)}/2)
  + \log\{\Gamma(\smhalf\xi_{\qDens(\sigma^2)})\} + \smhalf\lambda_{\qDens(\sigma^2)}\mu_{\qDens(1/\sigma^2)}
  \\
  & & - \smhalf \mbox{tr} ( \bM_{\qDens(\ASigma^{-1})} \bM_{\qDens(\bSigma^{-1})} )
  - \sum_{j=1}^{q} \log \Gamma (\smhalf (\nuSigma + 2q - j))
  - \smhalf \mbox{tr} (\bLambda_{\qDens(\bSigma)} \bM_{\qDens(\bSigma^{-1})})
  \\
  & & - \smhalf \mbox{tr} ( \bM_{\qDens(\ASigmad^{-1})} \bM_{\qDens(\left(\Sigmad\right)^{-1})} )
  - \sum_{j=1}^{q'} \log \Gamma (\smhalf (\nuSigmad + 2q' - j))
  - \smhalf \mbox{tr} (\bLambda_{\qDens(\Sigmad)} \bM_{\qDens(\left(\Sigmad\right)^{-1})})\\
   & &  - \smhalf\xi_{\qDens(\asigsq)}\,\log(\lambda_{\qDens(\asigsq)}/2)
  + \log\{\Gamma(\smhalf\xi_{\qDens(\asigsq)})\} + \smhalf\lambda_{\qDens(\asigsq)} \mu_{\qDens(1/\asigsq)}
  \\
  & & - \sum_{j=1}^{q} \log \Gamma (\smhalf (\xi_{\qDens(\bSigma)} + 2 - j))
      + \smhalf (\xi_{\qDens(\bSigma)} - q + 1) \log |\bLambda_{\qDens(\bSigma)}|
\end{eqnarray*}
}
{\setlength\arraycolsep{0.5pt}
\begin{eqnarray*}
  & & - \sum_{j=1}^{q'} \log \Gamma (\smhalf (\xi_{\qDens(\Sigmad)} + 2 - j))
      + \smhalf (\xi_{\qDens(\Sigmad)} - q' + 1) \log |\bLambda_{\qDens(\Sigmad)}|
  \\
  & & -\smhalf \mu_{\qDens (1/{\sigma^{2}})} \displaystyle{\sum_{i=1}^{m} \sum_{i'=1}^{m'}}
      \left\{ \left| \left| \by_{ii'} - \bX_{ii'} \bmu_{\qDens (\bbeta)} - \bZ_{ii'} \bmu_{\qDens(\bu_{i})}
      - \bZdash_{ii'} \bmu_{\qDens(\budash_{i'})}  \right| \right|^2  \right.
  \\
  & & \left. \quad \quad + \mbox{tr} \left( \bX_{ii'}^{T}\bX_{ii'} \bSigma_{\qDens(\bbeta)} \right)
      + \mbox{tr} \left( \bZ_{ii'}^{T}\bZ_{ii'} \bSigma_{\qDens(\bu_{i})} \right)
      + \mbox{tr} ((\bZ'_{ii'})^{T}\bZ'_{ii'} \bSigma_{\qDens(\bu'_{i'})}) \right.
  \\
  & & \left. \quad \quad + 2 \mbox{tr} \left[ \bZ_{ii'}^{T} \bX_{ii'} E_{\qDens} \{
      (\bbeta - \bmu_{\qDens(\bbeta)}) (\bu_{i} - \bmu_{\qDens(\bu_{i})})^{T}
      \} \right] \right.
  \\
  & & \left. \quad \quad
      + 2 \mbox{tr} \left[ \left(\bZ'_{ii'} \right)^{T} \bX_{ii'} E_{\qDens} \{
      (\bbeta - \bmu_{\qDens(\bbeta)}) (\bu'_{i'} - \bmu_{\qDens(\bu'_{i'})})^{T}
      \} \right] \right.
  \\
  & & \left. \quad \quad
      + 2 \mbox{tr} \left[ \bZ_{ii'}^{T} \bZ'_{ii'} E_{\qDens} \{
      (\bu_{i} - \bmu_{\qDens(\bu_{i})}) (\bu'_{i'} - \bmu_{\qDens(\bu'_{i'})})^{T}
      \} \right] \right\}.
\end{eqnarray*}
}
The $\log\underline{\pDens}(\by;\qDens)$ expression simplifies under product restrictions I and II,
since we have
$$
  E_{\qDens} \{ (\bbeta - \bmu_{\qDens(\bbeta)})(\bu'_{i'} - \bmu_{\qDens(\bu'_{i'})} )^{T} \}
  = \bO, \quad 1 \le i' \le m',
$$
and
$$
  E_{\qDens} \{ (\bu_{i} - \bmu_{\qDens(\bu_{i})}) (\bu'_{i'} -
  \bmu_{\qDens(\bu'_{i'})})^{T} \} = \bO, \quad 1 \le i \le m, 1 \le i' \le m'.
$$
Under product restriction I we also have
$$
   E_{\qDens} \{ ( \bbeta - \bmu_{\qDens(\bbeta)} ) ( \bu_{i} - \bmu_{\qDens(\bu_{i})} )^{T} \}
   = \bO, \quad 1 \le i \le m.
$$
From Theorem 1 of Nolan \myand Wand (2020), the  $\log|\bSigma_{\qDens(\bbeta, \bu, \bu')}|$
term has the following streamlined form:
\vskip2mm
\begin{equation*}
{\setlength\arraycolsep{1pt}
\begin{array}{lcl}
    \log|\bSigma_{\qDens(\bbeta, \bu, \bu')}| & = & \log \left| \mbox{$\bA^{11}$ component
    of $\Ssc$ from Algorithm 2} \right|
    \\[2ex]
    & & \quad \quad - \displaystyle{\sum_{i=1}^{m}} \log \left| \mu_{\qDens(1/{\sigma^2})} \Zuptri_i^{T} \Zuptri_i + \bM_{\qDens (\bSigma^{-1})} \right|,
\end{array}
}
\end{equation*}
under product restriction I, and
\begin{equation*}
{\setlength\arraycolsep{1pt}
\begin{array}{lcl}
   \log|\bSigma_{\qDens(\bbeta, \bu, \bu')}| & = & \log \left| \bSigma_{\qDens(\bbeta)} \right|
       + \displaystyle{\sum_{i=1}^{m}} \log \left| \bSigma_{\qDens(\bu_{i})} \right|
       - \displaystyle{\sum_{i=1}^{m}} \log \left| \mu_{\qDens(1/{\sigma^2})} \Zuptri_i^{T} \Zuptri_i +
      \bM_{\qDens (\bSigma^{-1})} \right|
\end{array}
}
\end{equation*}
under product restrictions II and III.

\section{Streamlined Computing for Frequentist Inference}

As an aside we point out that the approach used by Algorithm \ref{alg:MFVBforPRIII} for product restriction III,
in which the $\qDens$-density updates for the $(\bbeta,\buall)$ parameters are
embedded within the \SolveTwoLevelSparseLeastSquares\ infrastructure, can
also be used for streamlined \emph{frequentist} inference when $m'$ is moderate in size.
To the best of our knowledge, the results given here for efficient computation of 
the important sub-blocks of the relevant covariance matrix are novel.

The frequentist Gaussian response 
two-level linear mixed model with crossed random effects is
\begin{equation}
  \begin{array}{l}
    \by_{i\idash}|\bbeta, \bu_{i},\udashidash \simind N(\bX_{i\idash}\bbeta+\bZ_{i\idash}\,\bu_{i}
    +\Zdash_{i\idash}\,\udashidash,\sigma^2\,\bI),\\[2ex]
    \bu_{i} \simind N (\bzero, \bSigma), \ 
    \udashidash \simind N (\bzero, \Sigmad), \  1\le i\le m, \ 1\le i\le \mdash.
  \end{array}
\label{eq:crossedModel}
\end{equation}
The best linear unbiased predictor of $[\bbeta^T\ \bu^T]^T$ and
its corresponding covariance matrix are
\begin{equation}
  \setlength\arraycolsep{1pt}
  \begin{array}{rcl}
    \left[
    \begin{array}{c}
      \bbetahat\\
      \buhat
    \end{array}
    \right]&=&(\bC^T\RBLUP^{-1}\bC+\DBLUP)^{-1}\bC^T\RBLUP^{-1}\by\\[3ex]
    \mbox{and}\quad
    \mbox{Cov}\left(\left[
    \begin{array}{c}
      \bbetahat\\
      \buhat-\bu
    \end{array}
    \right]\right)&=&(\bC^T\RBLUP^{-1}\bC+\DBLUP)^{-1}
  \end{array}
\label{eq:blup_covs}
\end{equation}
where $\bC\equiv[\bX\ \bZ]$, with $\bX$ and $\bZ$ as defined in Section \ref{sec:additDataMats}.
$$
\DBLUP\equiv\left[
\begin{array}{cc}
\bO & \bO               \\[1ex]
\bO & \left[
\begin{array}{cc}
 \bI_{m}\otimes \bSigma^{-1} &  \bO \\
\bO                & \bI_{\mdash}\otimes(\Sigmad)^{-1}
\end{array}
                     \right]
\end{array}
\right]
\quad\mbox{and}\quad \RBLUP\equiv \sigma^2\bI.
$$
Note that the following sub-blocks are required for adding pointwise confidence
intervals to mean estimates:
\begin{equation}
\begin{array}{c}
\Cov(\bbetahat),\quad\Cov(\buhat_{i}-\bu_{i}),\quad\Cov(\buhat^{\prime}_{\idash}-\udashidash),\\[2ex]
E\{\bbetahat(\buhat_{i}-\bu_{i})^T\},\quad E\{\bbetahat(\buhat^{\prime}_{\idash}-\udashidash)^T\}
\quad\mbox{and}\quad E\{(\buhat_{i}-\bu_{i})(\buhat^{\prime}_{\idash}-\udashidash)^T\}
\end{array}
\label{eq:crossedCovSubBlocks}
\end{equation}
for $1\le i\le m$ and $1\le \idash\le \mdash$.

\begin{result} 
Computation of $[\bbetahat^T\ \ \buhat^T]^T$ and each
of the sub-blocks of $\mbox{Cov}([\bbetahat\ \ \buhat-\bu]^T)$
listed in (\ref{eq:crossedCovSubBlocks}) are expressible as the two-level sparse
matrix least squares form:
$$\left\Vert\bb-\bB\left[
\begin{array}{c}
\bbeta\\
\bu
\end{array}
\right]
\right\Vert^2
$$
where $\bb$ and the non-zero sub-blocks of $\bB$, according to the notation
in (\ref{eq:BandbFormsReprise}), are, for $1\le i\le m$,
$$
\bb_{i} \equiv \left[
  \begin{array}{c}
    \sigma^{-1} \by_{i} \\[1ex]
    \bzero \\[1ex]
    \bzero
  \end{array}
\right],
\quad\bB_{i}\equiv\left[\begin{array}{cc}
\sigma^{-1}\bX_{i} & \sigma^{-1}\Zdash_{i} \\[1ex]
\bO & m^{-1/2}(\bI_{\mdash} \otimes (\Sigmad)^{-1/2})\\[1ex]
\bO & \bO
\end{array}
\right]
\quad\mbox{and}\quad
\bBdot_{i} \equiv \left[
  \begin{array}{c}
    \sigma^{-1} \bZ_{i} \\[1ex]
    \bO \\[1ex]
    \bSigma^{-1/2}
  \end{array}
\right].
$$
Each of these matrices has $\mdash (n_{i \idash} + q^{\prime}) + q$ rows. The $\bB_{i}$ matrices each 
have $p + \mdash q^{\prime}$ columns and the $\bBdot_{i}$ each have $q$ columns. 
The solutions are
\begin{equation*}
  \begin{array}{c}
    \bbetahat = \mbox{ first $p$ rows of }\xveco,
    \ \
    \Cov(\bbetahat) = \mbox{ top left $p \times p$ sub-block of }\AUoo,
    \\[2ex]
    {\displaystyle\stack{1\le \idash\le \mdash}}\big(\buhat^{\prime}_{\idash}\big)
     = \mbox{ subsequent $(m'q^{\prime})\times 1$ entries of } \bx_{1} \mbox{ following $\bbetahat$},
    \\[2ex]
    E \{ \bbetahat( \buhat^{\prime}_{\idash} - \udashidash)^T\} =
    \mbox{ subsequent $p \times q^{\prime}$ sub-blocks of $\AUoo$ to the right of $\Cov(\bbetahat)$},
    \\[2ex]
    \Cov(\buhat^{\prime}_{\idash}-\udashidash)=\mbox{ subsequent $q^{\prime} \times q^{\prime}$
    diagonal sub-blocks of $\AUoo$ following $\Cov(\bbetahat)$},\ \ 1\le \idash \le \mdash,
    \\[2ex]
    \buhat_{i}= \xvectCi, \ \ \Cov (\buhat_{i} - \bu_{i}) =
    \bA^{22,i}, \ \ E\{\bbetahat(\buhat_{i} - \bu_{i})^T\}= \mbox{ first $p$ rows of $\bA^{12,i}$}
    \\[2ex]
    {\displaystyle\stack{1\le \idash\le \mdash}}
    \big[ E\{(\buhat_i-\bu_i)(\buhat^{\prime}_{\idash} - \udashidash)^T\} \big]= \mbox{ remaining $\mdash q^{\prime}$
    rows of $\bA^{12,i}$},\ \ 1\le i \le m,\\
  \end{array}
\end{equation*}
where the $\xveco$, $\xvectCi$, $\AUoo$, $\AUttCi$ and $\AUotCi$ notation
is given by (\ref{eq:AtLevInv}).
\label{res:BLUP}
\end{result}

Algorithm \ref{alg:crossedBLUPAlg} proceduralizes Result \ref{res:BLUP} to facilitate
computation of best linear unbiased predictors for the 
fixed and random effects parameters in (\ref{eq:crossedModel}) for 
fixed values of the covariance parameters. In practice, the covariance parameters
would need to be replaced by estimates obtained using an approach such as
restricted maximum likelihood. Algorithm \ref{alg:crossedBLUPAlg} also delivers
the matrices in (\ref{eq:crossedCovSubBlocks}). In the case where $m'$ is 
moderate but $m$ is potentially very large Algorithm \ref{alg:crossedBLUPAlg} 
performs efficient streamlined computing.

\begin{algorithm}[!th]
\begin{center}
\begin{minipage}[t]{154mm}
\begin{small}
\begin{itemize}
\setlength\itemsep{4pt}
\item[] Data Inputs: $\Big\{\left(\yuptri_i,\Xuptri_i,\Zuptri_i,\Zdblacksquare_i\right):\ 1\le i \le m\Big\}$\\[1ex]
        Covariance Matrix Inputs: $\sigma^{2} > 0,\ \ \Sigmad (q^{\prime} \times q^{\prime}), \ \ \bSigma (q \times q),
                  \mbox{ symmetric and positive definite.}$
\item[] For $i = 1, \ldots, m$:
\begin{itemize}
\setlength\itemsep{4pt}
\item[] $\bb_{i}\thickarrow\left[\begin{array}{c}
  \sigma^{-1}\by_{i}\\[1ex]
  \bzero \\[1ex]
  \bzero
  \end{array}
  \right], \
  \bB_{i}\thickarrow\left[\begin{array}{cc}
  \sigma^{-1}\bX_{i} & \sigma^{-1} \Zdash_{i} \\[1ex]
  \bO & m^{-1/2}(\bI_{\mdash} \otimes (\Sigmad)^{-1/2}) \\[1ex]
  \bO & \bO
  \end{array}
  \right],$
\item[] $\bBdot_{i}\thickarrow
  \left[\begin{array}{c}
  \sigma^{-1}\bZ_{i}\\[1ex]
  \bO \\[1ex]
  \bSigma^{-1/2}
  \end{array}
  \right]$
\end{itemize}
\item[] $\Ssc\thickarrow\SolveTwoLevelSparseLeastSquares
\Big(\big\{(\bveci,\Bmati,\Bmatdoti):1\le i \le m\big\}\Big)$
\item[] $\bbetahat\thickarrow\mbox{first $p$-rows of $\xveco$ component of $\Ssc$}$
\item[] $\Cov(\bbetahat)\thickarrow\mbox{top left $p \times p$ sub-block of $\AUoo$ component of $\Ssc$}$
\item[] $\iStt \thickarrow p + 1$
\item[] For $\idash = 1, \ldots, \mdash$:
\begin{itemize}
\setlength\itemsep{4pt}
\item[] $\iEnd \thickarrow \iStt + q^{\prime} - 1$
\item[] $\buhat_{\idash} \thickarrow \mbox{ sub-vector of $\bx_{1}$ component of $\Ssc$ with entries $\iStt$ to $\iEnd$}$
\item[] $\Cov(\buhat^{\prime}_{\idash} - \udashidash) \thickarrow \mbox{ diagonal sub-block of $\AUoo$ component
         of $\Ssc$ with rows }$
\item[] $ \hspace{28.5mm} \mbox{ $\iStt$ to $\iEnd$ and columns $\iStt$ to $\iEnd$}$
\item[] $E\{ \bbetahat (\buhat^{\prime}_{\idash} - \udashidash)^{T} \} \thickarrow \mbox{ sub-block of $\AUoo$
         component of $\Ssc$ with rows $1$ to $p$ and }$
\item[] $\hspace{33mm} \mbox{ columns $\iStt$ to $\iEnd$}$
\item[] $\iStt \thickarrow \iEnd + 1$
\end{itemize}
\item[] For $i=1,\ldots,m$:
\begin{itemize}
\setlength\itemsep{4pt}
\item[] $\buhat_{i}\thickarrow\mbox{$\bx_{2, i}$ component of $\Ssc$}$\ \ \ ;\ \ \
        $\Cov(\buhat_{i}-\bu_{i})\thickarrow\mbox{$\bA^{22,i}$ component of $\Ssc$}$
\item[] $E\{\bbetahat(\buhat_{i}-\bu_{i})^T\}\thickarrow
         \mbox{ sub-matrix of $\bA^{12, i}$ component of $\Ssc$ with rows $1$ to $p$}$
\item[] $\iStt \thickarrow p + 1$ 
\item[] For $\idash=1,\ldots,\mdash$:
\begin{itemize}
\setlength\itemsep{4pt}
\item[] $\iEnd \thickarrow \iStt + q' - 1$\ \ \ ;\ \ \ $\bOmega\thickarrow \mbox{$\bA^{12, i}$ component of $\Ssc$}$
\item[] $E\{(\buhat_{i} - \bu_{i})(\buhat^{\prime}_{\idash}-\udashidash)^T\} \thickarrow
         \mbox{sub-matrix of $\bOmega^T$ with columns $\iStt$ to $\iEnd$}$
\item[] $\iStt \thickarrow \iEnd + 1$
\end{itemize}
\end{itemize}
\item[] Outputs: $\bbetahat,\,\Cov(\bbetahat),
  \left\{(\buhat^{\prime}_{\idash},\,E\{\bbetahat(\buhat^{\prime}_{\idash}-\udashidash)^T\},
  \,\Cov(\buhat^{\prime}_{\idash}-\udashidash)):\ 1 \le \idash \le \mdash, \right. $
\item[] $\left. \quad \quad \quad \quad \left( \buhat_{i}, E\{\bbetahat(\buhat_{i}-\bu_{i})^T\},
   E\{(\buhat_{i}-\bu_{i})(\buhat^{\prime}_{\idash}-\udashidash)^T\},
   \Cov(\buhat_{i}-\bu_{i}) \right): 1 \le \idash \le \mdash, \right. $
\item[] $\left. \quad \quad \quad \quad 1 \le i \le m \right\}$
\end{itemize}
\end{small}
\end{minipage}
\end{center}
\caption{\it Streamlined algorithm for obtaining best linear unbiased predictions
and corresponding covariance matrix components for the linear mixed model with
crossed random effects.}
\label{alg:crossedBLUPAlg}
\end{algorithm}
%
%

\null\vfill\eject
\null\vfill\eject

\section{Full List of Items in the National Education Longitudinal Study}\label{eq:NELSitems}

Table \ref{tab:itemDefinitions} lists each of the 24 items within
the National Education Longitudinal Study data set used in 
Section \ref{sec:dataApplic}. Several of the measurements involve 
item response theory, which is abbreviated as IRT.

\begin{table}[!ht]
\centering
\begin{tabular}{rl}
\hline\\[-1.5ex]
item   & description   \\[0.9ex]
\hline\\[-0.5ex]
1   & \mbox{reading IRT-estimated number right}\\[0ex]
2   & \mbox{mathematics IRT-estimated number right}\\[0ex]
3   & \mbox{science IRT-estimated number right}\\[0ex]
4   & \mbox{history/citizenship/geography IRT-estimated number right}\\[0ex]
5   & \mbox{reading standardized score}\\[0ex]
6   & \mbox{mathematics standardized score}\\[0ex]
7   & \mbox{science standardized score}\\[0ex]
8   & \mbox{history/citizenship/geography standardized score}\\[0ex]
9   & \mbox{reading IRT estimate of ability}\\[0ex]
10  & \mbox{mathematics IRT estimate of ability}\\[0ex]
11  & \mbox{science IRT estimate of ability}\\[0ex]
12  & \mbox{history/citizenship/geography IRT estimate of ability}\\[0ex]
13  & \mbox{standardized test composite (reading, mathematics)}\\[0ex]
14  & \mbox{reading level 1: probability of proficiency}\\[0ex]
15  & \mbox{reading level 2: probability of proficiency}\\[0ex]
16  & \mbox{reading level 3: probability of proficiency}\\[0ex]
17  & \mbox{mathematics level 1: probability of proficiency}\\[0ex]
18  & \mbox{mathematics level 2: probability of proficiency}\\[0ex]
19  & \mbox{mathematics level 3: probability of proficiency}\\[0ex]
20  & \mbox{mathematics level 4: probability of proficiency}\\[0ex]
21  & \mbox{science level 1: probability of proficiency}\\[0ex]
22  & \mbox{science level 2: probability of proficiency}\\[0ex]
23  & \mbox{science level 3: probability of proficiency}\\[0ex]
24  & \mbox{science level 4: probability of proficiency}\\[1ex]
\hline
\end{tabular}
\caption{\it Descriptions of each of the 24 items in the National Education Longitudinal Study
data used in Section \ref{sec:dataApplic}. The abbreviation IRT stands for item response
theory. Fuller details are provided by Thurgood \textit{et al.} (2003).}
\label{tab:itemDefinitions} 
\end{table}

%
%
%
%
%
%
%
%
%

\end{document}